\documentclass[fleqn,usenatbib]{mnras}

\usepackage[T1]{fontenc}
\usepackage{ae,aecompl}

\usepackage{CJK}
\usepackage{multirow}
\usepackage{amssymb}
\usepackage{amsmath,amstext}
\usepackage{multirow}
\usepackage{booktabs}
\usepackage{graphicx}
\usepackage[caption=false]{subfig}
\usepackage{array}
\usepackage{threeparttable}
\usepackage{url}
\usepackage{hyperref}
\usepackage{newtxtext}
\usepackage[varvw]{newtxmath}

\renewcommand{\d}[1]{\ensuremath{\operatorname{d}\!{#1}}}

\newcommand{\lya}{\ensuremath{\rm Ly\alpha}}

\newcommand{\zla}{\ensuremath{z_{\rm Ly\alpha}}}
\newcommand{\zis}{\ensuremath{z_{\rm IS}}}
\newcommand{\zsys}{\ensuremath{z_{\rm sys}}}

\newcommand{\zfg}{\ensuremath{{z_{\rm fg}}}}
\newcommand{\zbg}{\ensuremath{{z_{\rm bg}}}}
\newcommand{\zneb}{\ensuremath{{z_{\rm neb}}}}

\newcommand{\kms}{\rm km~s\ensuremath{^{-1}\,}}

\newcommand{\dzfb}{\ensuremath{\Delta z_{\rm fb}}}

\newcommand{\lyb}{Ly$\beta$}

\newcommand{\nhi}{\ensuremath{N_{\rm HI}}}

\newcommand{\wlya}{\ensuremath{W_{\lambda}(\lya)}}

\newcommand{\minpoint}{\mbox{$'\mskip-4.7mu.\mskip0.8mu$}}
\newcommand{\secpoint}{\mbox{$''\mskip-7.6mu.\,$}}

\def\ltsima{$\; \buildrel < \over \sim \;$}
\def\simlt{\lower.5ex\hbox{\ltsima}}
\def\gtsima{$\; \buildrel > \over \sim \;$}
\def\simgt{\lower.5ex\hbox{\gtsima}}

\newcommand{\RNum}[1]{\uppercase\expandafter{\romannumeral #1\relax}}
\newcommand{\rnum}[1]{\lowercase\expandafter{\romannumeral #1\relax}}

\title[KBSS: Structure and Kinematics of \ion{H}{I}]{ The Keck Baryonic Structure Survey: \\ Using foreground/background galaxy pairs to trace 
the structure and kinematics 
of circumgalactic neutral hydrogen at $z \sim 2$} %

\date{Accepted 2020 September 10. Received 2020 August 31; in original form 2020 April 30}

\pubyear{2020}

\hypersetup{draft}
\begin{document}
\begin{CJK*}{UTF8}{gbsn}

\author[Y. Chen et al.]{
Yuguang Chen (陈昱光),$^{1}$\thanks{Email: yuguangchen@astro.caltech.edu}
Charles C. Steidel,$^{1}$
Cameron B. Hummels,$^{1}$ \newauthor
Gwen C. Rudie,$^{2}$ 
Bili Dong (董比立),$^{3}$
Ryan F. Trainor,$^{4}$
Milan Bogosavljevi\'{c},$^{5}$ \newauthor
Dawn K. Erb,$^{6}$
Max Pettini,$^{7, 8}$
Naveen A. Reddy,$^{9}$
Alice E. Shapley,$^{10}$ \newauthor
Allison L. Strom,$^{2}$ 
Rachel L. Theios,$^{1}$
Claude-Andr\'{e} Faucher-Gigu\`{e}re,$^{11}$ \newauthor
Philip F. Hopkins,$^{1}$
and Du\v{s}an Kere\v{s}$^{3}$
\\
$^{1}$Cahill Center for Astronomy and Astrophysics, California Institute of Technology, 
MC249-17, 
Pasadena, CA 91125, USA\\
$^{2}$The Observatories of the Carnegie Institution for Science, 
813 Santa Barbara Street, 
Pasadena, CA 91101, USA\\
$^{3}$Department of Physics, Center for Astrophysics and Space Sciences, University of California at San Diego, 9500 Gilman Drive,\\La Jolla, CA 92093, USA\\
$^{4}$Department of Physics and Astronomy, Franklin \& Marshall College, 
637 College Ave., 
Lancaster, PA 17603, USA\\
$^{5}$Division of Science, New York University Abu Dhabi, P.O. Box 129188, Abu Dhabi, UAE\\
$^{6}$The Leonard E. Parker Center for Gravitation, Cosmology and Astrophysics, Department of Physics, University of Wisconsin-\\Milwaukee, 3135 North Maryland Avenue, Milwaukee, WI 53211, USA\\
$^{7}$Institute of Astronomy, University of Cambridge, Madingley Road, Cambridge CB3 0HA, UK\\
$^{8}$Kavli Institute for Cosmology, University of Cambridge, Madingley Road, Cambridge CB3 0HA, UK\\
$^{9}$Department of Physics and Astronomy, University of California, Riverside, 900 University Avenue, Riverside, CA 92521, USA\\
$^{10}$Department of Physics and Astronomy, University of California, Los Angeles, 430 Portola Plaza, Los Angeles, CA 90095, USA\\
$^{11}$Department of Physics and Astronomy and Center for Interdisciplinary Exploration and Research in Astrophysics (CIERA),\\Northwestern University, 2145 Sheridan Road, Evanston, IL 60208, USA\\
}
\label{firstpage}
\pagerange{\pageref{firstpage}--\pageref{lastpage}}
\maketitle
\begin{abstract}
We present new measurements of the spatial distribution and kinematics of neutral hydrogen in the circumgalactic and intergalactic medium surrounding star-forming galaxies at $z\sim 2$. 
Using the spectra of $\simeq 3000$ galaxies with redshifts $\langle z \rangle =2.3\pm0.4$ from the Keck Baryonic Structure Survey,
we assemble a sample of more than 200,000 distinct foreground-background pairs with projected angular separations of $3\arcsec-500\arcsec$
and spectroscopic redshifts, with $\langle z_{\rm fg}\rangle = 2.23$ and $\langle z_{\rm bg} \rangle= 2.57$ (foreground, background redshifts, respectively.) 
The ensemble of sightlines and foreground galaxies is used to construct 
a 2-D map of the mean excess \ion{H}{I} \lya\ optical depth relative to the intergalactic mean
as a function of projected galactocentric distance ($20 \simlt D_{\rm tran}/{\rm pkpc} \simlt 4000$) 
and line-of-sight velocity. 
We obtain accurate galaxy systemic redshifts, providing significant information on the line-of-sight kinematics of \ion{H}{I} gas as a function of projected distance $D_{\rm tran}$.
We compare the map with cosmological zoom-in simulation, finding qualitative agreement between them.  A simple two-component (accretion, outflow) 
analytical model 
generally reproduces the observed line-of-sight kinematics and projected spatial distribution of \ion{H}{I}. The best-fitting model
suggests that galaxy-scale outflows with initial velocity $v_{\rm out} \simeq 600$ \kms\ dominate the kinematics of circumgalactic \ion{H}{I}  
out to $D_{\rm tran} \simeq 50$ kpc, while \ion{H}{I} at $D_{\rm tran} \gtrsim 100$ kpc is dominated by infall with characteristic $v_{\rm in} \simlt$ circular velocity.
Over the impact parameter range  $80 \simlt D_{\rm tran}/{\rm pkpc} \simlt 200$, the \ion{H}{I} line-of-sight velocity range reaches a minimum, with 
a corresponding flattening in the rest-frame \lya\ equivalent width. 
These observations can be naturally explained as the transition between outflow-dominated and accretion-dominated flows. 
Beyond $D_{\rm tran} \simeq 300$ pkpc ($\sim 1$ cMpc), 
the line of sight kinematics are dominated by Hubble expansion.  
\end{abstract}


\section{Introduction}

\label{sec:intro}

Galaxy formation involves a continuous competition between gas cooling and accretion on the one hand, and
feedback-driven heating and/or mass outflows on the other. The outcome of this competition, as a function of 
time, controls 
nearly all observable properties
of galaxies: e.g., the star-formation rate, the fraction of galactic baryons converted to stars over
the galaxy lifetime, and  
the fraction of baryons that remain bound to the galaxy. This competition eventually halts star formation and the growth of 
supermassive black hole mass. The exchange of gaseous baryons between the diffuse intergalactic medium (IGM) and 
the central regions of galaxies (the interstellar medium; ISM) involves an intermediate baryonic reservoir that has come to be
called the ``circumgalactic medium'' (CGM) (e.g., \citealt{steidel2010,rudie12b,tumlinson2017}.)    

Although there is not yet a consensus, one possible working definition of the CGM is the region containing gas
that is outside of the interstellar medium of a galaxy, but that is close enough
that the physics and chemistry of the gas and that of the central galaxy are causally
connected. 
For example, the CGM may be 1) the baryonic reservoir that supplies gas, via accretion, to the central regions of the galaxy, providing fuel
for star formation and black hole growth; 2) the CGM may also consist of gas that has already been part of the ISM at some point in the past, but has since
been dispersed or ejected to large galactocentric radii; or 3) the physical state of the gas can be otherwise affected
by energetic processes (mechanical or radiative) originating in the galaxy's central regions, e.g., 
via galactic winds, radiation pressure, ionization, etc. Therefore, the CGM represents a galaxy's evolving ``sphere of influence''. 

Since being postulated by \citet{bahcall+spitzer69} more than 50 years ago, 
evidence for extended ($\sim 100$ pkpc) halos of highly-ionized, metal-enriched gas around galaxies has continuously accumulated. 
In recent years, there has been increasing attention given to 
understanding the physics and
chemistry of CGM gas as a function of galaxy properties, e.g., environment \citep{johnson2015, burchett2016, nielsen2018}, mass and star-formation rate \citep{adelberger05,chen2010, tumlinson2011,rakic12,johnson2017, rubin2018}, and cosmic epoch \citep{nelson19,hafen2019,hummels19}. 
In large part, the increased focus on the CGM is attributable to a growing appreciation that diffuse gas outside of galaxies
is a laboratory where many of the most important, but poorly understood, baryonic processes can be observed and tested.

Redshifts near the peak of cosmic star formation history, at $z \simeq 2-3$ \citep{madau14}, are especially attractive for observations
of galaxies and their associated diffuse CGM/IGM gas, due to the  accessibility of spectroscopic diagnostics 
in the rest-frame far-UV (observed optical) and rest-frame optical (observed near-IR) using large ground-based telescopes (see, e.g., \citealt{steidel14}). 
The most sensitive measurements of neutral hydrogen and metals in diffuse gas in the outer parts of galaxies along the line of sight require high-resolution (FWHM $\simlt 10$ \kms), 
high signal-to-noise ratio (SNR) of bright background continuum sources -- i.e., quasi-stellar objects (QSOs). 
However, QSOs bright enough to be observed in this
way are extremely rare, thereby limiting the number of galaxies whose CGM can be probed. Moreover, each sightline to a suitable background
QSO provides at most a single sample, at a single galactocentric distance, for any identified foreground galaxy. This inefficiency
make the assembly of a statistical picture of the CGM/IGM around galaxies at a particular redshift, or having particular properties, very
challenging.

Improved efficiency for such QSO sightline surveys can be realized by conducting deep galaxy surveys in regions of the sky {\it selected} to include the lines of sight
to one or more background QSOs, with emission redshifts chosen to optimise the information content of absorption lines in the QSO spectrum given
the galaxy redshift range targeted by the survey (e.g., \citealt{lanzetta95,chen2001,adelberger03,adelberger05,morris06,prochaska11b,crighton11}.) 
The Keck Baryonic Structure Survey (KBSS\footnote{The complete spectroscopic catalogs of the galaxies used in this paper and the processed data can
be found at the KBSS website: \href{http://ramekin.caltech.edu/KBSS}{http://ramekin.caltech.edu/KBSS}.}; \citealt{rudie12a,steidel14,strom17}) was designed along these lines, specifically 
to provide a densely-sampled spectroscopic survey
of star-forming galaxies in the primary redshift range $1.9 \simlt z_{\rm gal} \simlt 2.7$ in 15 survey regions, each of which is
centered around the line of sight
to a very bright QSO with $z \sim 2.7-2.8$.  The Keck/HIRES spectra of the QSOs, together with the positions and redshifts of the galaxies
in each survey region, have been analysed in detail to measure neutral hydrogen (\ion{H}{I}) and metals associated with the foreground
galaxies. Absorption has been measured  as a function of projected galactocentric distance to the QSO sightline and as a function
of line-of-sight velocity with 
respect to the galaxy systemic redshift, using both Voigt profile fitting \citep{rudie12a,rudie13,rudie2019} and ``pixel optical depth''
techniques \citep{rakic12,rakic13,turner14,turner15}. These studies have shown that there is \ion{H}{I} and \ion{C}{IV} significantly in excess of the 
intergalactic mean extending to at least 2.5 physical Mpc around identified galaxies, but with the most prominent excess of both
\ion{H}{I} and metals lying within $D_{\rm tran} \sim 200-300$ pkpc and $\Delta v_{\rm LOS} \simlt 300-700$ \kms. The statistical inferences
were based on $\sim 900$ QSO/galaxy pairs with projected separation $D_{\rm tran} < 3$ Mpc, but only (90,26,10) sample the CGM within
$D_{\rm tran} \le (500,200,100)$ pkpc. Thus, in spite of the large observational effort behind KBSS, 
the statistics of diffuse gas surrounding $z \simeq 2-2.7$ galaxies
is limited to relatively small samples within the inner CGM.  

Alternatively, as shown by \citet{steidel2010} (S2010; see also \citealt{adelberger05}), it is also possible to use 
the grid of background {\it galaxies} -- which comes ``for free'' with a densely sampled spectroscopic survey -- 
to vastly increase the number of lines of sight sampling the CGM of foreground galaxies, particularly for 
small transverse distances (or impact parameter, $D_{\rm tran} \simlt 500$ pkpc.) The penalty for increased spatial
sampling is, unavoidably, the vastly reduced spectral resolution and SNR -- and the associated loss of the ability to resolve 
individual components and measure column densities along individual sightlines, compared to 
the HIRES QSO spectra.  
S2010 used a set of $\sim 500$ galaxy foreground/background
angular pairs with separation $\theta \le 15\arcsec$ to trace the rest-frame equivalent width of \lya\ and several strong metal lines as
a function of impact parameter over the range $20 \le D_{\rm tran}/{\rm pkpc}  \le 125$ at $\langle z \rangle = 2.2$. 
In this paper, we extend the methods of S2010, with significant improvements in both the size and quality of the galaxy sample, 
to characterize \ion{H}{I} absorption over the full range of $20-4000$ pkpc. Compared to the earlier KBSS QSO/galaxy pairs, the new
galaxy/galaxy analysis includes $\sim 3000$ galaxies, with a factor $> 100$ increase in the number of sightlines sampled with $D_{\rm tran} \le 500$ pkpc. 

As discussed by S2010, background galaxies are spatially extended\footnote{Typical galaxies in 
the spectroscopic sample have physical sizes of $d \simeq 4$ kpc. The diameter of the beam as it traverses a galaxy with $z_{\rm fg} \simeq z_{\rm bg} - 0.3$ would 
have a similar physical extent.},  unlike QSOs, 
and thus each absorption line probe is in effect averaging over a spatially extended line of sight through the circumgalactic gas associated with
the foreground galaxies.  CGM gas is known to be clumpy, with indications that the degree of ``clumpiness'' (i.e., the size scale on which
significant variations of the ionic column density are observed) depends on ionization level, with low-ionization species having smaller 
coherence scales (see \citealt{rauch99, rudie2019}). In general, this means that the strength of an absorption feature produced by gas in a foreground galaxy
as recorded in the spectrum of a background galaxy will depend on three factors: the fraction of the beam covered by a 
significant column of the species, the column density in the beam, and the range of line-of-sight velocity ($v_{\rm LOS}$) 
sampled by the roughly cylindrical volume through the CGM. The dynamic range in total \ion{H}{I} column density measurable using
stacks of background galaxy spectra is much smaller (and less quantitative) than could be measured from high-resolution, high SNR QSO spectra. 
However, using galaxy-galaxy pairs provides much more rapid convergence to the mean CGM absorption as a function of impact parameter, where
samples of QSO-galaxy pairs would be limited by sample variance. This improvement -- along with the larger sample size -- allows us to probe more details in the kinematics and the spatial distribution of \ion{H}{I} compared to that obtained from QSO sightlines (e.g. \citealt{rudie12a, turner14, tummuangpak14, bielby17}; and \citealt{ryanweber06,tejos14} at lower redshifts).

This paper is organized as follows. In \S\ref{sec:sample}, we describe the KBSS galaxy spectroscopic sample and the steps used in the analysis; \S\ref{sec:obs} presents
the principal results of the analysis. We discuss the implications of the results in \S\ref{sec:discussion}. Particularly, in \S\ref{sec:discussion_zoomin}, we compare the results with cosmological zoom-in simulations, and in \S\ref{sec:model}, we develop a simple analytic model
to describe the 2-D spatial and kinematic distribution of \ion{H}{I} on scales $0.020 - 4.0$ pMpc ($\simeq 0.06-12.0$ cMpc) surrounding typical star-forming galaxies
at $z \sim 2$. We summarize our conclusions in \S\ref{sec:summary}.

Unless stated otherwise, throughout the paper we assume a $\Lambda$CDM cosmology with $\Omega_{\rm m} = 0.3$, $\Omega_\Lambda = 0.7$, and $h=0.7$. 
Units of distance are generally given in terms of physical kpc (pkpc) or physical Mpc (pMpc). 

\section{Sample and Analysis} \label{sec:sample}

\begin{table*}
\centering
\caption{  Field-by-Field Summary of the Properties of the KBSS Galaxy Pair Sample.\label{tab:summary}}
\begin{threeparttable}
\begin{tabular}{lcccccll}
\hline \hline
Field & RA\tnote{a} & DEC\tnote{a} & Area\tnote{b} & $N_\mathrm{gal}$ & $N_\mathrm{pair}$\tnote{c} & $\langle z_\mathrm{fg} \rangle$ & $\Delta z_\mathrm{fb} / (1+z_\mathrm{fg})$ \\
Name & (J2000.0) & (J2000.0) & (arcmin$^2$) & ($z>1.9$) & (Full/$z_\mathrm{neb}$)\tnote{d} & (Full/$z_\mathrm{neb}$)\tnote{d} & (Full/$z_\mathrm{neb}$)\tnote{d} \\\hline
Q0100 & 01:03:11 & +13:16:27 & $7.6\times 5.6$ & 153 & 762/441 & 2.11/2.11 & 0.108/0.111\\
Q0105 & 01:08:08 & +16:35:30 & $7.4\times 5.3$ & 137 & 519/296 & 2.15/2.12 & 0.103/0.108\\
Q0142 & 01:45:15 & -09:45:30 & $7.2\times 5.2$ & 131 & 572/320 & 2.21/2.22 & 0.113/0.116\\
Q0207 & 02:09:52 & -00:05:22 & $7.0\times 5.4$ & 133 & 480/296 & 2.15/2.15 & 0.130/0.140\\
Q0449 & 04:52:14 & -16:40:29 & $6.5\times 5.0$ & 128 & 561/297 & 2.28/2.23 & 0.108/0.105\\
Q0821 & 08:21:05 & +31:07:42 & $7.3\times 5.5$ & 124 & 413/242 & 2.35/2.35 & 0.112/0.113\\
Q1009 & 10:11:55 & +29:41:36 & $7.2\times 5.2$ & 141 & 552/325 & 2.44/2.31 & 0.125/0.129\\
Q1217 & 12:19:33 & +49:40:46 & $6.9\times 5.1$ & ~93 & 242/90~ & 2.19/2.19 & 0.101/0.109\\
GOODS-N\tnote{e} & 12:36:52 & +62:14:20 & $14.3\times 10.4$ & 249 & 590/209 & 2.27/2.31 & 0.108/0.142\\
Q1307 & 13:07:54 & +29:22:24 & $10.0\times 11.0$ & ~71 & 93/--- & 2.13/--- & 0.113/---\\
GWS & 14:17:47 & +52:28:49 & $15.1\times 14.8$ & 228 & 270/12~ & 2.82/2.92 & 0.081/0.084\\
Q1442 & 14:44:54 & +29:19:00 & $7.3\times 5.1$ & 137 & 613/373 & 2.29/2.34 & 0.124/0.122\\
Q1549 & 15:51:55 & +19:10:53 & $7.1\times 5.2$ & 144 & 605/297 & 2.36/2.29 & 0.111/0.131\\
Q1603 & 16:04:57 & +38:11:50 & $7.2\times 5.4$ & 112 & 354/166 & 2.27/2.28 & 0.083/0.080\\
Q1623 & 16:25:52 & +26:47:58 & $16.1\times 11.6$ & 284 & 781/239 & 2.18/2.24 & 0.108/0.104\\
Q1700 & 17:01:06 & +64:12:02 & $11.5\times 11.0$ & 210 & 585/337 & 2.29/2.29 & 0.103/0.103\\
Q2206 & 22:08:54 & -19:43:35 & $7.5\times 5.4$ & 119 & 457/183 & 2.15/2.16 & 0.113/0.116\\
Q2343 & 23:46:20 & +12:47:28 & $11.5\times 6.3$ & 224 & 859/610 & 2.17/2.17 & 0.096/0.103\\
Q2346 & 23:48:31 & +00:22:42 & $11.8\times 10.3$ & ~44 & 43/7~ & 2.07/2.03 & 0.067/0.070\\\hline
All & & & 1447 & 2862 & 9351/4741 & 2.23/2.22 & 0.106/0.113  \\
\hline\addlinespace[1ex]
\end{tabular}
\begin{tablenotes}\footnotesize 
\item[a]{Mean coordinates for galaxies with spectroscopic redshifts $z>1.9$. }
\item[b]{Angular size of field over which spectroscopy was performed; (long axis $\times$ short axis, both in arc minutes).}
\item[c]{The number of distinct foreground/background galaxy pairs with $D_\mathrm{tran}<500 \mathrm{~pkpc}$. Numbers for other ranges of $D_\mathrm{tran}$ can be estimated based on Figure \ref{fig:pairnum}.}
\item[d]{The KGPS-Full sample and the KGPS-$z_\mathrm{neb}$ subsample. See \S\ref{sec:redshift} for definitions.}
\item[e]{Full catalog published in \cite{reddy06}; the majority of nebular redshifts in the 
GOODS-N region are as reported by the MOSDEF survey (\citealt{kriek15}).}
\end{tablenotes}
\end{threeparttable}
\end{table*}

Table \ref{tab:summary} provides a summary of the KBSS galaxy pairs sample, described in more detail in the remainder of this section. 

The KBSS galaxy pairs sample (hereafter KGPS) is drawn from 2862 galaxies in 19 densely sampled survey regions (Table~\ref{tab:summary}), 
of which 15 comprise the nominal KBSS survey \citep{rudie12a,steidel14}
of bright QSO sightlines. 
KGPS includes 4 additional fields (GOODS-N, Q1307, GWS, and Q2346) observed using the same selection criteria 
and instrumental configurations as the KBSS fields, and thus
have a similar redshift selection function and similarly-dense spectroscopic sampling. 
GWS and GOODS-N\footnote{Referred to as ``Westphal'' and ``HDF-N'', respectively, by \citet{steidel03}.} were 
observed as part of a Lyman break galaxy (LBG) survey targeting primarily the redshift range $2.7 \simlt z \simlt 3.4$ 
\citep{steidel03}, but were subsequently supplemented by observations favoring the slightly lower redshift range $1.9 \simlt z \simlt 2.7$ 
selected using a different set of rest-UV color criteria. For all 19 fields, two groups of photometric pre-selection of
candidates are included: at $z \sim 3.0\pm0.4$, using the MD, C, D, and M criteria described by \citet{steidel03}; and at $z\sim 2.3\pm0.4$, using
the BX and BM criteria described by \citet{adelberger04,steidel04}, as well as the RK criteria from \citet{strom17}. The limiting apparent magnitude of the photometric selection is ${\cal R} \le 25.5$ (AB). The galaxies were observed spectroscopically
over the period 2002-2016, with the goal of achieving the densest-possible sampling of galaxies in the redshift range $2 \simlt z \simlt 3$. 

The survey regions listed in Table~\ref{tab:summary} are identical to those included in the analysis of galaxy-galaxy pairs 
by \citet{steidel2010}; however the current spectroscopic catalog 
is larger by $\sim 30$\% in terms of the number of galaxies with spectroscopic redshifts in the most useful range ($1.9 \simlt z \simlt 3.0$), increasing
the number of pairs sampling angular scales of interest by $\simgt 70$\%. More importantly, as
detailed in \S\ref{sec:redshift} below, $\sim 50$\% of the foreground galaxies in KGPS pairs have precisely-measured systemic redshifts ($z_\mathrm{neb}$)
from nebular emission lines observed in the near-IR, compared with only a handful available for the S2010 analysis.  

We use the substantial subset of galaxies with nebular emission line measurements to improve the calibrations of systemic redshifts inferred from
measurements of spectral features in the rest-frame UV (observed frame optical) spectra (\S\ref{sec:redshift}).  
The much improved redshift precision and accuracy\footnote{The number of galaxies with insecure or incorrect redshifts is also greatly reduced compared
to S2010.} -- as well as a more careful construction of composite (stacked) spectra (\S\ref{sec:stacking}) -- 
allow us to extend the technique using galaxy foreground-background pairs to angular separations far beyond the $\theta = 15$ arcsec ($D_{\rm tran} \simeq 125$ pkpc)  
used by \citet{steidel2010}. 
The various improvements represented by KGPS significantly increase the 
sampling density and S/N ratio (SNR) of the \ion{H}{I} absorption measurements as a function of impact parameter ($D_\mathrm{tran}$). 
As we show in the next section, this leads to a major improvement compared to S2010, 
allowing us to resolve and model details of the kinematic structure of the \ion{H}{I} 
with respect to the galaxies. 
   
Subsets of the KGPS sample have figured prominently in many previous investigations involving galaxies and the CGM/IGM at $1.9 \simlt z \simlt 3.5$. In what follows
below, we direct the reader to the most relevant references for more information on some of the details.  All of the rest-UV spectra were obtained 
using the Low Resolution Imaging Spectrograph (LRIS; \citealt{oke95}) on the Keck I telescope; the vast majority were obtained after June 2002, 
when LRIS was upgraded to a dual-channel configuration (see \citealt{steidel04}).

A small subset of the $z \sim 2$--$2.6$ galaxies 
in earlier catalogs in some of the fields listed in Table~\ref{tab:summary} was observed in the near-IR using Keck/NIRSPEC \citep{erb+06a,erb+06b,erb+06c};
the sample was used to calibrate UV measurements of systemic redshifts by \citet{steidel2010}. However, the vast majority of the nebular redshifts
used in this paper were obtained using the Multi-Object Spectrometer for InfraRed Exploration (MOSFIRE; \citealt{mclean12,steidel14}) beginning in 
2012 April. MOSFIRE observations in all but the GOODS-N field\footnote{Nebular redshifts of 89 galaxies in 
our catalog were obtained by the MOSFIRE Deep Evolution Field Survey (MOSDEF;\citealt{kriek15}.)}  
were obtained as part of KBSS-MOSFIRE (see \citealt{steidel14,strom17} for details.)

The statistical properties of the galaxies in the KGPS sample are as described in previous work:  
stellar masses 
${\rm 8.6 \simlt log(M_{\ast}/M_{\odot}) \simlt 11.4}$ (median~$\simeq 10.0$),  
star formation rates ${\rm 2 \simlt SFR/(M_{\odot} yr^{-1}) \simlt 300}$   
(median~$\simeq 25$) (\citealt{shapley05,erb+06a,erb+06b,reddy08a,reddy12,steidel14,strom17,theios19}),
and clustering properties indicate host dark matter halos of typical mass 
$\langle {\rm log(M_{h}/M_{\odot})\rangle = 11.9\pm0.1}$ \citep{adelberger05a,trainor12}.

\subsection{Rest-Frame Far-UV Spectra} 
\label{subsec:lris}
All of the rest-UV spectra used in this work were obtained with the Low Resolution Imaging Spectrometer (LRIS; \citealt{oke95,steidel04}) on the
Keck \RNum{1} 10m telescope; most were obtained between 2002 and 2016, after LRIS was upgraded to a dual-beam spectrograph. Most of the spectra used
here were obtained using the blue channel (LRIS-B), with one of two configurations: a 400 line/mm grism blazed at 3400 \AA\ in first order, covering
3200-6000 \AA, or a 600 line/mm grism blazed at 4000 \AA, typically covering 3400-5600 \AA. Approximately half of the slitmasks were observed
with each configuration. Further details on the observations and reductions with LRIS-B are given in, e.g., \citet{steidel04,steidel2010,steidel18}.  

\begin{figure*}%
\centering
\includegraphics[width=12cm]{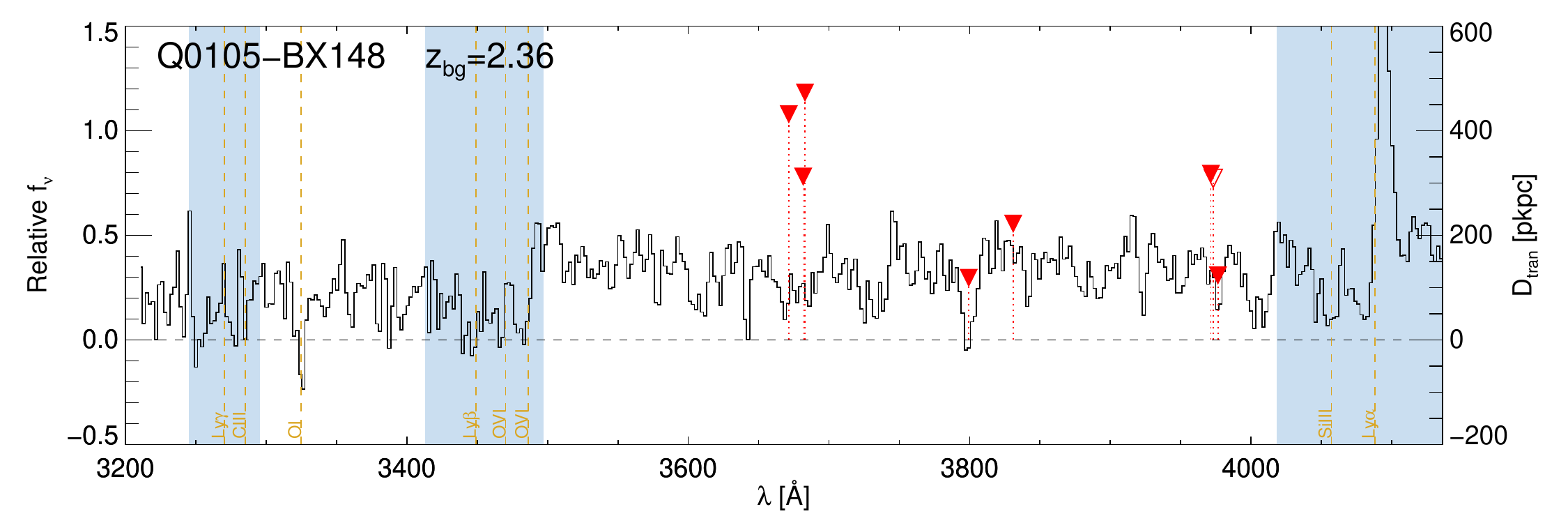} \\
\includegraphics[width=12cm]{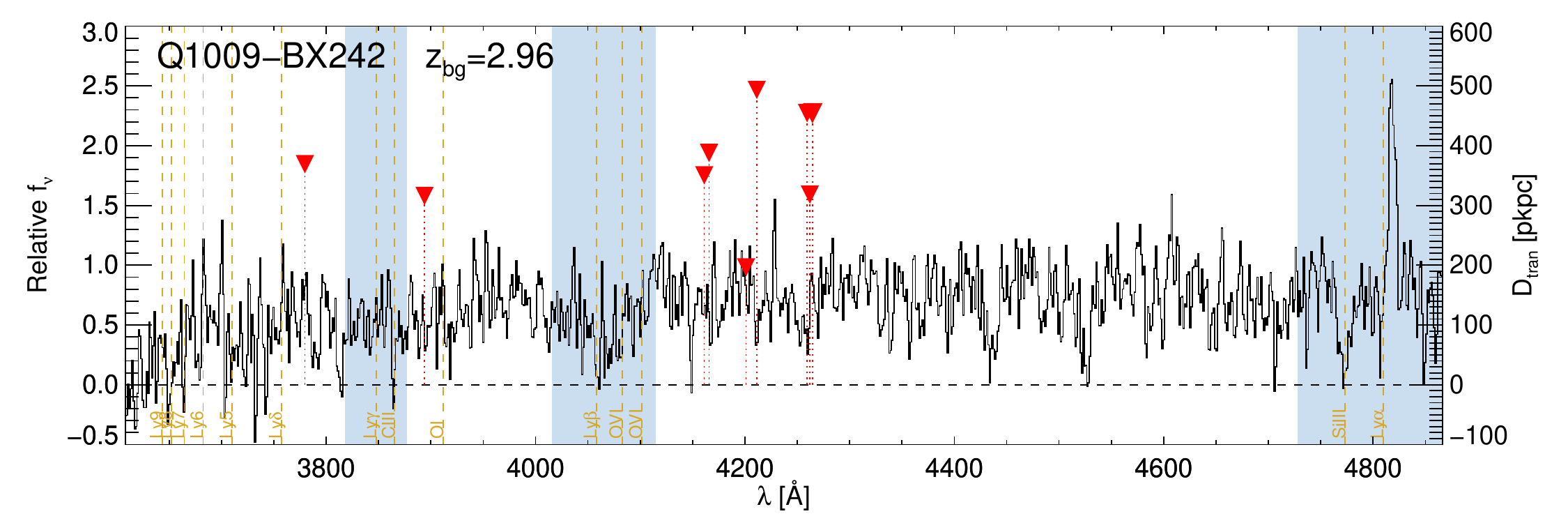} \\
\includegraphics[width=12cm]{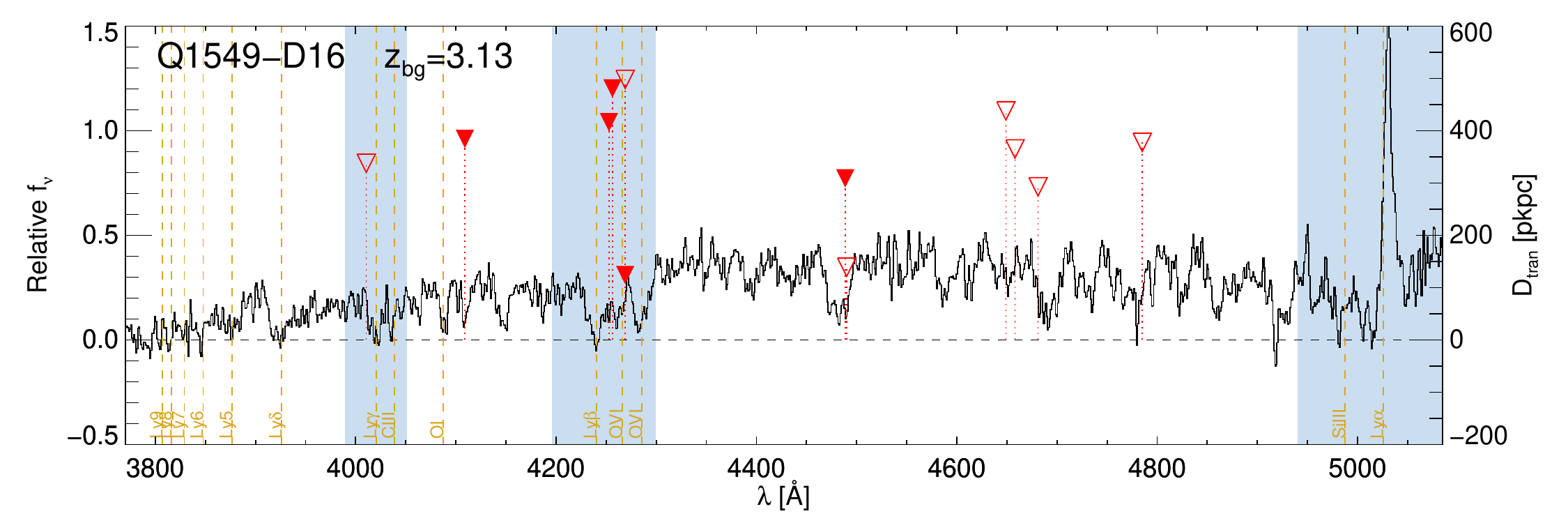} \\
\includegraphics[width=12cm]{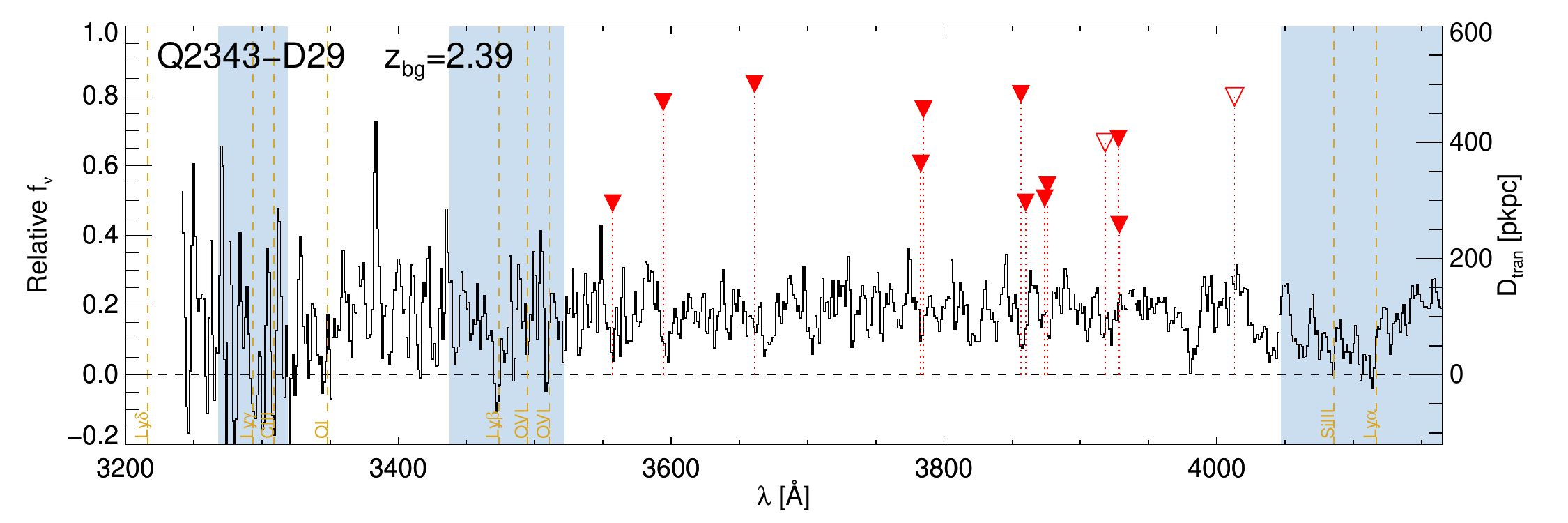}
\caption{   Randomly-selected examples of individual rest-UV spectra of background galaxies in KGPS. Wavelengths are in the observed frame, 
with red triangles marking the position of Ly$\alpha$ $\lambda 1215.67$ at wavelengths $1215.67 (1+z_{\rm fg})$ \AA\ for each foreground galaxy with 
projected distance $D_\mathrm{tran} \le 500 \mathrm{~pkpc}$. The solid (open) triangles correspond to foreground galaxies whose systemic redshifts are based on $z_{\rm neb}$ ($z_\mathrm{UV}$). Their typical redshift uncertainties, in terms of observed wavelength, are
$\sim$ 0.2 \AA\ (1.8 \AA) at $z=2.2$ as estimated in \S\ref{sec:redshift}.
The y-coordinate of each triangle indicates $D_\mathrm{tran}$ for the foreground galaxies, with reference to the scale marked on the righthand side of each plot.  
The light shaded regions are those that would
be masked prior to using the spectrum to form composites stacked in the rest frame of foreground galaxies (see \S\ref{sec:stacking} \& Table~\ref{tab:mask}), to minimise
contamination by spectral features at $z = z_{\rm bg}$. The yellow vertical lines indicate UV absorption lines arising in the ISM of the background galaxy. 
Note that some foreground galaxies have clear counterparts in the \lya\ forest, even in low-resolution spectra. } 
\label{fig:egspec}
\end{figure*}

The total integration time for individual objects ranges from 5400s to $>$54,000s. About $40\%$ of the galaxies were observed with two or more masks, particularly in the KBSS fields for which the field size is comparable to the 5\minpoint5 by 7\minpoint5 field of
view of LRIS. 
Examples of typical reduced 1-D spectra are shown in Figure~\ref{fig:egspec}. 
The wavelength solutions for the LRIS-B spectra were based on polynomial fits to arc line lamp observations using the same mask and instrument
configuration, which have typical residuals of $\simeq 0.1$ \AA. Small shifts between the arc line observations and each 1800s science exposure 
were removed during the reduction process with reference to night sky emission features in each science frame. The
wavelength calibration uncertainties make a negligible contribution to the redshift measurement errors (\S\ref{sec:redshift}).  

Each reduced 1-D spectrum\footnote{For galaxies observed on multiple masks, each independent 1-D spectrum was examined separately.} 
was examined interactively, in order to mask regions of very low SNR, poor background subtraction, unphysical 
flux calibration, or previously unmasked artifacts (e.g., cosmic rays, bad pixels) that were not recognized during data reduction. 
A total of 280 spectra (out of nearly 10,000 in total) were entirely discarded because of generally poor quality or unphysical continuum shape. 
Spectra of the same object observed on multiple masks and/or with multiple spectroscopic setups 
were assigned individual weights according to spectral quality, based on a combination of visual inspection and exposure time. 
They were then resampled onto a common wavelength grid with the finest sampling of the individual spectra to preserve the spectral resolution 
(using cubic-spline interpolation) and averaged together to create a single spectrum for each object.  

\subsection{Calibration of Systemic Redshifts}\label{sec:redshift}

Spectral features commonly observed in the far-UV spectra
of high redshift star-forming (SF) galaxies -- \lya\ emission, when present, and interstellar (IS) absorption from strong resonance lines
of (e.g.) \ion{Si}{II}, \ion{Si}{IV}, \ion{C}{II}, \ion{C}{IV}, \ion{O}{I} -- are rarely found at rest with respect to the stars in the same
galaxy due to gas motions and radiative transfer effects (e.g., \citealt{steidel96,franx97,lowenthal97,pettini01,shapley03,erb+06b}).  
Clearly, measuring the kinematics of diffuse gas in the CGM of foreground galaxies benefits from the most accurate available measurements of
each galaxy's systemic redshift ($z_{\rm sys}$).  

The centroids of nebular emission lines from ionized gas (i.e., \ion{H}{ii} regions) are  
less strongly affected by galaxy-scale outflows and radiative transfer effects, and are generally 
measured with significantly higher precision, than 
the rest-FUV features. 
As previously noted, 50\% of the foreground galaxies used in this work have measurements of one or more strong nebular emission lines in the
rest-frame optical (observed frame J, H, K bands) using MOSFIRE.  
Independent observations of the same galaxies with MOSFIRE  
have demonstrated  
redshift precision (i.e., rms repeatability) of $\sigma_v \simeq 18$ \kms\ \citep{steidel14}.  

\begin{figure}%
\includegraphics[width=8.5cm]{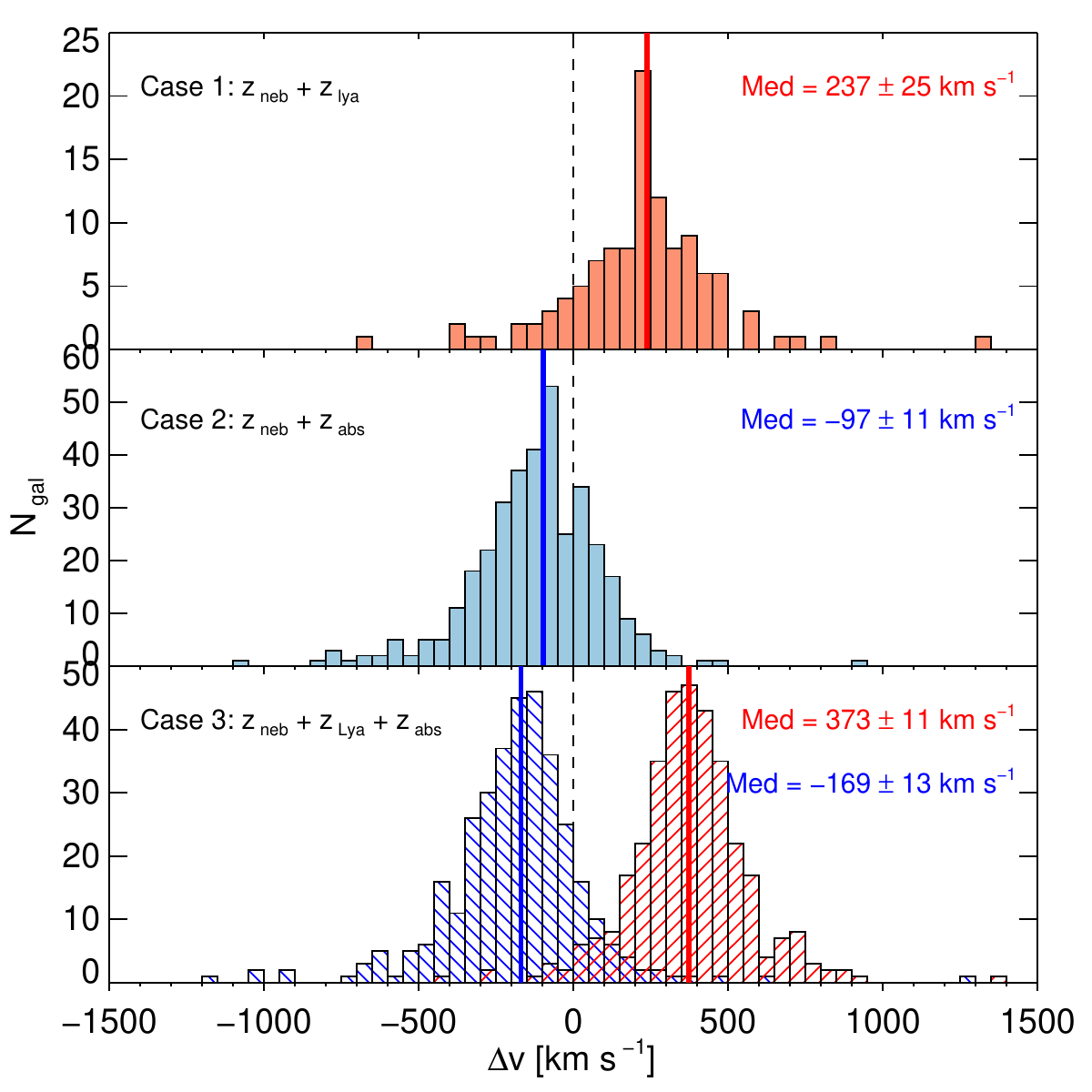}
\caption{ Distribution of velocity differences between \zla\ ($\Delta v_{\lya}$, in red) or \zis\ ($\Delta v_{\rm IS}$, in blue) 
and the systemic redshift measured from $z_\mathrm{neb}$. Top-right of each panel shows the median shift in 
velocity and its error estimated by dividing the standard deviation by the square-root of the number of galaxies. These distributions have been
used to calibrate the systemic redshifts of galaxies in equations \ref{eqn:case1}--\ref{eqn:case3}, for which only the UV spectral features are available. 
} 
\label{fig:zcal}
\end{figure}

For the 50\% of foreground galaxies lacking nebular emission line measurements, we used estimates of $z_{\rm sys}$ based
on the full KBSS-MOSFIRE sample (\citealt{steidel14,strom17}) with $z_{\rm neb} > 1.9$ and existing rest-UV LRIS spectra.
These were used to calibrate relationships
between $z_{\rm sys}$ and redshifts measured from features in the rest-frame FUV spectra, 
strong interstellar absorption lines ($z_{\rm IS}$) and/or the centroid of Lyman-$\alpha$ emission (\zla).     
As for previous estimates of this kind (e.g., \citealt{adelberger03,steidel2010,rudie12a}), we adopt rules that
depend on the particular combination of features available in each spectrum. 
Figure \ref{fig:zcal} shows the distribution of velocity offsets of the UV redshift measurements relative to \zneb, $\Delta v_{\mathrm{Ly}\alpha} = c(\zla - \zneb)/(1+\zneb)$ and/or $\Delta v_\mathrm{IS} = c(\zis - \zneb)/(1+\zneb)$ for the three cases below. 
The median velocity offsets (see Figure~\ref{fig:zcal}) for the three sub-samples were then used to derive the following relationships that map UV redshift measurements
to an estimate of $z_{\rm sys}$:
\begin{itemize}
\item Case 1: \zla\ only,
\begin{eqnarray}
z_\mathrm{sys}=z_{\mathrm{Ly}\alpha}-\frac{237 \mathrm{~km ~s}^{-1}}{c}(1+z_{\mathrm{Ly}\alpha}),
\label{eqn:case1}
\end{eqnarray}
\item Case 2: \zis\ only,
\begin{eqnarray}
z_\mathrm{sys}=\zis + \frac{97 \mathrm{~km ~s}^{-1}}{c}(1+\zis). 
\label{eqn:case2}
\end{eqnarray}
\item Case 3: \zla\ and \zis,

if $\zla > \zis$,
\begin{eqnarray}
\label{eqn:case3}
z_\mathrm{sys} & = & \frac{1}{2}\{[z_{\mathrm{Ly}\alpha}-\frac{373 \mathrm{~km ~s}^{-1}}{c}(1+z_{\mathrm{Ly}\alpha})]+ \nonumber \\
&&[z_\mathrm{IS}+\frac{169 \mathrm{~km ~s}^{-1}}{c}(1+z_\mathrm{IS})]\};
\end{eqnarray}
otherwise, 
\begin{eqnarray}
\label{eqn:case4}
z_\mathrm{sys} & = & \frac{1}{2}(z_{\mathrm{Ly}\alpha} + z_\mathrm{abs}).
\end{eqnarray}
\end{itemize}

When the above relations are used to estimate $z_{\rm sys,uv}$ on an object by object basis, 
applied to the UV measurements of the redshift calibration sample on an object-by-object basis, 
the outlier-clipped mean and rms of the velocity difference between $z_{\rm neb}$ and $z_{\rm sys,uv}$ are 
\begin{eqnarray}
\langle c (z_{\rm sys,uv} - z_{\rm neb})/(1+z_{\rm neb}) \rangle = 
-5 \pm 143~ \kms, 
\end{eqnarray}
implying that we expect negligible systematic offset in cases where only $z_{\rm sys,uv}$ is available, with $\sigma_{\rm z} \simeq 140$ \kms.

In what follows below, we define the subsample of galaxy foreground-background pairs for which $z_{\rm sys,fg}$ is based on $z_{\rm neb}$ 
as the ``KGPS-$z_{\rm neb}$'' sample, with $\sigma_{\rm z}\simeq 18 \kms$; the ``KGPS-Full'' sample includes all of KGPS-$z_{\rm neb}$ plus the remaining pairs for which $z_{\rm sys,fg}$ is estimated 
from the rest-UV spectra according
to equations~\ref{eqn:case1}--\ref{eqn:case3}. The fraction of pairs with $z_\mathrm{sys,fg}$ estimated from rest-UV features is 
consistently $\simeq 50\%$ at $D_\mathrm{tran}<1 \mathrm{~pMpc}$, but gradually decreases to 
$\simeq 40\%$ from $D_\mathrm{tran} \simeq 1 \mathrm{~pMpc}$ to $4\mathrm{~pMpc}$.

\subsection{Assembly of Galaxy Foreground-Background Pairs}
\label{sec:pairs}

\begin{figure}
\includegraphics[width=8.5cm]{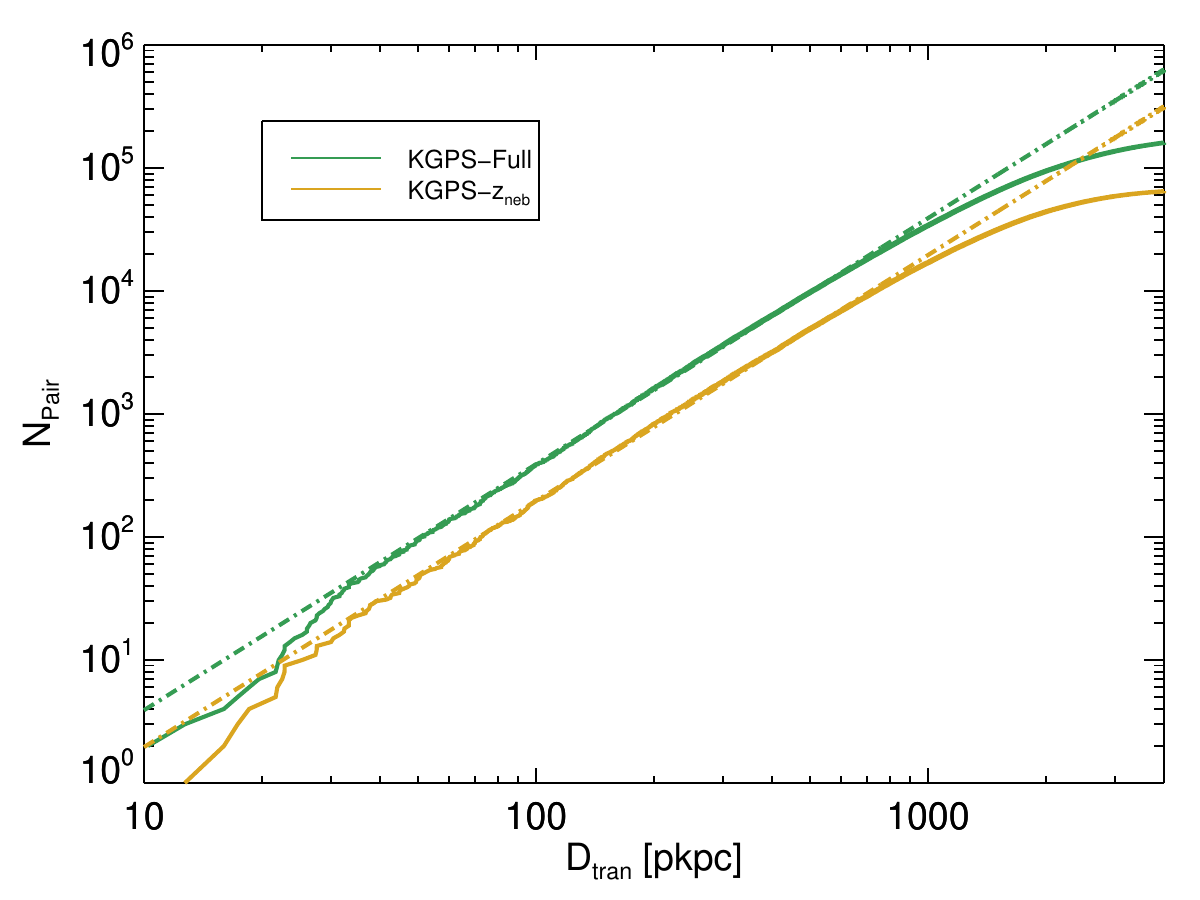}
\caption{The cumulative number of pairs as a function of impact parameter 
$D_{\rm tran}(z_{\rm fg})$ for the KGPS-Full sample (green), the KGPS-$z_\mathrm{neb}$ sample (yellow).  
The solid curves are the actual number of pairs, and the dashed-dotted lines are quadratic fits over 
the range $50 \mathrm{~pkpc}<D_\mathrm{tran} < 500 \mathrm{~pkpc}$  
for each of the KGPS samples. The smaller number of pairs (relative to the quadratic extrapolation) 
at very small $D_{\rm tran}$ results from observational biases caused
by geometrical slitmask constraints and finite angular resolution of the ground-based images used for target selection. 
At large $D_{\rm tran}$, the pair count
falls below the quadratic fit as the angular separation approaches the size of the KBSS survey regions. 
\label{fig:pairnum} }
\end{figure}

We assembled samples of KGPS galaxy pairs ($z_{\rm fg},z_{\rm bg}$) according to several criteria designed to optimise the measurement of weak absorption lines 
near the redshift $z_{\rm fg}$ in the spectrum of the background galaxy:  
\begin{enumerate}
\item The background galaxy has one or more LRIS spectra 
\item The foreground galaxy does not host a known Type I or Type II active galactic nucleus\footnote{Pairs for which the foreground object harbors an AGN
will be considered in a separate paper (Y. Chen et al 2020, in prep.)}.  
\item The paired galaxies have redshifts (\zfg, \zbg) such that
\begin{eqnarray}
\label{eqn:zfb}
0.017(1+\zfg)  < \dzfb < 0.3 (1+\zfg), 
\end{eqnarray}
where $\Delta z_{\rm fb} = \zbg - \zfg$. This is equivalent to $5100 \mathrm{~km~s}^{-1}<\Delta v_\mathrm{LOS}<90000 \mathrm{~km~s}^{-1}$, or $37 \mathrm{~pMpc} < D_\mathrm{LOS} < 520 \mathrm{~pMpc}$ at $z=2.2$, assuming pure Hubble flow, where $\Delta v_\mathrm{LOS}$ and $D_\mathrm{LOS}$ are velocity difference and distance in the line-of-sight direction.
\end{enumerate}

The lower limit on \dzfb\ 
ensures that the two galaxies are not physically associated, 
so that any absorption features detected near \zfg\ are well-separated from features at \zbg\ and are not part of the same large scale structure in which
the background galaxy resides. 
An upper limit on $\dzfb/(1+z_{\rm fg})$ was set by S2010 to maximise the detectability of \ion{C}{IV}$\lambda\lambda 1548$,1550 
at \zfg\ by ensuring that it would fall longward of the \lya\ forest in the spectrum of the background galaxy, i.e.  
\begin{eqnarray}
(1+\zfg)1549 > (1+\zbg)1215.67 \mathrm{\AA} 
\end{eqnarray}
for typical $\zbg \sim 2.4$. In the present case, using a more empirical approach, 
we tested different upper limits on $\dzfb/(1+z_{\rm fg})$ in order to optimise the SNR of the final stacks. 
In principle, if one is interested in detecting \lya\ absorption, large $\dzfb/(1+\zfg)$ would increase the relative contribution of
the shorter wavelength, noisier portions of the background galaxy spectra, particularly
when regions shortward of \lyb\ at $z=\zbg$ [i.e., $\lambda \le (1+\zbg)1025.7$] are included. 
On the other hand, choosing a small upper limit on $\dzfb/(1+\zfg)$ would increase the noise by significantly decreasing the number of spectra contributing.  
Depending on the strength of the Ly$\alpha$ absorption, we found that the SNR does not depend strongly on the upper limit so long as it is close to
$\dzfb/(1+z_{\rm fg}) \simeq 0.30$, similar to the upper limit used by \citet{steidel2010} ($\dzfb/(1+\zfg) \simeq 0.294$ for $\zbg \simeq 2.4$).

Figure~\ref{fig:pairnum} shows the cumulative number of galaxy pairs as a function of $D_\mathrm{tran}$ between the
foreground galaxy and the
line of sight to the background galaxy, evaluated at $\zfg$. The cumulative number of distinct pairs
varies as $D_{\rm tran}^{2}$ (dashed lines in Figure~\ref{fig:pairnum}; as expected for uniform sampling of a constant surface density of galaxies) 
over the range $50 \le D_{\rm tran}/\mathrm{pkpc} \le 500$. The departure of the observed number of pairs falls below the quadratic extrapolation for $D_{\rm tran} < 30$~pkpc (angular
scales of $\theta \simlt$3\secpoint6 at $z \sim 2.2$) due to a combination of limited spatial resolution of the ground-based images used to select targets, and
the constraints imposed by slit assignment on LRIS slitmasks. For $D_\mathrm{tran} \gtrsim 1\mathrm{~pMpc}$ ($\theta \simgt 2$~arcmin at $z=2.2$), 
the number of pairs begins to be limited by the size of individual survey regions (see Table~\ref{tab:summary}). Over the range $30< D_{\rm tran}/\mathrm{pkpc} < 1000$, 
the number of pairs is well represented by a quadratic function, 
\begin{eqnarray}
\textrm{KGPS-Full:}\quad N_\mathrm{pair}(<D_\mathrm{tran})  = 39069 \times \left( \frac{D_\mathrm{tran}}{1 \mathrm{~pMpc}} \right)^2, 
\end{eqnarray}
and 
\begin{eqnarray}
\textrm{KGPS-\zneb :}\quad N_\mathrm{pair}(<D_\mathrm{tran})  = 19577 \times \left( \frac{D_\mathrm{tran}}{1\mathrm{~pMpc}} \right)^2. 
\end{eqnarray}

\begin{figure}%
\centerline{\includegraphics[width=8.5cm]{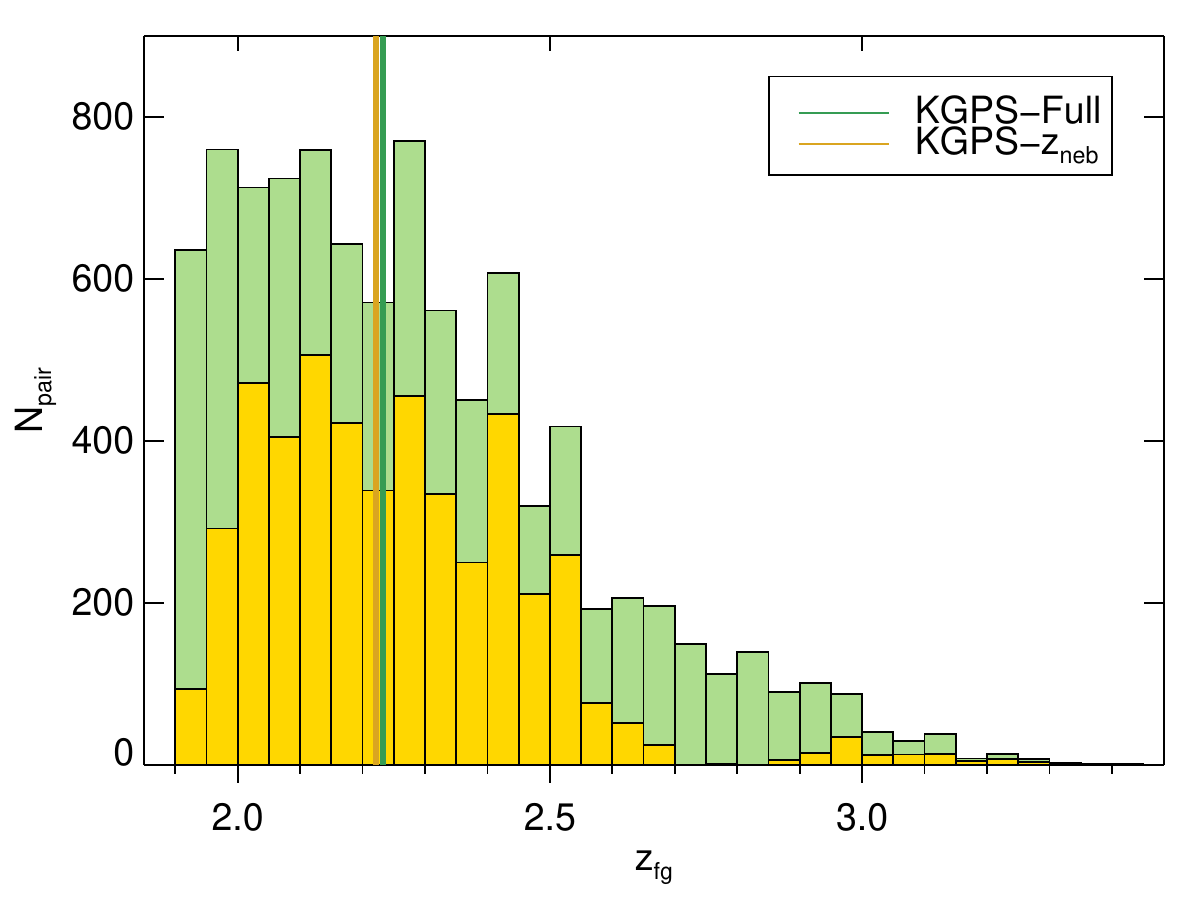}} 
\centerline{\includegraphics[width=8.5cm]{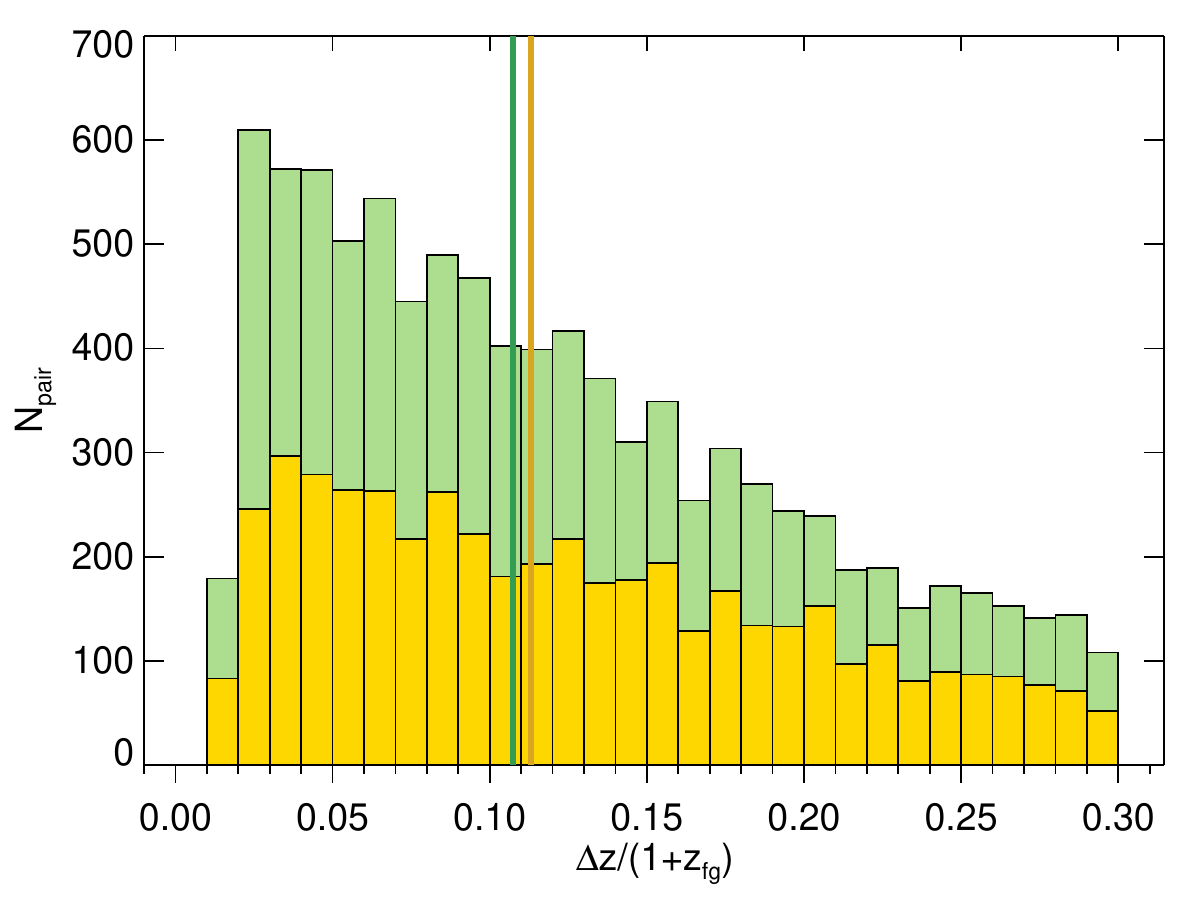}}
\caption{  Distribution of $z_\mathrm{fg}$ (top) and $\dzfb/(1+z_\mathrm{fg})$ (bottom) for foreground-background galaxy pairs
with with $D_\mathrm{tran} < 4.7 \mathrm{~pMpc}$. The vertical lines mark the median values of each distribution. The KGPS-Full and KGPS-$z_{\rm neb}$ samples have 
similar distributions of both $z_{\rm fg}$ and $\Delta z_{\rm fb}/(1+z_{\rm fg})$.  \label{fig:histpair}}
\end{figure}

Figure~\ref{fig:histpair} shows the distributions of $\zfg$ and $\dzfb/(1+\zfg)$ for both the KBSS-Full and KBSS-\zneb\ samples. The difference in the $\zfg$ distributions
is caused by gaps in redshift for the KBSS-\zneb\ sample that correspond to regions of low atmospheric transmission between the J, H, and K bands. However, 
in spite of this, the distributions of $\dzfb/(1+z_\mathrm{fg})$ remain very similar. The median foreground galaxy redshift is $\langle z_\mathrm{fg} \rangle_{\rm med} = 2.23$ 
for the KGPS-Full sample, and $\langle \zfg \rangle_{\rm med} = 2.22$ for the KGPS-$z_\mathrm{neb}$ sample; the median value
of the redshift differences between the foreground and background galaxies are $\dzfb/(1+\zfg) = 0.106$ and 0.113 respectively.

\subsection{Composite Spectra} \label{sec:stacking}

\label{sec:composites}

The typical spectrum of an individual galaxy in the KGPS sample has SNR per spectral resolution element of only 1--6 
in the region shortward of \lya\ in the galaxy rest frame.  Individual spectra also include absorption from other spectral lines due to the interstellar and circumgalactic
medium of the galaxy itself -- whose locations are predictable --  and intervening absorption caused by gas at redshifts different from
the foreground at which a measurement of \lya\ absorption is made \footnote{In the case of QSO sightlines, the problem of contamination also exists, but is partially overcome by observing at very high spectral resolution.}.
Both problems -- limited SNR of faint background galaxy spectra, and contamination from absorption at other redshifts -- are mitigated by forming ``stacks'' of many
spectra sampling a particular range of $D_{\rm tran}$ for an ensemble of foreground galaxies probed by background galaxies.  

A distinct advantage of stacking, particularly when it comes to detecting absorption lines arising from gas at a particular redshift, is
that one naturally suppresses small-wavelength-scale noise caused by contamination from absorption lines at redshifts other than that of the foreground
galaxies of interest. With a suitable number of spectra comprising a stack, unrelated (stochastic) absorption will produce a new, lower, effective continuum level 
against which the \lya\ absorption due to the foreground galaxy ensemble can be measured. The amount by which the continuum is lowered depends on $z_{\rm fg}$,
and is expected to be close to the mean \lya\ forest flux decrement $D_{\rm A}(z)$ (\citealt{oke82}). At $z \sim 2$--$2.5$ most relevant for the KGPS sample,
$D_{\rm A} \simeq 0.2$, i.e., a reduction in the apparent continuum level near \lya\ of $\simeq 20$\%. Thus, any residual \lya\ absorption is equivalent to
an \ion{H}{I} ``overdensity'', in the sense that it signals the amount by which \ion{H}{I} gas associated with the foreground galaxy exceeds the mean
IGM absorption at the same redshift.  

\begin{table}
\caption{  Masked Background Spectral Regions \label{tab:mask}}
\centering
\begin{tabular}{ccl}
\hline\hline
$\lambda$ (\AA) & $\Delta z_{\rm fb}/(1+z_\mathrm{fg})$ & Spectral Features \\\hline
965 -- 980 & 0.2405 -- 0.2598 & Ly$\gamma$, \ion{C}{III} $\lambda$977\\
1015 -- 1040 & 0.1689 -- 0.1977 & Ly$\beta$, \ion{O}{VI} $\lambda$1031,1036\\
1195 -- 1216 & 0 -- 0.0173 & \ion{Si}{III} $\lambda$1206, Ly$\alpha$ \\\hline
\end{tabular}
\end{table}

We arranged galaxy pairs into bins of $D_\mathrm{tran}$, and stacked the spectra of all of the background galaxies in the same bin of $D_{\rm tran}$ 
 after (1) normalizing the flux-calibrated background galaxy spectra to have unity median flux density evaluated over the \zbg-frame rest wavelength
interval $1300 \le \lambda_{\rm bg,0}/\textrm{\AA} \le 1400$; (2) masking regions of the background galaxy spectra corresponding to the 
locations of strong absorption lines
or sets of absorption lines at $z=\zbg$; the 
relevant spectral ranges are given in Table~\ref{tab:mask}; (3) shifting the 
result to the rest-frame of the foreground galaxy, i.e., 
\begin{eqnarray}
\label{eqn:fgbg}
\lambda_{\rm fg,0} = \lambda_{\rm bg,obs}/(1+\zfg)~~; 
\end{eqnarray}
and (4) resampling the normalised, masked, and \zfg-shifted spectra onto a common range of rest wavelength and combining to form 
a composite rest-frame spectrum representing the bin in $D_{\rm tran}$. 

The normalisation in (1) was performed in order to give roughly equal weight to each galaxy pair in a given bin of $D_{\rm tran}$ without requiring an actual 
continuum fit to each low-SNR spectrum. 
We found that continuum fitting is less prone to large systematic errors \citep[e.g.,][]{fg08} when it is performed for the composite spectra after stacking, rather than for individual spectra, particularly
within the \lya\ forest where it would be difficult to perform the fit reliably in the face of both shot noise and real \lya\ forest absorption. 
Error spectra for each stack were generated using bootstrap resampling of the galaxy ensemble, which should account for both sample variance and random (Poisson) errors. 
Step (2) was implemented in order to reduce contamination by unrelated absorption features near \zbg\ by excluding pixels known to be contaminated; 
the wavelength ranges listed in Table~\ref{tab:mask} were adopted based on tests using larger or smaller 
masked intervals\footnote{Note that masking a particular range of rest-wavelength in the frame of each background galaxy is similar to eliminating pairs having
particular range of $\Delta z_{\rm fb}/(1+z_\mathrm{fg})$. These intervals are also provided in Table~\ref{tab:mask}.}. For step (3), we used the
best available systemic redshift of the foreground galaxy (\S\ref{sec:redshift}).  

We experimented with several different methods for accomplishing step (4) before adopting a straight median of unmasked pixels at each dispersion point;  details of the tests are summarized in Appendix~\ref{app:stack}. The median
algorithm produces composite spectra with SNR comparable to the best
$\sigma$-clipped mean algorithm, but has the added
benefit of computational and conceptual simplicity.  

\begin{figure*}
\centering
\includegraphics[width=17cm]{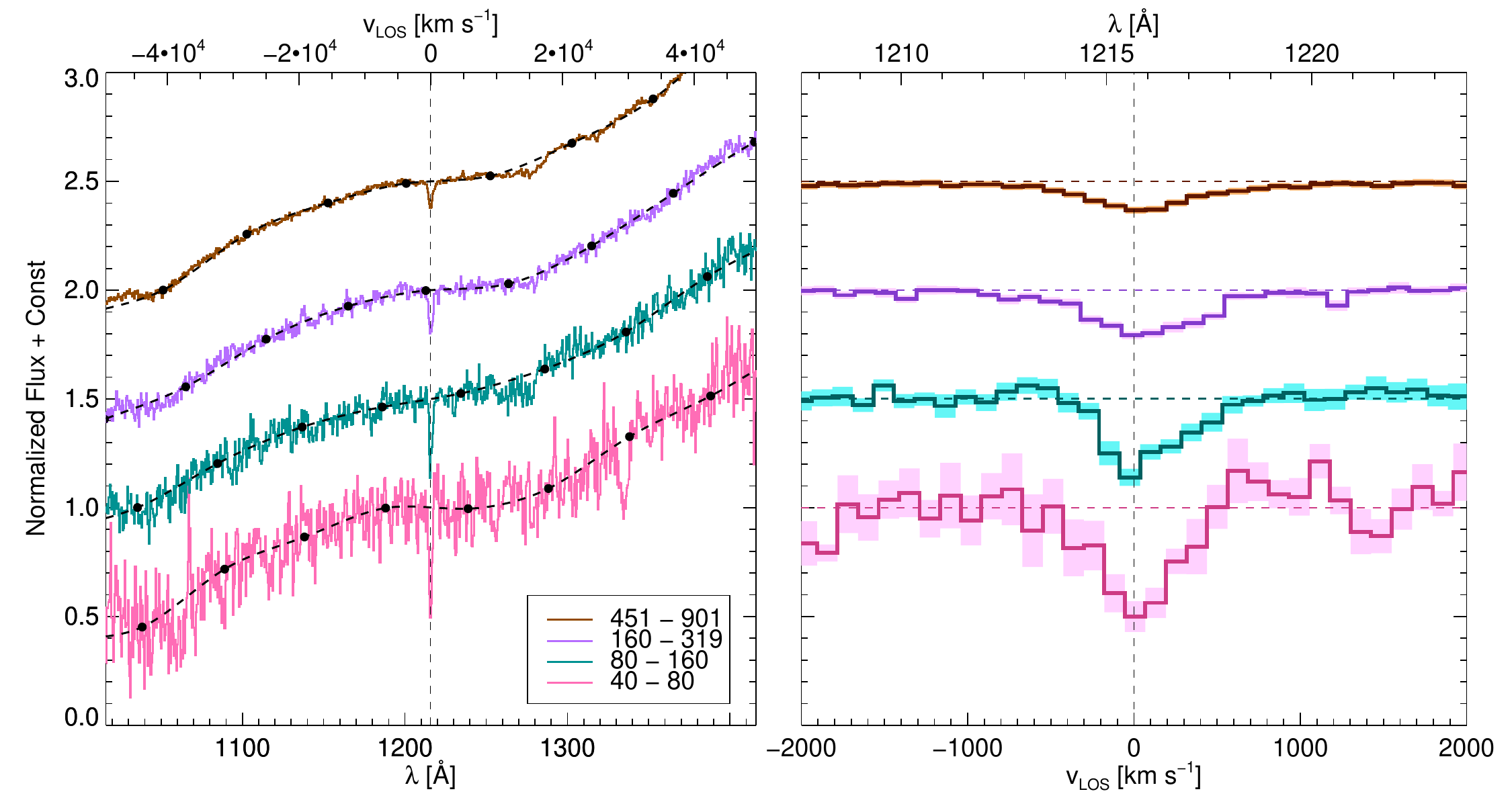}
\caption{  Example composite spectra near \lya\ in the rest frame of foreground galaxies for four different bins of $D_\mathrm{tran}$. The left panel shows the spectra before continuum-normalisation. 
The black curves with dots are the fitted continua and spline points used for cubic interpolation. The legend indicates the range in pkpc for the bin in $D_\mathrm{tran}$. 
The right-hand panel shows normalised spectra near Ly$\alpha$, along with their 1$\sigma$ uncertainty (shaded histogram). In both panels the spectra shown have
been shifted relative to one another by 0.5 in y for display purposes. 
The vertical dashed line is the rest wavelength of Ly$\alpha$, 1215.67 \AA.
The \lya\ absorption profiles in the composite spectra clearly vary in both depth and width with $D_{\rm tran}$.
The clear asymmetry in the velocity profiles of the two middle spectra ([50-100] and [160-319] pkpc) is discussed in \S\ref{sec:asymmetry} below.}
\label{fig:egstack}
\end{figure*}

To remove the continuum, each composite spectrum was divided into 200-\AA\ segments; for each segment, we calculated the mean value 
(with 2.5-$\sigma$ clipping applied) 
of the flux density; the mean flux density and mean wavelength within each segment were then used to constrain a cubic spline fit to the continuum flux density as
a function of wavelength, as illustrated in Figure~\ref{fig:egstack}.  
The composite spectra were then divided by the initial continuum fit, after which the continuum level of each normalised spectrum was further adjusted 
by dividing by the linear interpolation of pixels within two windows fixed in velocity relative to the nominal rest-wavelength of \lya\ : 
$3000 \mathrm{~km ~s}^{-1} < |\Delta v_{\mathrm{Ly}\alpha}| < 5000 \mathrm{~km ~s}^{-1}$. 
Figure~\ref{fig:egstack} shows examples of the $z_\mathrm{fg}$-frame stacked spectra before and after dividing by a fitted continuum. 
Both the velocity width and the depth of the excess Ly$\alpha$ absorption associated with the foreground galaxies clearly varies with $D_\mathrm{tran}$. 

\section{H{\sc I} Absorption} 
\label{sec:obs}

\subsection{Ly$\alpha$ Rest Equivalent Width [\wlya]} 
\label{sec:ew}

Expressing the total strength of \lya\ absorption in terms of the rest-frame equivalent width [hereafter \wlya] is appropriate in the present case, where  
the velocity structure is only marginally resolved. Because \wlya\ does not depend on the spectral resolution of the observed spectra (provided it is sufficient
to allow accurate placement of the continuum level), it is also useful for comparisons among samples obtained with different spectral resolution.  \wlya\ is also
entirely empirical, and does not depend on any assumptions regarding the fine-scale kinematics or component structure of the absorbing gas.

In KGPS, which uses composite spectra of many background galaxies, \wlya\ is 
modulated by a potentially complex combination of the mean integrated covering fraction of absorbing gas,
its line-of-sight (LOS) kinematics for an ensemble of sightlines falling  
within a range of $D_{\rm tran}$ relative to foreground galaxies, and the total column density of \ion{H}{I}. As discussed above (see also S2010), the finite ``footprint'' of 
the image of a background galaxy projected
onto the gas distribution surrounding a foreground galaxy also means that the \lya\ absorption profile may depend on the spatial variations on scales of a few
pkpc\footnote{The typical effective radius of the background galaxies in the KBSS sample is $r_{\rm e} \simeq 1.5$ pkpc \citep{law12}.} for CGM sightlines near individual galaxies.  However, given a sample of foreground galaxies, the average dependence of \wlya\ on $D_{\rm tran}$ should
be identical for extended (galaxy) and point-like (QSO) background sources so long as the number of sightlines is large enough to overcome sample variance within
each bin of $D_{\rm tran}$.

For a sample of galaxy pairs as large as KGPS, where both the width and depth of the \lya\ profile varies with $D_{\rm tran}$ (Figure~\ref{fig:egstack}), 
it is desirable to develop a robust method for automated measurement of \wlya\ while maximizing the SNR. 
For spectra of limited continuum SNR -- particularly where the absorption line profile has a spectral shape that is unknown {\it a priori}, the size 
of the measurement aperture has a significant effect on the SNR of the \lya\ line; it should not be unnecessarily large, which would contribute unwanted noise without
affecting the net signal, 
nor so small that it would exclude significant absorption signal. We set the integration aperture using   
2-D maps of apparent optical depth (Figure \ref{fig:2dabs}; to be discussed in detail in \S\ref{sec:kinematics})): when $D_\mathrm{tran} < 100$~pkpc, 
we set the aperture width to $\Delta v = 1400 \mathrm{~km~s}^{-1}$ ($\Delta \lambda_{0} \simeq 11.35~\textrm{\AA}$), centered on the nominal rest wavelength of \lya; 
otherwise, the width of the aperture is set to, 
\begin{eqnarray}
\Delta v= 1000~ \langle \log (D_{\rm tran}/{\rm pkpc}) \rangle - 600~~\kms~, 
\end{eqnarray}
also centered on rest-frame \lya.

\begin{figure}
\includegraphics[width=8.5cm]{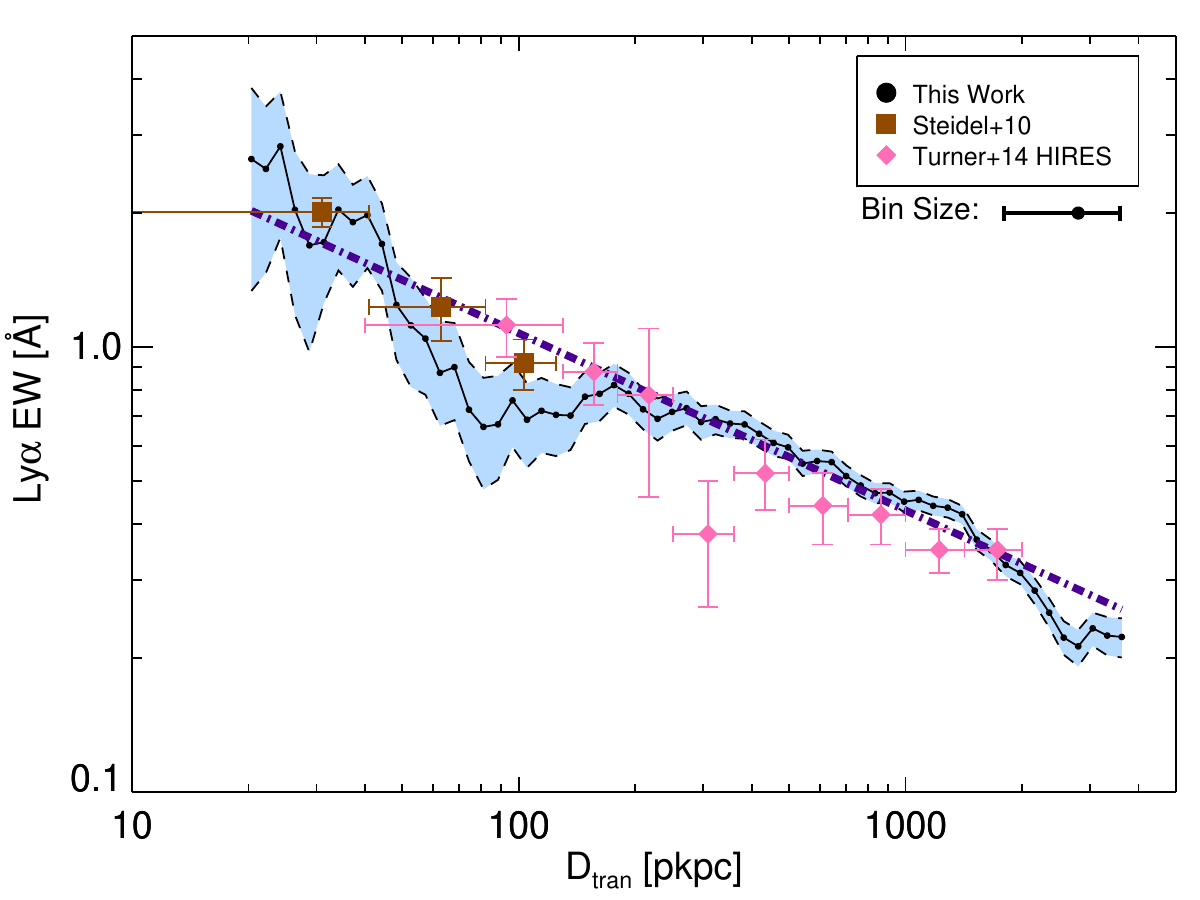}
\caption{ \wlya\ as a function of $D_\mathrm{tran}$ measured from the KGPS-Full sample. Each black dot (connected with the solid line segments) is a measurement
made from a composite spectrum in a bin spanning a factor of 2 in $D_{\rm tran}$ where the point marks the geometric mean $D_{\rm tran}$ within the bin. The
light blue shaded region represents the $\pm 1$$\sigma$ uncertainty based on bootstrap resampling of the spectra comprising each composite in each bin. 
The single dot with error bars below the legend box shows the bin size used to make composite spectra; i.e., the bins were evaluated at intervals smaller than
the bin size, thus adjacent points on smaller than the bin size are correlated. 
The purple dash-dotted line is the best single-power-law fit to the \wlya-$D_\mathrm{tran}$ relation, with slope $\beta = -0.40\pm 0.01$.
The new measurements are compared with \citet[][brown squares]{steidel2010} and \citet[][magenta diamonds]{turner14}, where the latter are based on HIRES QSO/galaxy pairs in KBSS. }
\label{fig:ew}
\end{figure}

Figure~\ref{fig:ew} shows \wlya\ measured from the KGPS-Full sample as a function of $D_\mathrm{tran}$ with the bin size in $D_\mathrm{tran}$ set to be 0.3 dex, together with measurements from S2010 for galaxy-galaxy
foreground/background pairs and from \citet{turner14} for foreground galaxy-background QSO pairs from the KBSS survey. The points from \cite{turner14} were
measured from the spectra of only 17 background QSOs in the 15 KBSS fields, evaluated at the redshifts of foreground galaxies within $\simeq$4\minpoint2 drawn from
essentially the same parent galaxy sample as KGPS. For $D_{\rm tran} > 400$ pkpc, where the sample variance of QSO-galaxy sightlines is relatively small, 
there is excellent agreement between KGPS-Full and \citet{turner14}, as expected. At smaller $D_{\rm tran}$, the QSO sightline measurements are not as detailed, although
they remain statistically consistent given the larger uncertainties. 
Although the QSO spectra used by \citet{turner14} are far superior to the KGPS galaxy spectra
in both resolution ($\sigma_{\rm  v} \simeq 8$ \kms\ versus $\sigma_v \simeq 190$ \kms) and SNR ($\simeq 100$ versus $\simeq$a few), the QSO-based measurements
are less precise for the ensemble.
This is because both the local continuum level and the net absorption profile contribute to the uncertainty. In the case of the QSO sightlines, the stochastic variations in 
the mean \lya\ forest opacity in the QSO spectra in the vicinity of \zfg\ modulate the apparent continuum against which excess \lya\ absorption at \zfg\ is evaluated, and have a large sample variance in the absorption strength at fixed $D_{\rm tran}$.
The points in Figure~\ref{fig:ew} from S2010 were measured using stacked LRIS spectra of a subset of the current KGPS sample, with a comparable range of $\zfg$ and overall
 galaxy properties, and are consistent with our measurements within the uncertainties.

Figure~\ref{fig:ew} clearly shows that excess \lya\ absorption is detected to transverse distances of at least $D_{\rm tran} \simeq 3500$ pkpc. 
A single power law reasonably approximates the dependence of \wlya\ on $D_{\rm tran}$, with power law index $\beta = -0.40 \pm 0.01$:
\begin{eqnarray}
\label{eqn:wlya_vs_D}
    \mathrm{\wlya\ } = (0.429 \pm 0.005 \textrm{~\AA}) \left( \frac{D_\mathrm{tran}}{\mathrm{~pMpc}} \right)^{(-0.40\pm 0.01)}. 
\end{eqnarray}
This power law is also shown in Figure \ref{fig:ew}. 
There is evidence from the new KGPS results for subtle differences in slope over particular ranges of $D_{\rm tran}$. Specifically, at $D_{\rm tran} < 100$ pkpc, $\beta = -1.0\pm 0.1$; for $100 < D_{\rm tran}/{\rm pkpc}  \le 300$, $\beta = 0.0 \pm 0.1$; and 
$300 < D_{\rm tran}/{\rm pkpc}  \le 2000$, $\beta = -0.48 \pm 0.02$. Details are discussed in \S\ref{sec:discussion} below. 

\subsection{Kinematics}
\label{sec:kinematics}

In order to interpret observations of the kinematics of \lya\ absorption, one must first develop a detailed understanding of the
effective spectral resolution, including the net contribution of redshift uncertainties. 
We showed in \S\ref{sec:redshift} that redshift uncertainties are negligible for the KGPS-\zneb\ galaxies, but that $\simeq 50$\%
of the KGPS-Full sample whose \zfg\ was estimated using equations~\ref{eqn:case1}--\ref{eqn:case3} have larger redshift uncertainties, $\sigma_{\rm z} \simlt 140$ \kms.   
In the latter case, redshift uncertainties would make a non-negligible contribution to the effective spectral resolution of the stacked spectra. 
We determined the effective spectral resolution applicable to composites formed from the KGPS-\zneb\ and KGPS-Full samples separately, 
using procedures whose details are described in Appendix~\ref{app:resolution}. 

In fact, Appendix~\ref{app:resolution} concludes that the effective spectral resolution, including the contribution of
redshift uncertainties, is nearly identical for the two subsamples:  $\sigma_{\rm eff} = 189$~\kms\ (KGPS-\zneb) and $\sigma_{\rm eff} = 192$~\kms
(KGPS-Full). As explained in the appendix, this suggests that the contribution of redshift uncertainties to the 
effective spectral resolution is small compared to that of the instrumental resolution\footnote{The KGPS-Full and KGPS-\zneb\ samples comprise a comparable mix of the two LRIS instrumental configurations.}: 
$\sigma_{\rm z,eff} \simeq \sqrt{\sigma_{\rm eff}^2({\rm Full}) - \sigma_{\rm eff}^2(z_{\rm neb})} \simlt  60$ \kms.  
This implies that the additional degradation in the effective spectral resolution caused by the use of calibrated $z_\mathrm{UV}$ redshifts
for the $\sim 50$\% of the KGPS-Full sample lacking measurements of \zneb\ is not significant. The contribution is
smaller than the redshift error estimated in \S\ref{sec:redshift} for galaxies with only rest-UV measurements; we
suggest that the reason for the apparent discrepancy is that the earlier 
estimate included both the uncertainty in the mean offsets between $z_\mathrm{UV}$ and $z_\mathrm{neb}$, and the noise associated with 
the measurement of spectral features in individual spectra, which are effectively averaged out in applying the fits in equations 1-4. 
Nevertheless, we retain the KGPS-$z_\mathrm{neb}$ sample as a sanity check to eliminate unknown systematic uncertainties that may be present in the KGPS-Full sample.

As discussed above (\S \ref{sec:ew}), for composite spectra of modest spectral resolution, much of the physical information that could be measured from 
high-resolution QSO spectra through the same ensemble of sightlines is sacrificed in order to increase the spatial sampling. However, from
the smaller samples of QSO sightlines through the CGM of a subset of the KGPS galaxies \citep{rudie12a},  
we know that the \lya\ absorption profile resolves into a number of individual components of velocity width $\sigma_v \simeq 20-50$ \kms\ 
that do not fully occupy velocity space within $\pm 700$ \kms\ of the galaxy's systemic redshift. The total absorption strength 
tends to be dominated by components with \nhi$\simgt 10^{14}$ cm$^{-2}$, whose \lya\ transitions are saturated. Thus, \lya\ profiles in stacked spectra
at low resolution can be usefully thought of as smoothed, statistical averages of a largely bimodal distribution of pixel intensities that is modulated by whether or not a saturated absorber is present.  

Nevertheless, the \lya\ line profiles in the KGPS composites encode useful information on the total \lya\ absorption as a function of line-of-sight velocity
relative to the galaxy systemic redshifts, and the large number of pairs allows us to map these parameters as a 
function of $D_{\rm tran}$. To describe the absorption profiles with sufficient dynamic range, we use the apparent optical depth, defined as
\begin{eqnarray}
\tau_\mathrm{ap}(v) = -\ln \frac{F(v)}{F_\mathrm{cont}(v)}, 
\end{eqnarray}
where $F(v)$ is the flux density of the composite spectrum as a function of velocity, $F_{\rm cont}(v)$ is the continuum level, and $v$ is the line-of-sight velocity relative to \zsys.
For continuum-normalised spectra, $F_{\rm cont}(v)=1$. Since we know that \ion{H}{I} gas is clumpy and normally saturated, $\tau_\mathrm{ap}$ is relatively weakly dependent on $N_\mathrm{HI}$ for $N_\mathrm{HI} > 10^{14} \mathrm{~cm}^{-2}$. Rather, $\tau_\mathrm{ap}(v)$ is modulated
by both the covering fraction ($f_c$) of the clumps, and the typical line of sight velocity range over which significant absorption is present.  

\begin{figure*}
\centering
\includegraphics[width=8.5cm]{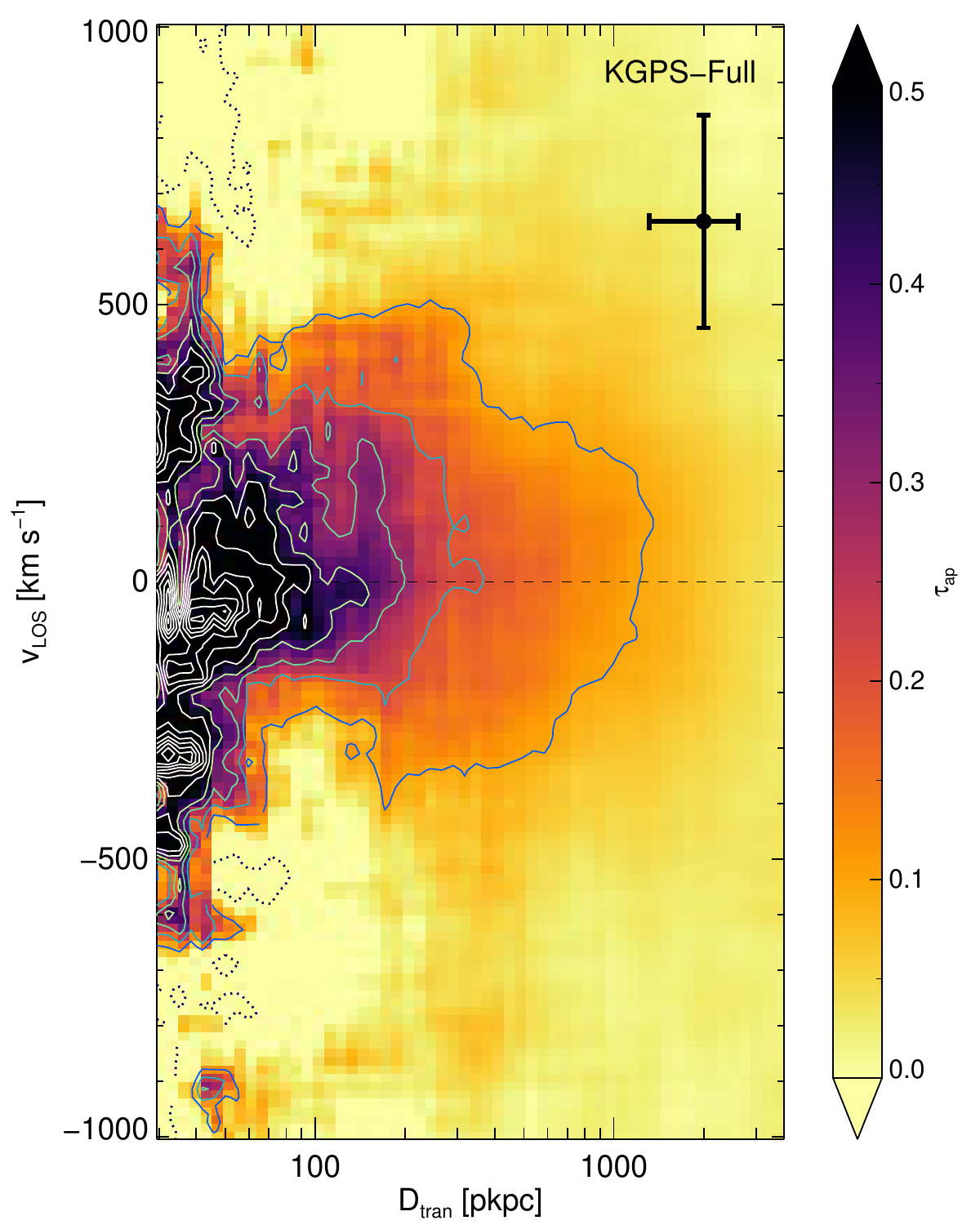}
\includegraphics[width=8.5cm]{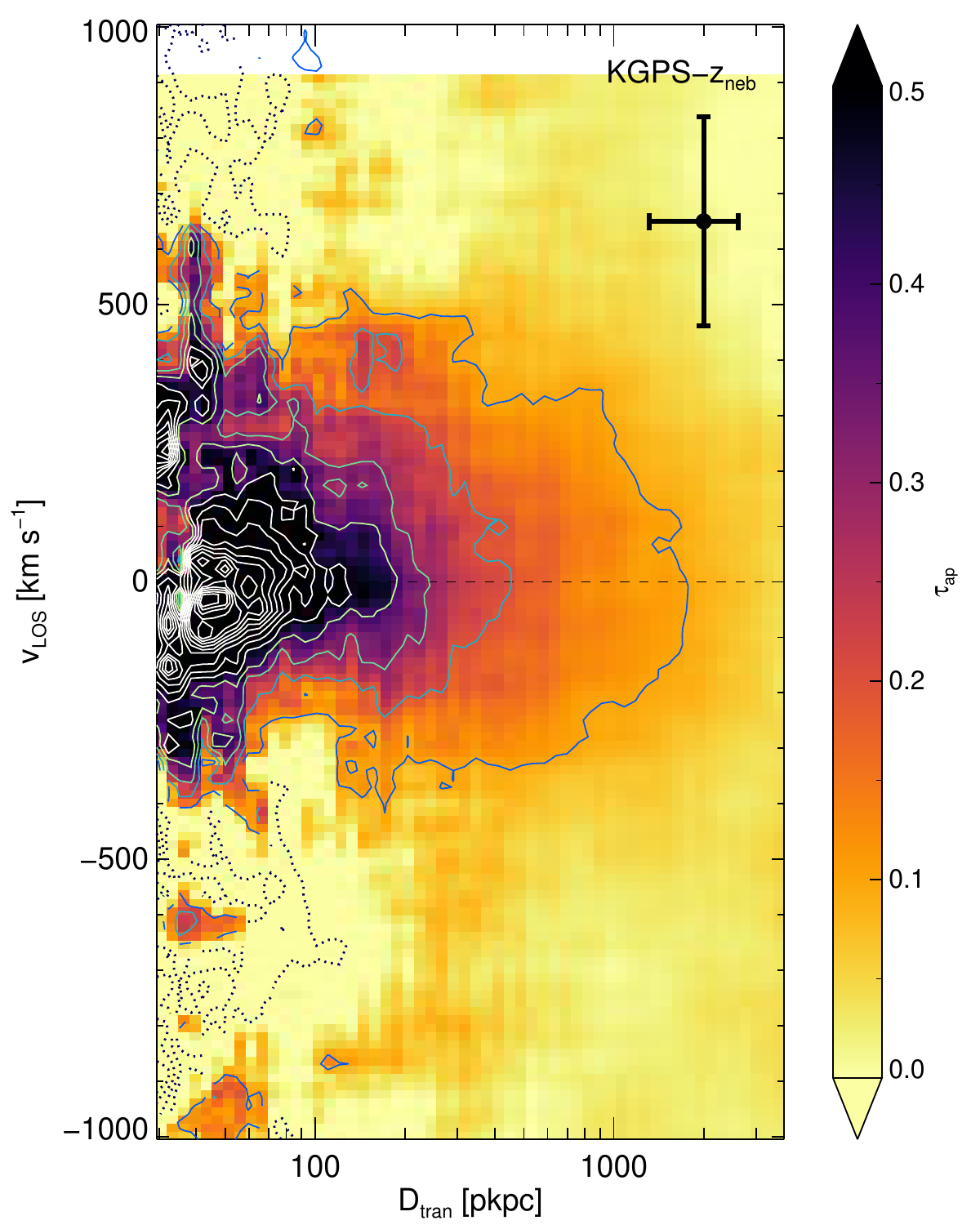}
\caption{  The line-of-sight velocity structure of \ion{H}{I} absorbers around foreground galaxies. Each column of pixels in each plot corresponds to a measurement
of $\tau_{\rm ap}(v_\mathrm{LOS})$ evaluated from the corresponding composite spectrum in the bin of $D_{\rm tran}$ made from  
the KGPS-Full sample (left) and the KGPS-$z_\mathrm{neb}$ sample (right). The black dots with horizontal error bars show the range of $D_{\rm tran}$ used to make each column of the map; the effective 
velocity resolution of the map is shown as a vertical error bar. Solid contours 
correspond to positive optical depth $\tau_\mathrm{ap}$, with dotted contours indicating negative values, 
which are consistent with the 1-$\sigma$ uncertainties. 
The contour levels are separated by $\Delta\tau_{\rm ap} = 0.1$.}
\label{fig:2dabs_unfold}
\end{figure*}

Figure~\ref{fig:2dabs_unfold} shows the map of $\tau_\mathrm{ap}$ in the $v_\mathrm{LOS}$ - $D_\mathrm{tran}$ plane for the KGPS-Full and the KGPS-$z_\mathrm{neb}$ sample. 
Each column in the maps represents a stacked spectrum for sightlines within a bin of $D_{\rm tran}$, each of which 
represents an equal logarithmic interval $\Delta(\log D_\mathrm{tran}) = 0.3$, as in Figure \ref{fig:ew}. 
Assuming that the cumulative number of pairs is proportional to $D_\mathrm{tran}^2$ (see Figure~\ref{fig:pairnum}), 
the average $D_\mathrm{tran}$ weighted by number of sightlines 
within a bin is $\sim 0.18 \mathrm{~dex}$ greater than the lower bin edge, and $\sim 0.12 \mathrm{~dex}$ smaller than the higher bin edge. 
The map of $\tau_\mathrm{ap}$ vs. $D_{\rm tran}$ is symmetric between the blue and red sides across all $D_\mathrm{tran}$ except between $\sim 50\mathrm{~pkpc}$ and $\sim 200 \mathrm{~pkpc}$, where the total $\tau_{\rm ap}$ for $v_{\rm LOS} > 0$ (i.e., the side redshifted with respect to the galaxy systemic redshift) is larger by $>50\%$
than that of the blueshifted side. The possible origin of the asymmetry is discussed in \S\ref{sec:asymmetry}.  

\begin{figure*}%
\centering
\includegraphics[width=8.5cm]{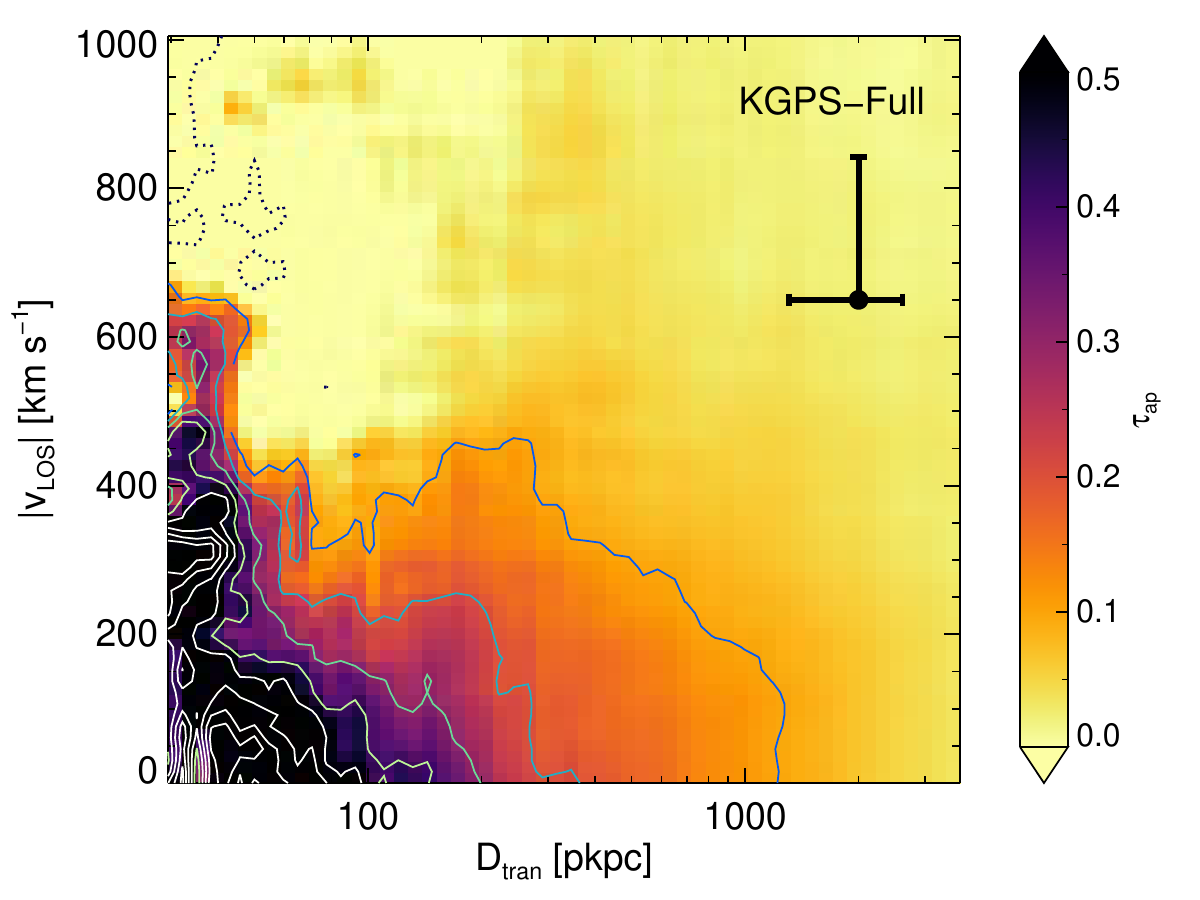}
\includegraphics[width=8.5cm]{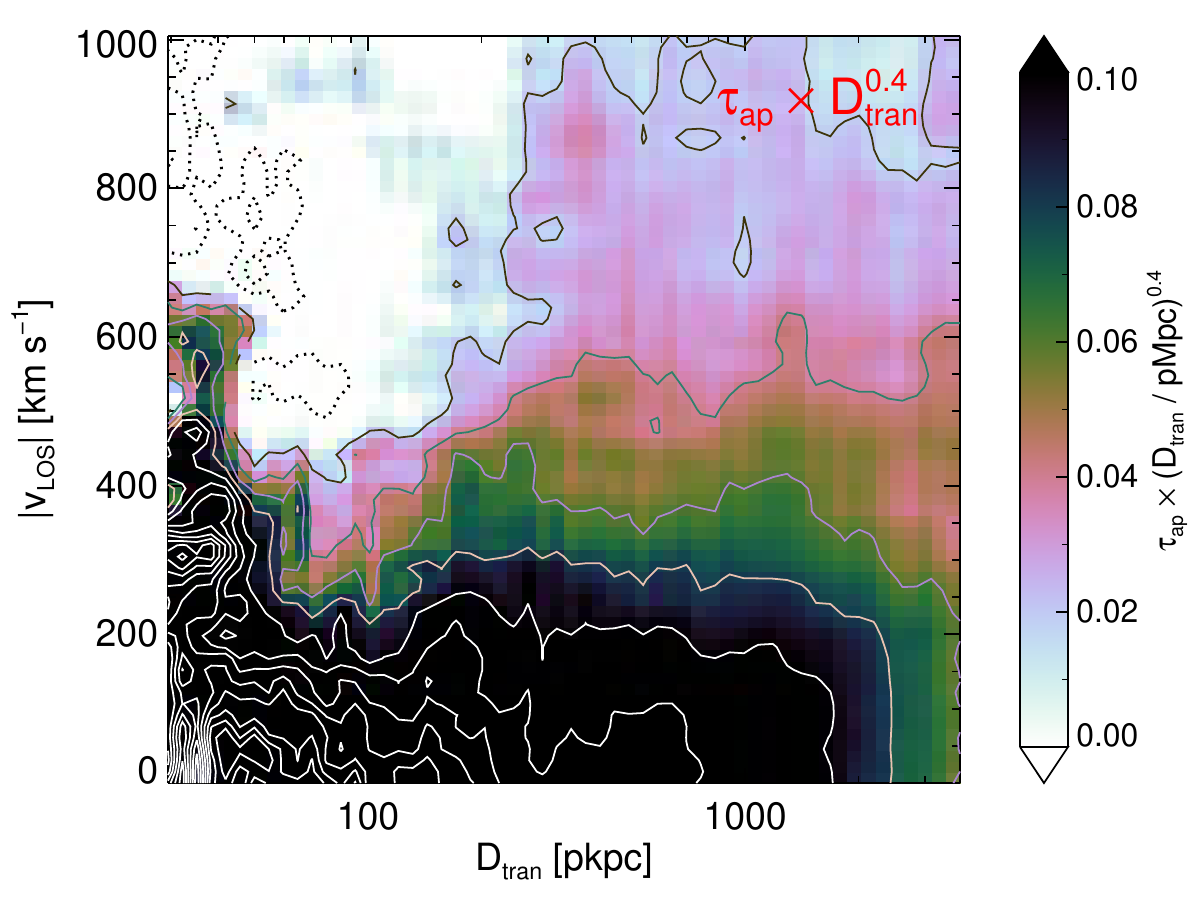}\\
\includegraphics[width=8.5cm]{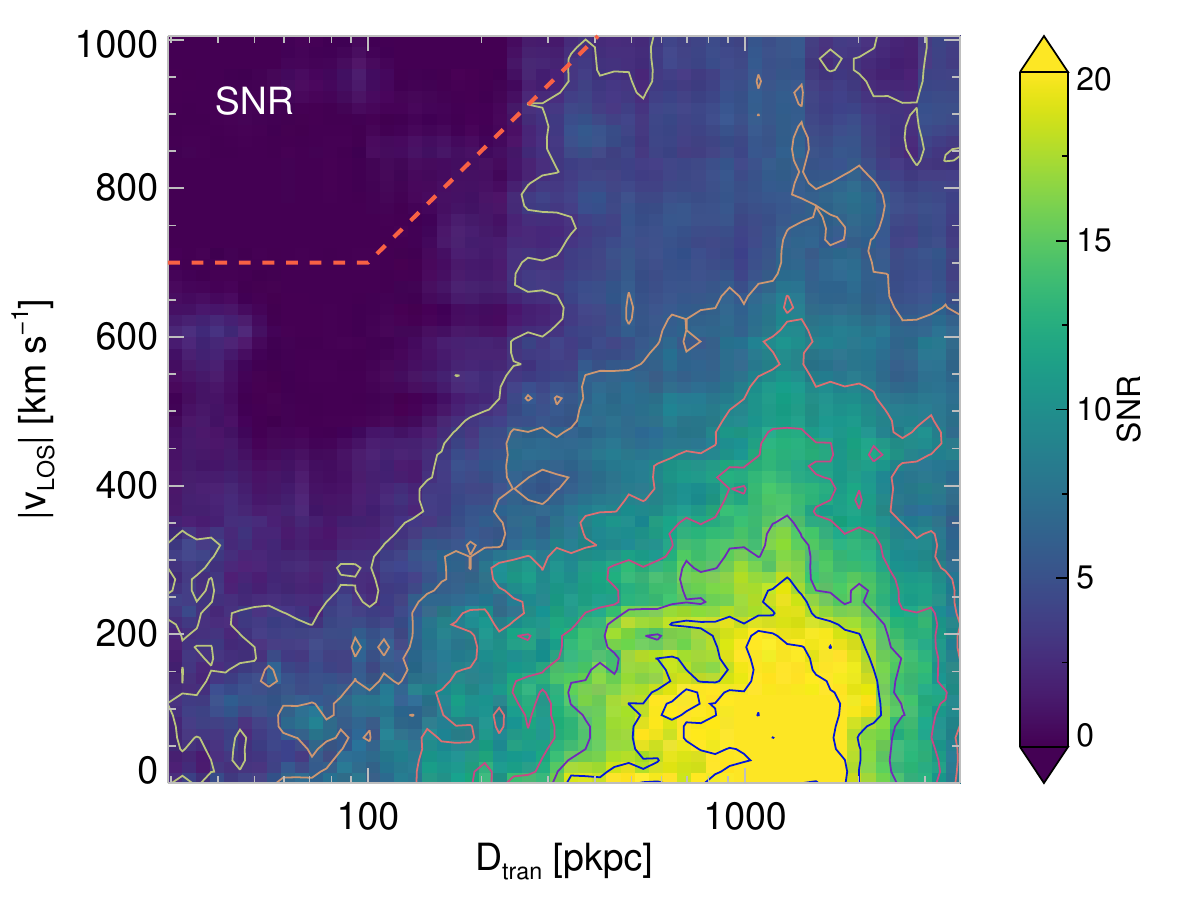}
\caption{  Maps of $\tau_\mathrm{ap}$ with the blue and red sides folded to increase the signal-to-noise ratio. The three maps are (left to right, top to bottom): $\tau_\mathrm{ap}(|v_{\rm los}|,D_{\rm tran})$,  
$\tau_\mathrm{ap}\times D_\mathrm{tran}^{0.4}$ (to better illustrate the structure at large $D_{\rm tran}$), and the map of the SNR per \AA\ 
of the $\tau_{\rm ap}$ measurement. The contour decrements are 0.1 for the $\tau_\mathrm{ap}$ map, 0.02 for the 
$\tau_\mathrm{ap}\times D_\mathrm{tran}^{0.4}$ map, and 3 for the SNR map. The half-aperture in which the \wlya\ is measured in \S\ref{sec:ew} is shown as the orange dashed line in the SNR map.} 
\label{fig:2dabs}
\end{figure*}

\begin{figure}
\centering
\includegraphics[width=8.5cm]{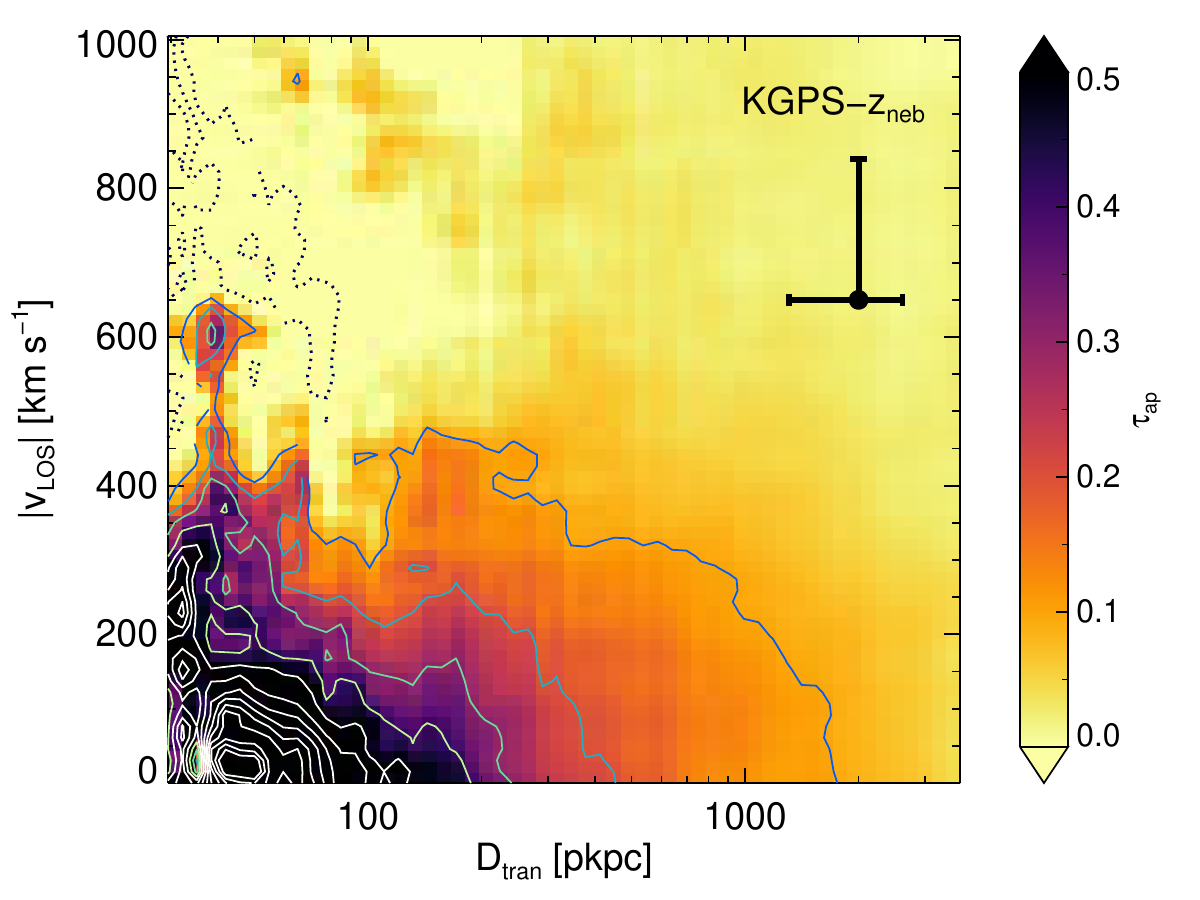}
\caption{Same as the top-left plot of Figure \ref{fig:2dabs}, but for the KGPS-$z_\mathrm{neb}$ sample.}
\label{fig:2dabs_zneb}
\end{figure}

To further increase the SNR, the red and blue sides of the $\tau_\mathrm{ap}$ map are folded together in Figure \ref{fig:2dabs} for the KGPS-Full sample and Figure \ref{fig:2dabs_zneb} for the KGPS-$z_\mathrm{neb}$ sample. 
Also shown in Figure \ref{fig:2dabs} (top-right) is the map of $\tau_\mathrm{ap}\times D_\mathrm{tran}^{0.4}$, in order to
accentuate the structure at large $D_\mathrm{tran}$; the power $0.4$ was chosen to approximately counteract the overall decrease in absorption strength with 
increasing $D_{\rm tran}$ (Equation~\ref{eqn:wlya_vs_D}). 
Equal size bins in $\Delta \log D_{\rm tran}$ means that fewer sightlines are being averaged when $D_\mathrm{tran}$ is smaller, so that the uncertainties are larger
for pixels on the lefthand side of the $\tau_{\rm ap}$ maps: the SNR (per \AA) maps are shown in the bottom panels of Figure~\ref{fig:2dabs}. The SNR was calculated
by bootstrap resampling the ensemble of spectra contributing to each $D_{\rm tran}$ bin 2000 times, and then rescaling so that the units are SNR per \AA\ 
in the rest frame of $z_\mathrm{fg}$. 
The values in the uncertainty estimation are also supported by the strength of apparent negative $\tau_\mathrm{ap}$ (dotted contours) in the $\tau_{\rm ap}$ maps, 
which are consistent with the 1-$\sigma$ errors from the bootstrap determinations. 

The maps of the KGPS-Full and KGPS-$z_\mathrm{neb}$ samples show similar features and are consistent within the uncertainties after convolution of the KGPS-$z_\mathrm{neb}$ map with a $\sigma_z=60 \mathrm{~km~s}^{-1}$ Gaussian kernel in the $v_\mathrm{LOS}$ direction. 

Figure~\ref{fig:2dabs} confirms trends in \lya\ absorption strength versus $D_{\rm tran}$ based on QSO sightlines in KBSS \citep{rudie12a,rakic12, turner14} -- a sample with
much smaller number of foreground galaxies --  and similar surveys at comparable \citep{adelberger03,adelberger05,tummuangpak14, bielby17} or lower redshifts \citep{ryanweber06, tejos14}. It is also consistent with analytic modeling from \citet{kakiichi2018} and the mock observations of QSO sightlines from the Evolution and Assembly of Galaxies and their Environments (EAGLE) simulations \citep{rakic13, turner17}, who proposed measuring the typical halo mass of the host galaxies based on matching the observed
pixel optical depth as a function of $D_{\rm tran}$ to dark matter halos in the simulation. 

The KGPS maps show that, at small $D_{\rm tran}$, excess Ly$\alpha$ absorption reaches velocities $|v_\mathrm{LOS}| \simgt 500 \mathrm{~km ~s}^{-1}$ relative to
 the systemic redshifts of foreground galaxies, and that excess absorption over and above that of the average IGM extends to transverse distances of at least 3.5 pMpc in $D_\mathrm{tran}$ direction. 
Figure~\ref{fig:2dabs} suggests several interesting features in the 2-D maps of $\tau_{\rm ap}$, moving from small to large $D_{\rm tran}$:  
\begin{enumerate}
    \item A region with high line-of-sight velocity spread ($\langle |v_{\rm LOS}|\rangle \simgt 300-400$ \kms) 
on transverse distance scales $D_{\rm tran} \simlt 50$ pkpc. 
    \item An abrupt compression of the $v_\mathrm{LOS}$ profile beginning near $D_{\rm tran} \simeq 70$ pkpc, extending to $D_{\rm tran} \sim 150$ kpc, with
a local minimum near $D_{\rm tran} \simeq 100$ pkpc. 
    \item A gradual broadening of the Ly$\alpha$ velocity profile beginning at $D_\mathrm{tran} \simgt 150 \mathrm{~pkpc}$ 
with $|v_\mathrm{LOS}| \sim 500 \mathrm{~km~s}^{-1}$ and extending to $ > 4 \mathrm{~pMpc}$ with $\sim 1000 \mathrm{~km~s}^{-1} $ (most evident in 
the upper righthand panel of Figure~\ref{fig:2dabs}.) 
\end{enumerate}

We show in \S\ref{sec:model} that feature (iii) is a natural consequence of the Hubble expansion coupled with decreasing \ion{H}{I} overdensity, causing 
the absorption to become weaker and broader as $D_\mathrm{tran}$ increases. Eventually, one would expect that the profile would broaden and weaken until
it becomes indistinguishable from the ambient IGM -- recalling that $\tau_{\rm ap}$ is the {\it excess} \lya\ optical depth over the intergalactic mean. 

We address the likely origin of each of the enumerated features in \S\ref{sec:discussion}.

\section{Discussion} 
\label{sec:discussion}

\subsection{Comparison to Cosmological Zoom-In Simulations}
\label{sec:discussion_zoomin}

To aid in the interpretation of the observed $\tau_\mathrm{ap}$ maps, we compared them 
to the distribution of \nhi\ as a function of $v_{\rm LOS}$ 
in a subset of simulations taken from the Feedback In Realistic Environments (FIRE) cosmological zoom-in simulations \citep{hopkins14, hopkins18}. 
The selected simulations are intended to
reproduce LBG-like galaxies at $z\sim 2$; they were originally run at lower spatial resolution using the FIRE-1 feedback model \citep{fauch15}, but have since
been migrated to FIRE-2 with improved mass resolution of $m_\mathrm{gas} \sim m_\mathrm{star} \sim 700$ M$_{\odot}$ (Dong et al 2020, in prep.). 
We randomly selected a single main simulated
galaxy whose halo mass ($M_{\rm h}$), stellar mass ($M_{\ast}$), and star formation rate (SFR) roughly match values inferred for a typical
galaxy in the KGPS sample:  $\log M_{\rm h} / M_\odot \sim 12$, $\log M_{\ast}/ M_{\odot}  \sim 10.5$, and $\langle {\rm SFR}/(M_{\odot} \mathrm{~yr}^{-1}) \rangle \sim 30$ 
at $z \sim 2.2$.

\begin{figure*}%
	\centering
	\includegraphics[width=16cm]{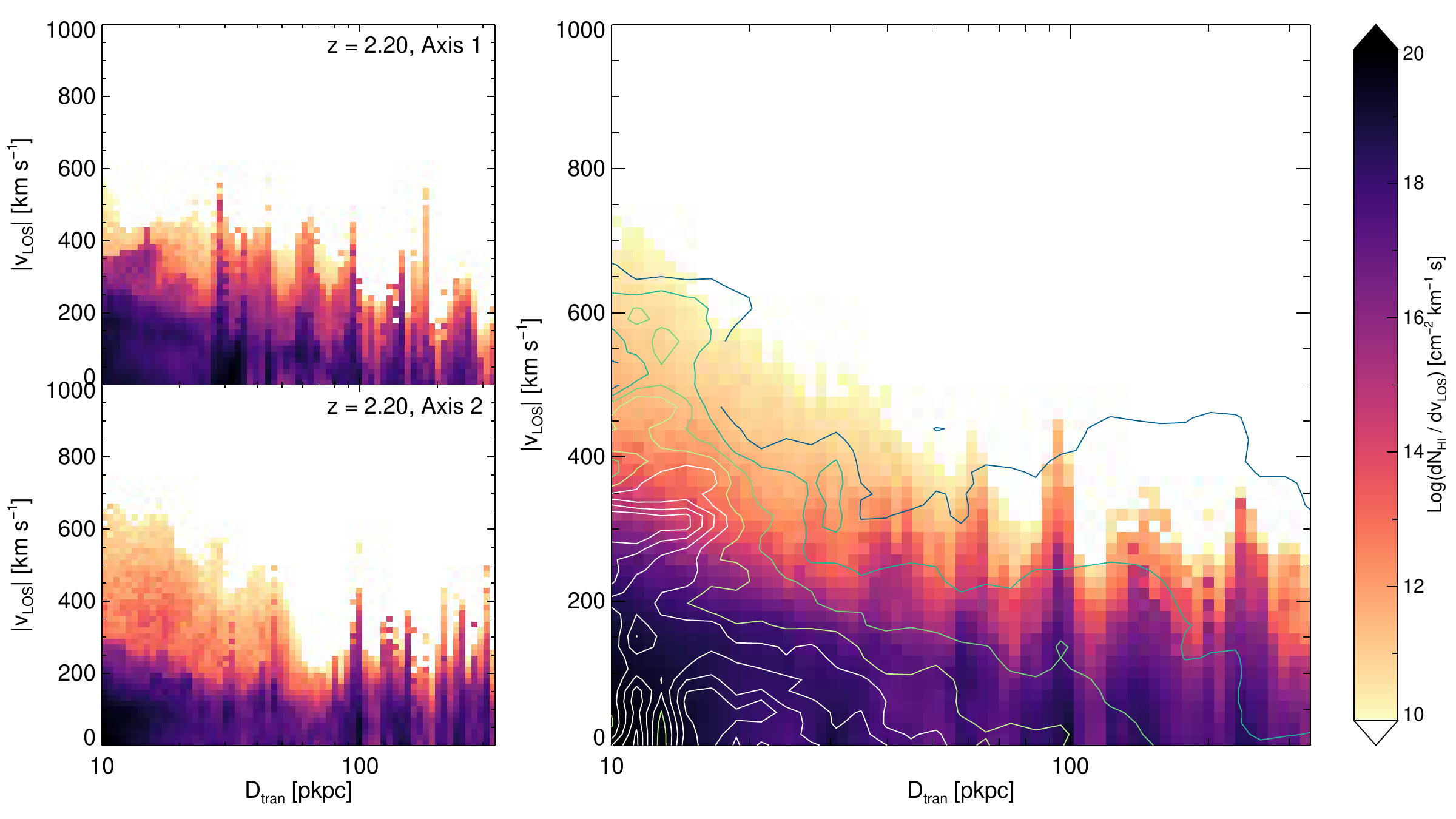}
	\caption{ Maps of $\d N_\mathrm{HI} / \d v_\mathrm{LOS}$ in the FIRE-2 simulation (h350), as seen by an observer, projected
onto the $v_{\rm LOS}$-$D_{\rm tran}$ plane. {\it Left:} Snapshots from
a single time step ($z=2.20$) as viewed from two orthogonal viewing angles. {\it Right:} The median-stack of 22 such maps: 11 time steps
at intervals of $\delta z = 0.05$ for $2.0 \le z \le 2.4$ from each of two
orthogonal viewing angles. Contours based on the observed map of $\tau_\mathrm{ap}$ for the KGPS-Full sample (Figure~\ref{fig:2dabs}) are overlaid for comparison. 
Note that the $D_{\rm tran}$-axis has been zoomed in from that of Figure~\ref{fig:2dabs} because of the limited size (a few times $r_\mathrm{vir}$) of the high-resolution zoom-in region of the FIRE-2 simulation. } 
\label{fig:fire}
\end{figure*}

We selected 11 snapshots evenly distributed in redshift between $z=2.4$ and $z=2.0$\footnote{The interval $\delta z =0.05$ corresponds to intervals of
cosmic time in the range $\delta t \simeq 70 \pm 10$ Myr over the redshift range included in our analysis.}. In each time step, we chose two random (but
orthogonal) viewing angles, and calculated the projected quantity  $\d N_\mathrm{HI} / \d v_\mathrm{LOS}$ (a proxy for the observed $\tau_{\rm ap}$) projected
onto the ``observed'' 
$D_\mathrm{tran}$ -- $v_\mathrm{LOS}$ plane; example maps for the $z=2.20$ time step are shown in the lefthand panels of Figure~\ref{fig:fire}.    
We then median-stacked the 22 $N_\mathrm{HI}$ maps (11 snapshots for each of two viewing angles); the result is shown in the righthand panel 
of Figure~\ref{fig:fire}). 
Note that the range shown on the $D_{\rm tran}$ axis is smaller than that used for the observations (e.g., Figure~\ref{fig:2dabs}) because 
of the limited size of the full-resolution volume of the zoom-in simulation.

Thanks to the large dynamic range in the simulation, one can see 
that the highest $v_{\rm LOS}$ values ($v_{\rm LOS} \sim 400-700$ \kms) are found within $D_{\rm tran} \lesssim 50$ pkpc and are associated 
with material having total \nhi\ $\simeq \d\nhi/\d v\times \Delta v \simeq 10^{14.5-15.0}$ cm$^{-2}$, whereas the slower 
material with $v_{\rm LOS} < 200$ \kms\ has \nhi\ typically 100-1000 times larger. While essentially all pixels in the map that are
significantly above the background in Figure~\ref{fig:fire} would give rise to \lya\ absorption detectable in KGPS, the low spectral
resolution coupled with the inherent loss in dynamic range in $\tau(\lya)$ for log\nhi$\simgt 14.5$ makes a detailed quantitative comparison 
of the KGPS observed map and the simulation map more challenging. Part of the difficulty stems from the limited depth along the line of sight ($\sim 1$ pMpc) 
over which the high-resolution zoom-in simulations were conducted, meaning they could be missing potential high-velocity material in the IGM. 
However, we conclude that the general morphology of the \nhi\ map from the averaged zoom-in simulations is qualitatively 
consistent with that of the observed $\tau_{\rm ap}$ map. 

Individual snapshots provide insight into the physical origin of features in the time-averaged map. 
For example, the most prominent features in Figure~\ref{fig:fire} are the vertical ``spikes'' evident 
at $D_{\rm tran}/{\rm pkpc} \simgt 50$ -- these are due to the passing of satellite galaxies through the simulation
box, giving rise to line-of-sight components of velocity due both to their motion relative to the central galaxy and
their internal gas motions, including outflows; e.g., \citet{faucher16,angles17,hafen2019}. 
Some of these features remain even in the median of 22 snapshots (righthand panel of Figure~\ref{fig:fire}),
caused by gradual decay of orbits of galaxies destined to merge with the central galaxy.  
Such features would not be expected to remain in an average over many galaxies, each observed at a particular time (as in the KGPS data), 
except insofar as there might
be a characteristic range of galactocentric radius and relative velocity affected by satellites, so that a net signal might be 
detected statistically.

With the above {\it caveats} in mind, the envelope of $v_{\rm LOS}$ as a function of $D_{\rm tran}$ in the simulation
resembles that of the KGPS observations in several respects: both the highest velocities and the highest optical depths occur within
the central 50 pkpc, with a local minimum in the range of $v_{\rm LOS}$ over which significant \ion{H}{I} is present somewhere in
the range $50$--$100$ pkpc, reminiscent of the expected virial radius for a dark matter halo of $\log(M_{\rm h}/M_{\odot}) \sim 12$ of
$r_{\rm vir} \simeq 75$--$95$ pkpc. 

\begin{figure}
\centering
\includegraphics[width=8.5cm]{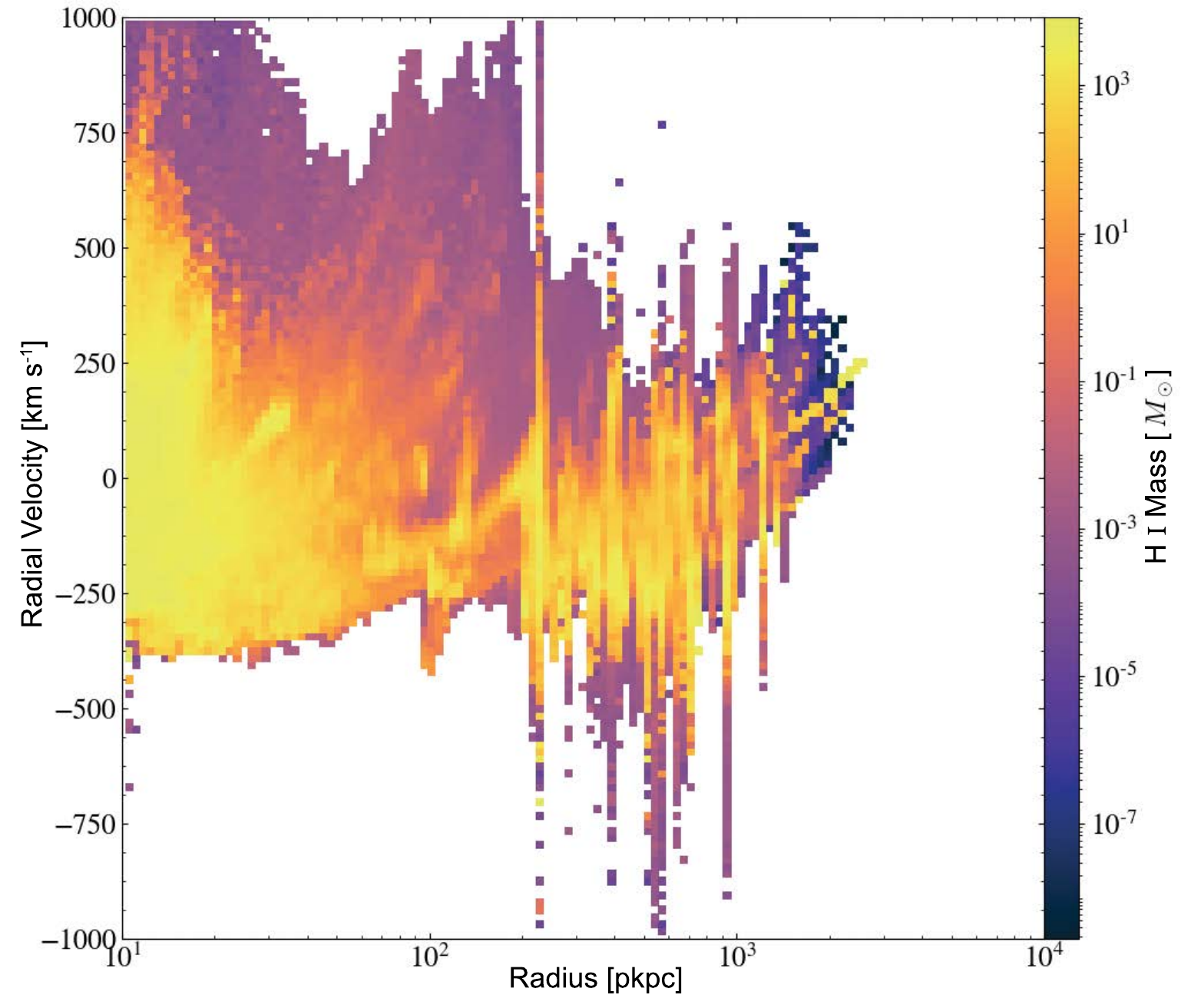}
\caption{The galactocentric radial velocity versus galactocentric radius for neutral hydrogen 
in the same $z=2.20$ snapshot as shown in the lefthand panels of Figure~\ref{fig:fire}. Positive and negative values
indicate net outward and net inward radial motion, respectively. The vertical strips are due to the outflow of the satellite galaxies. The colourbar represents the total \ion{H}{I} mass in each pixel. The gradual decrease of \ion{H}{I} content beyond 500 pkpc is artificial because of the limited volume within which the gas simulation is conducted. }
\label{fig:fire2}
\end{figure}  

Further insight can be gleaned with reference to Figure~\ref{fig:fire2}, which is essentially a histogram of \ion{H}{I} as a function
of galactocentric radius and radial velocity\footnote{The center of the galaxy is determined by the center of mass of the dark matter halo.}. 
Negative (positive) radial velocities indicate motion toward (away from) the center of the galaxy.  
This type of diagram makes it easier to distinguish gas with substantial outflow velocities, which cause the clear asymmetry of the radial
velocity distribution relative to the galaxy center of mass rest frame. In this particular case, there is neutral hydrogen with $v_{r} > 300$ \kms\
at galactocentric radii from $r=0$ to $r \simeq 200$ pkpc, but essentially no gas with $v_{\rm r} < -300$ \kms\ except for that due to satellite
galaxies. If one looks at the same diagram in successive time steps, it is clear that the high velocities are associated with episodes of
high SFR, and that this particular galaxy is experiencing such an episode at the present time step, which produces most of the
high velocity gas with $r \simlt 50$ kpc; the ``plume'' of high velocity material that peaks near $r=100$ pkpc is a remnant of a similar
episode that occurred $\sim 100$ Myr earlier that has propagated to larger galactocentric radii with somewhat reduced $v_{\rm r}$.
Such episodic star formation and outflows are typical in high-$z$ galaxies in the FIRE simulation \citep{muratov15}. Such plumes are also seen in IllustrisTNG simulations \citep{nelson19}, although they are generally due to the implementation of AGN feedback in higher-mass
halos, and thus have even larger velocities.
One can also see an accretion stream that has $v_{\rm r} \simeq 0$ at $r \simeq 200$ kpc, and evidently has accelerated to $v_{\rm r} \simeq -200$
\kms\ by $r \sim 50$ pkpc. However, the bulk of the \ion{H}{I} mass within the virial radius is not obviously accreting or outflowing, 
with $|v_{\rm r}| \simlt 250$ \kms.

A halo with $M_{\rm h} = 10^{12}$ M$_{\odot}$ and $r_{\rm vir} \simeq 90$ kpc has a circular velocity of $\simeq 220$ \kms, so that if gas were
on random orbits one would expect to measure a 1-D velocity dispersion (more or less independent of radius, for realistic mass profiles) 
of $\sigma_{\rm 1D} \simeq 220/\sqrt{3} \simeq 130$ \kms,
with $\sim 95$\% of particle velocities expected to lie within $|v_{\rm LOS}| \simlt 260$ \kms. These expectations could be modulated by the prevalence
of mostly-circular or mostly-radial orbits, of course. However, in a statistical sense, gravitationally-induced motions could contribute a fraction of the most extreme velocities (i.e., $|v_{\rm r}| > 300$ \kms\ within $\simlt 2R_{\rm vir}$), but are not enough to make up the whole.

\subsection{A Simple Analytic Model}
\label{sec:model}

\subsubsection{Context}
\label{sec:context}

Inspired by comparisons to the cosmological zoom-in simulations, and to offer an explanation for the general shape of the $\tau_\mathrm{ap}$ map before discussing the details, 
we constructed a two-component analytic model intended to capture the salient features of Figure~\ref{fig:2dabs}.

To construct a model in 3-D physical space, a parameter must be chosen to represent \lya\ absorption strength. 
As discussed in \S \ref{sec:stacking}, $\tau_{\rm ap} (v_{\rm LOS})$ is affected by a combination of the total \nhi, which is known
to depend on $D_{\rm tran}$ (e.g., \citealt{rudie12a}), the average distribution of $v_{\rm LOS}$ at a given $D_{\rm tran}$, and   
the fraction of sightlines that give rise to detectable absorption at a given $D_{\rm tran}$. 
The relative importance of these effects depends on galactocentric distance (i.e., $D_{\rm tran}$) in a complex manner that cannot
be resolved from an ensemble of low-resolution spectra, which also cannot be expected to reveal the level of detail that could be measured from individual
sightlines observed at very high spectral resolution. However, some general statements about the ``sub-grid'' behavior of \lya\ absorption may be
helpful in providing some intuition.

At large $D_\mathrm{tran}$, given the minimum total equivalent width detected of
$\simeq 0.2$ \AA\ (see Figure~\ref{fig:ew}), the lowest column density that could be measured for a single \lya\ absorption line is $\log \nhi \simeq 13.6$, 
assuming a linear curve of growth; for 
a typical value of the Doppler parameter $b_{\rm d} \simeq 25$ \kms ($\sigma_{\rm d} \equiv b_{\rm d}/\sqrt{2} \simeq 17.7$ \kms), the single line
would have an optical depth at line center of $\tau_0 \simeq 2.4$, which if resolved would produce a minimum flux density relative to the continuum
of $F_{\nu,0} \sim 0.09$. However, given the effective spectral resolution of $\sigma_{\rm eff} \simeq 190$ \kms, the observed line profile of the
same line would have $\sigma_{\rm LOS} \simeq 190$ \kms\ and central {\it apparent} optical depth of only $\tau_{\rm ap,0} \simeq 0.2$, or 
$F_{\nu,0}/F_{\nu,{\rm cont}} \simeq 0.82$. In fact, at large $D_{\rm tran}$ we measure 
$\sigma_{\rm LOS} \simeq 600$ \kms, and a maximum apparent optical depth $\tau_{\rm ap}\simlt 0.06$.  This suggests that at large $D_{\rm tran}$ we 
are measuring a small total \nhi\ excess [log(\nhi/cm$^{-2}) < 14$] spread over a large range of $v_{\rm LOS}$; the apparent line width 
is best interpreted as a probability distribution in $v_{\rm LOS}$ of the small excess absorption over that of the general IGM.

At small $D_{\rm tran}$, the situation is very different; in most cases, if observed at high spectral resolution, one
would see several components within a few hundred \kms\ of the galaxy systemic velocity, most of which would be strongly saturated ($\log \nhi/{\rm cm}^{-2} \simgt 14.5$)
and complexes of absorption could often produce large swaths of velocity space with $F_{\nu}=0$ \citep[e.g.,][]{rudie12a}. At $D_{\rm tran} \simlt 50$ kpc, the total $\wlya \simeq 2$\AA\, 
with maximum $\tau_{\rm ap,0} \simeq 1$ and $\sigma_{\rm LOS} \sim 320$ \kms. Once saturation occurs, the equivalent width contributed by individual absorbers 
grows very slowly with increasing \nhi\ until Lorentzian damping wings begin to become important ($\log(\nhi/{\rm cm}^{-2}) \simgt 19.0$). At high spectral
resolution, ${\rm exp}({-\tau_0}) \simeq 0$ independent of \nhi\, so that additional absorbers in the same range of $v_{\rm LOS}$ might have little or no effect
on the absorption profile, especially when observed at low spectral resolution.  However, a relatively small amount of \ion{H}{I} with larger $|v_{\rm LOS}|$ --
if it is a common feature of $D_{\rm tran} \simlt 50$ pkpc sightlines -- would be easily measured. 

\subsubsection{Parametrisation} 
\label{sec:parameters}

\begin{figure}
    \centering
    \includegraphics[width=7cm]{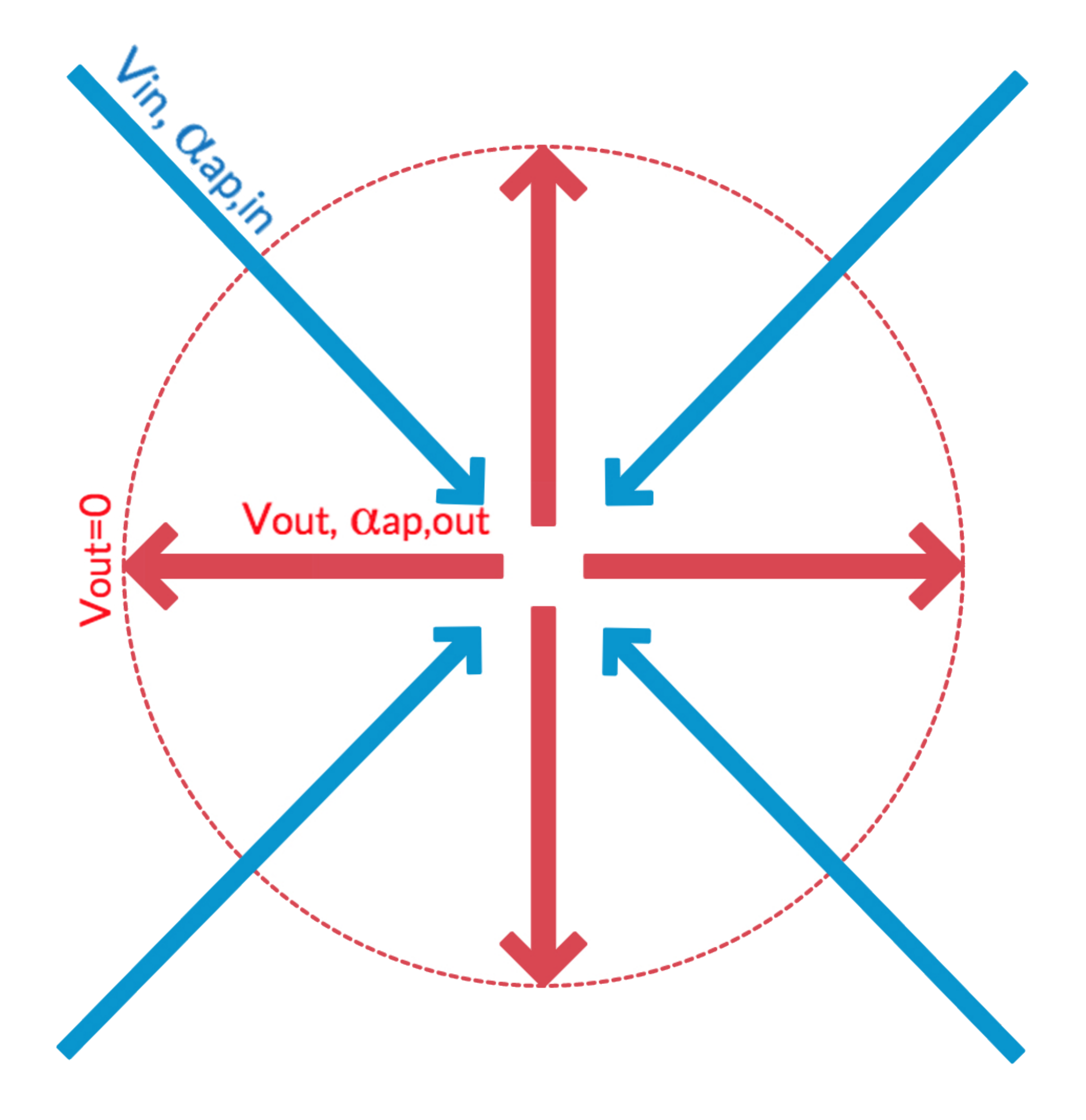}
    \caption{ A cartoon illustration of the parametrization of the analytic model described in \S\ref{sec:parameters}. 
The model comprises two isotropic, non-interacting, purely radial components: ``outflow'' (red) and ``inflow'' (blue). 
Each component has free parameters describing its radial velocity profile $v(r)$ and apparent absorption coefficient $\alpha_\mathrm{ap}(r)$ (defined
in equations~\ref{eqn:alpha_ap},\ref{eqn:beta}, and \ref{eqn:beta2}. The outflow component is truncated at the point that it slows to $v_\mathrm{out}=0$.}
  \label{fig:diagram}
\end{figure}

Figure \ref{fig:diagram} illustrates the parametrisation of our model. In the interest of simplicity, our model makes no 
attempt to capture the detailed radiative processes \citep[e.g.,][]{kakiichi2018} of individual absorption components that would be revealed by 
high resolution QSO spectra. Instead, in order to  
unify the two extreme scenarios described in \S\ref{sec:context} above, 
we treated \ion{H}{I} in the CGM as a continuous medium in which the absorption strength   
per unit pathlength is represented by an absorption coefficient, 
\begin{eqnarray}
\label{eqn:alpha_ap}
	\alpha_\mathrm{ap}  = \frac{\d \tau_\mathrm{ap}}{\d l},
\end{eqnarray}
where $\d l$ is the differential path length. In essence, $\alpha_{\rm ap}$ represents the overdensity of \ion{H}{I} relative to the average in the IGM, accounting 
for some of the non-linearity that results from curve-of-growth effects that remain unresolved by the data.  

We assumed that the \ion{H}{I} surrounding galaxies is composed of two isotropic (non-interacting) components: one moves radially outward (``outflow''), 
the other radially inward (``inflow''). Each has $\alpha_{\rm ap}$ parametrised as an independent radial power law:   
\begin{eqnarray}
\label{eqn:beta}
\alpha_\mathrm{ap,out}(r) & = & \alpha_{0,\rm out} r_{100}^{-\gamma_{\rm out}} \\
\alpha_\mathrm{ap,in}(r) & = & \alpha_{0,\rm in} r_{100}^{-\gamma_{\rm in}},
\label{eqn:beta2}
\end{eqnarray}
where $\alpha_{0,{\rm out}}$ and $\alpha_{0,{\rm in}}$ are normalisation constants, $\gamma_{\rm out}$ and $\gamma_{\rm in}$ are power law indices,
and $r_{100} = r/\mathrm{100~pkpc}$ is the galactocentric radius.    

For simplicity, we assumed that the velocity fields of the two components are also isotropic and purely radial; clearly this is unrealistic. However, 
since only the line-of-sight component of gas velocity is measured, random motions of gas moving in the galaxy potential is partly degenerate with 
our treatment of gas accretion. In the context of the simplified model, one can think of the ``inflow'' component as a proxy for all gas motions that are induced by the galaxy's potential. 
Under these assumptions, there is a simple geometric relationship between the line-of-sight component of velocity $v_{\rm LOS}(D_{\rm tran},l)$ at each point along a sightline through
the CGM and the radial velocity $v_{\rm r}(r)$
\begin{eqnarray}
\label{eq:vlos}
    v_\mathrm{LOS}(D_{\rm tran},l)  =  \frac{l}{r}v_{\rm r}(r)~ ,
\end{eqnarray}
where $l$ is the line of sight coordinate distance  
measured from the tangent point where $v_{\rm LOS} = 0$, i.e., where $r=D_{\rm tran}$,   
and in general, $r^2 = l^2+D_\mathrm{tran}^2$. 
Within this paradigm, specification of $v_{\rm out}(r)$, $v_{\rm in}(r)$, $\alpha_{\rm ap,out}(r)$, and $\alpha_{\rm ap,in}(r)$ can be transformed to  
maps of $\tau_{\rm ap}(D_{\rm tran},v_{\rm LOS})$ directly analogous to those in Figure~\ref{fig:2dabs}.   

For the outflow component velocity field $v_{\rm out}(r)$, we assume that the gas has been accelerated  
to an initial ``launch'' velocity $v_1$ at a galactocentric radius $r=1$ pkpc, 
beyond which its trajectory is assumed to be  
purely ballistic (i.e., no pressure gradient or mass loading is accounted for) within an NFW halo \citep{nfw1996} density profile. 
The outflow component is truncated (i.e., $\alpha_{\rm ap, out} = 0$) at the radius where $v_{\rm out}(r)\rightarrow 0$.  
With these assumptions,  
\begin{eqnarray}
\centering
\label{eqn:vout}
    v_{\rm out} (r) & = & \sqrt{v_{1}^2 + A 
{\left(-\ln\frac{R_{\rm s}+1}{R_{\rm s}}+ \frac{1}{r} \ln \frac{R_{\rm s}+r}{R_{\rm s}} \right)}},
\end{eqnarray}
where $R_{\rm s}$ is the NFW scale radius in pkpc, and 
\begin{eqnarray}
\centering
	A & = & \frac{8\pi G\rho_0 R_{\rm s}^3}{1 \mathrm{~pkpc}} \\
	& \simeq & 1.2\times 10^7 \mathrm{~km}^2\mathrm{~s}^{-2}.
\end{eqnarray}
Following \citet{klypin16}, for a $M_h=10^{12} M_\odot$ NFW halo with concentration parameter $c=3.3$ at $z=2.3$, 
$R_s=27 \mathrm{~pkpc}$ (i.e. $R_\mathrm{vir} \simeq 90 \mathrm{~pkpc}$).

For the inflow component, we make the simplifying assumption that $v_{\rm in}(r)$ is just 
a constant velocity offset relative to the Hubble expansion (similar to \citealt{kaiser87}),  
\begin{eqnarray}
\centering
\label{eqn:vin}
	v_\mathrm{in}(r)  =  v_\mathrm{offset} + H(z)r, 
\end{eqnarray}
where $H(z)$ is the Hubble parameter at redshift $z$. 
For our assumed $\Lambda$CDM cosmology and given the median redshift of the KGPS foreground galaxies, $\langle z \rangle =2.2$,
we set $H(z)= 227$ \kms\ pMpc$^{-1}$. 

Given the parametrisation in equations~\ref{eqn:beta}, \ref{eqn:beta2}, \ref{eqn:vout}, and \ref{eqn:vin}, for each realisation of the MCMC we  
projected both components independently to the $|v_\mathrm{LOS}|$-$D_\mathrm{tran}$ plane, in the process of which $\alpha_\mathrm{ap}$ 
was converted to $\tau_\mathrm{ap}$ by integration. 
Subsequently, $\tau_\mathrm{ap,in}$ and $\tau_\mathrm{ap,out}$ are added in $D_{\rm tran} - v_{\rm LOS}$ space, and convolved with 
the effective resolution of the observed $\tau_\mathrm{ap}$ maps as determined in \S\ref{sec:kinematics}. 

\subsubsection{Results}
\label{sec:results} 
The best fit model parameters 
were estimated using a Markov-Chain Monte Carlo (MCMC) method to fit to the observed maps of $\tau_\mathrm{ap}$.
Prior to fitting, in order to reduce pixel-to-pixel correlations, the observed maps  were resampled to a grid with $\Delta v_\mathrm{LOS} = 101~\kms$ and $\Delta \log (D_\mathrm{tran}) = 0.126$. These pixel dimensions represent one standard deviation of the fitted 2-D Gaussian 
covariance profile determined from 
bootstrap resampling of over-sampled maps.
 
Even with the simplified model, we found it was necessary to fit the inflow and outflow serially, rather
than simultaneously, to achieve convergence in the MCMC. 
Specifically, the inflow component was fit in the region of 
$D_\mathrm{tran} > 400 \mathrm{~pkpc}$, 
assuming that this part of the $\tau_\mathrm{ap}$ map is dominated by inflow\footnote{The region was chosen by eye based on Figure \ref{fig:2dabs} to minimize the contamination from outflow.}. 
Once the inflow parameters ($\alpha_\mathrm{0,in}, \gamma_\mathrm{in}, v_\mathrm{offset}$) were obtained, 
they were held fixed and combined with the as-yet-undetermined outflow model to fit the whole $\tau_\mathrm{ap}$ map. 
For the inflow MCMC, the priors were assumed to be flat in linear space 
with positive values (except for $v_\mathrm{offset}$, which is negative for inflow). The
outflow MCMC, however, adopts the most probable posterior of the inflow parameters and fixes them; this may be viewed as a
strong prior  on the 
resulting outflow parameters, which are otherwise assumed to be flat and positive-valued. 

\begin{table}
\caption{Best Fit Model Parameters \label{tab:mcmc}}
\centering
\begin{tabular}{llcc}
\hline\hline
&Parameter & KGPS-Full & KGPS-$z_\mathrm{neb}$ \\\hline
&$v_{\rm offset}$ ($\mathrm{km~s}^{-1}$) & $-84_{-6}^{+6}$ & $-110_{-9}^{+8} $ \\
Inflow & $\alpha_{\rm 0,in}$ ($\mathrm{pkpc}^{-1}$) & $0.0083_{-0.0008}^{+0.0011}$ & $0.0095_{-0.0013}^{+0.0018}$  \\
&$\gamma_\mathrm{in}$ & $0.58_{-0.02}^{+0.03}$ & $0.62_{-0.03}^{+0.04}$ \\\hline
&$v_{\rm 1}$ ($\mathrm{km~s}^{-1}$) & $603_{-11}^{+5}$ & $575_{-3}^{+7} $  \\
Outflow& $\alpha_{\rm 0,out}$ ($\mathrm{pkpc}^{-1}$) & $0.031_{-0.008}^{+0.010} $ & $0.034_{-0.012}^{+0.057}$ \\
&$\gamma_{\rm out}$ & $2.0_{-0.1}^{+0.1}$ & $1.1_{-0.1}^{+0.4}$ \\\hline
\end{tabular}
\end{table}

\begin{figure}
    \centering
    \includegraphics[width=\columnwidth]{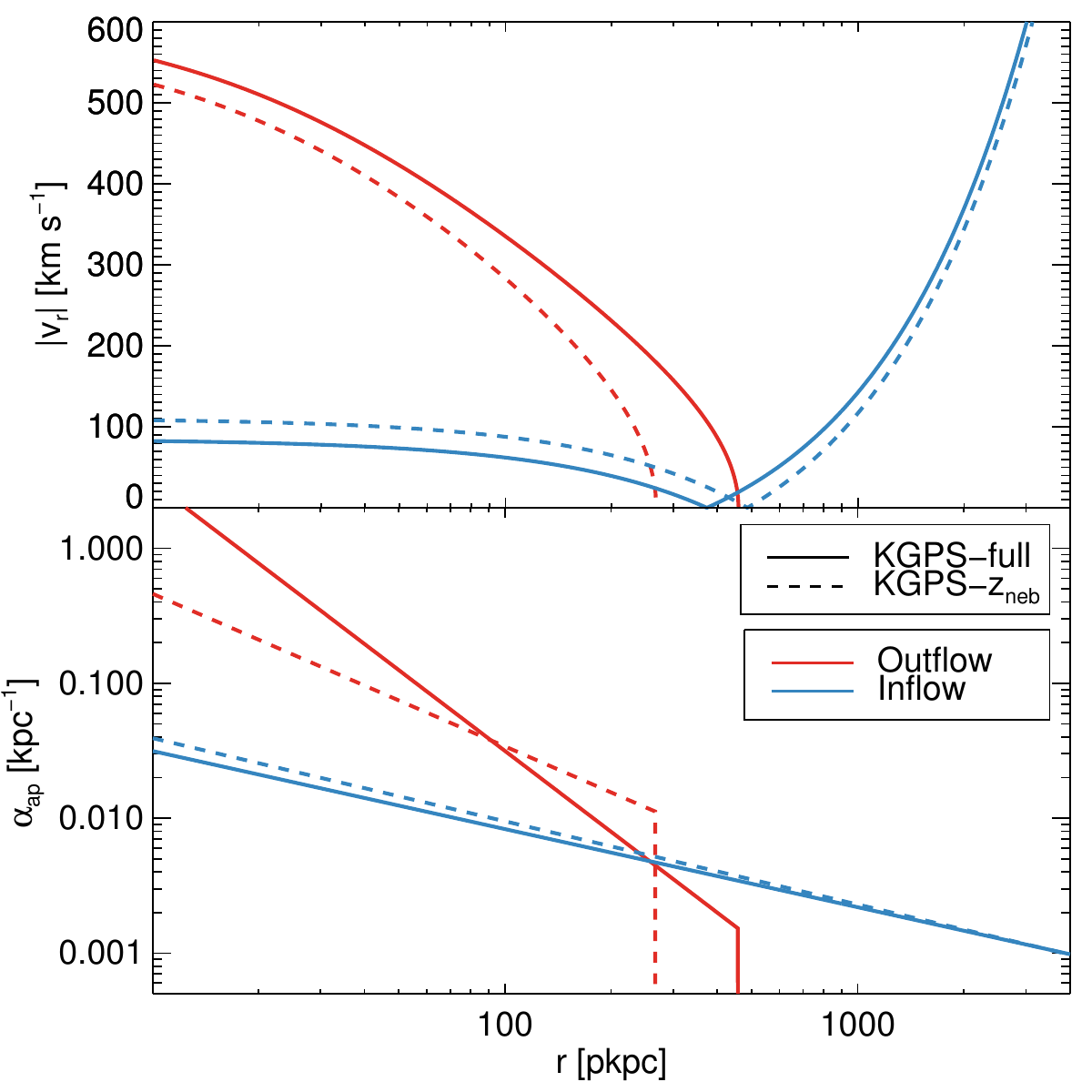}
    \caption{The best-fit radial profiles of $v_r$ and $\alpha_\mathrm{ap}$ as functions of galactocentric radius (i.e., before projection to the
observed $v_{\rm LOS}-D_{\rm tran}$ plane). 
The model parameters are as in Table \ref{tab:mcmc}. The red curves correspond to the outflow component and blue curves to the inflow. 
The best-fit model for the KGPS-Full sample is shown with solid lines, 
while the best-fit model for the KGPS-$z_\mathrm{neb}$ is shown with dashed lines. }
  \label{fig:rprofile}
\end{figure}
\begin{figure*}
\centering
\includegraphics[width=0.99\textwidth]{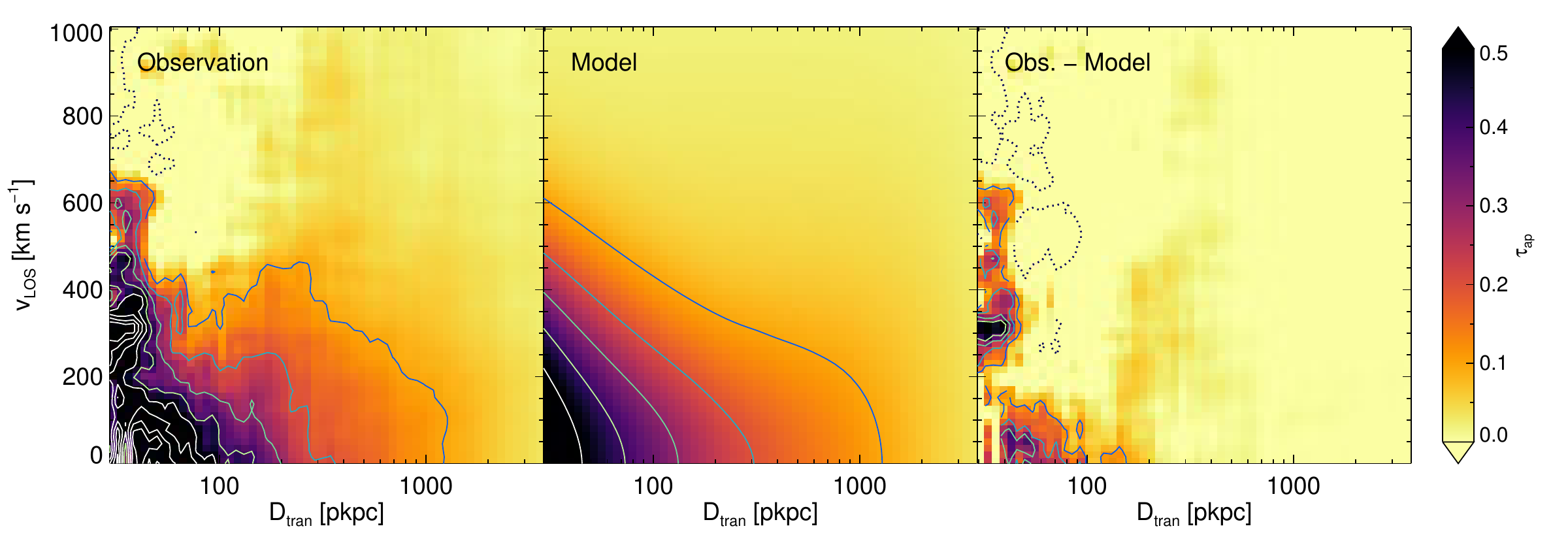}\\
\includegraphics[width=0.99\textwidth]{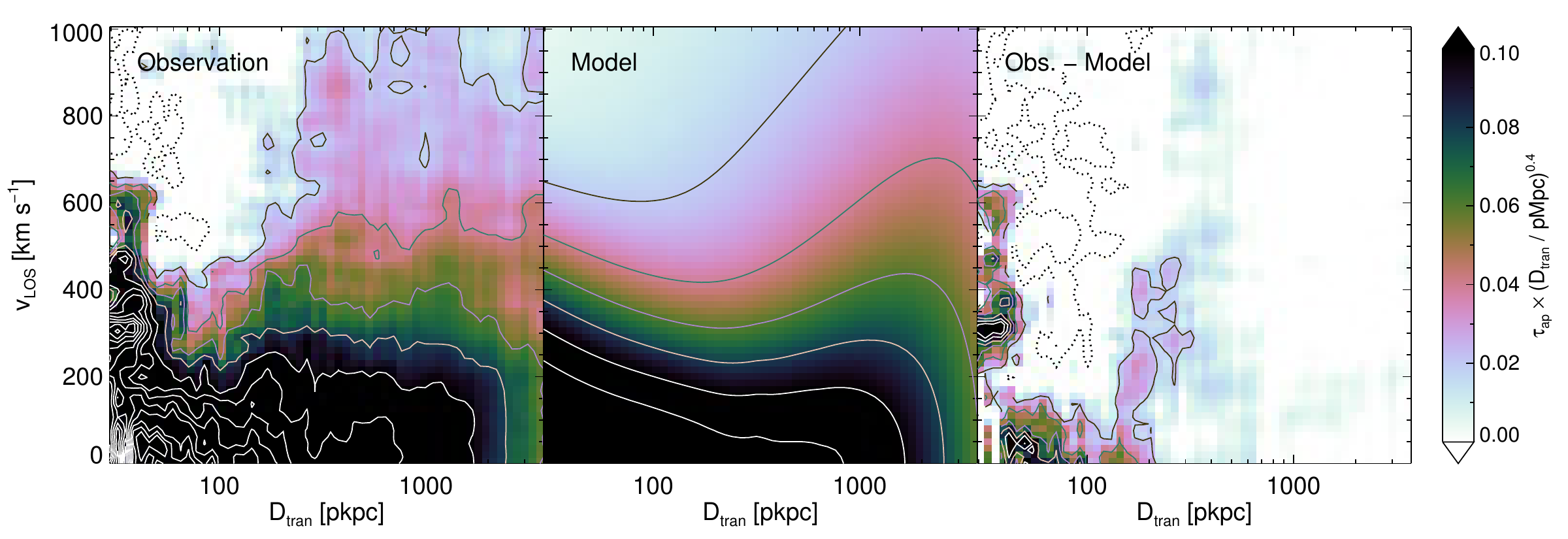}
\caption{Comparison of the observed KGPS-Full map (left) with the best-fit model (middle). The right-most plots are residuals after subtraction of the model from
the observed map. 
The top panel shows $\tau_\mathrm{ap}$ as a function of $D_{\rm tran}$ and $v_{\rm LOS}$, while the bottom panel shows $\tau_{\rm ap}$ multiplied
by $D_{\rm tran}^{0.4}$. 
The colourbars and contour levels are identical to those shown in Figure \ref{fig:2dabs}.} 
\label{fig:model}
\end{figure*}

The best-fit model parameters are summarized in Table~\ref{tab:mcmc}, and the $\alpha(r)$ and $v_r(r)$ profiles 
for the KGPS-Full and the KGPS-$z_\mathrm{neb}$ samples are 
shown in Figure~\ref{fig:rprofile}. In general, the fitted parameters for the two maps are consistent with one another within the
estimated uncertainties; the main difference is in the exponent $\gamma_{\rm out}$, the inferred radial dependence of the outflow
component (see Figure~\ref{fig:rprofile}). In the case of the sparser KGPS-\zneb\ sample, $\gamma_{\rm out}$ is not very well constrained, likely due
to sample variance at small $D_{\rm tran}$ caused by the relatively small number of independent sightlines. 
Figure~\ref{fig:model} compares the 
observed and modeled $\tau_\mathrm{ap}$ maps and the residual map for the KGPS-Full sample. The best-fit model reproduces the general 
features of Figure \ref{fig:2dabs} reasonably well.

The best-fit outflow models above are, on the face of it, inconsistent with the kinematic outflow model of S2010, in which the radial velocity was
constrained primarily by ``down-the-barrel'' (DTB) profiles of blue-shifted low-ionizaton metal lines (rather than \ion{H}{I}) in galaxy spectra.  
In the S2010 models,  
most of the acceleration of outflowing material was inferred to occur within $r \simlt 5-10$ pkpc of the galaxy center, with few constraints on $v(r)$ at
larger radii; however, we note that
the asymptotic velocity of the fastest-moving material in the S2010 models was $\sim 600-700$ \kms, similar to the highest detected 
$v_{\rm LOS}$ at small $D_{\rm tran}$ in the KGPS optical depth maps (e.g., Figure~\ref{fig:2dabs}). 
S2010 did not use information from measurements of $v_{\rm LOS}$ profiles as a function of $D_{\rm tran}$, which has provided the most
important new constraints in the present work. 

Additionally, although various hydrodynamical simulations of galaxy formation differ substantially in their spatial resolution and 
feedback implementation, 
our best-fit outflow velocity of $\sim 600$ \kms\ is consistent with the upper range of gas velocities found for simulated
galaxies with similar halo mass and redshift.  For example, in the FIRE zoom-in simulations, the 95th percentile gas-phase velocity is 
$\sim$ 600 \kms\ at $0.25 R_\mathrm{vir}$ ($r\simeq 25$ pkpc) \citep{muratov15}; for larger volume simulations, which depend to a greater
extent on   
``sub-grid'' treatment of feedback physics, the level of agreement depends on the simulation suite: in the IllustrisTNG simulations, 
the 95th-percentile 
gas-phase velocity is $\sim$ 650 \kms\ at $r=10$ pkpc \citep{nelson19},  whereas in the EAGLE simulations,  
the 90-percentile gas velocity between 0.1 and 0.2 $r_{\rm vir}$ (equivalent to $r \sim 10-20$ pkpc for the galaxies in our observed
sample) is smaller, $\sim$ 350 \kms\ \citep{mitchell20}. It is important to note that the comparison of our observations 
with the FIRE simulation in \S\ref{sec:discussion_zoomin} is 
more direct, since it was confined to neutral H in the simulation box, while the results quoted above pertain
to {\it all} outflowing gas, regardless of physical state. It will be important in future to examine simulation results in terms of
parameters that
are most directly comparable to the available observational constraints.

The simple model framework we adopted is not intended to reproduce all aspects of what is undoubtedly a more complex situation in reality. 
There are clearly features of the observations that are not successfully captured by the model; these are discussed in \S\ref{sec:detailed} below.
Here we examine whether its basic assumptions --  that both infall and outflows are required to reproduce the
general behavior of the observed $\tau_{\rm ap}$ maps -- are justified.

One questionable assumption we adopted for the fiducial model is the kinematic nature of the outflow, which was launched from $r=1$~pkpc at high velocity and subsequently
affected only by gravity.  It neglects the possible effects of a pressure gradient in the halo, through which 
buoyancy forces could conceivably counter-act gravity
in determining the velocity of outflowing material as it moves to larger radii \citep[e.g.,][]{ji2019}. It also assumes 
that no ambient gas is entrained by the outflow (i.e., no additional mass loading) as it moves outward, which 
could reduce the outflow velocity $v_{\rm out}(r)$ more rapidly than the model presented above. 

\begin{figure*}
\centering
\includegraphics[width=8.5cm]{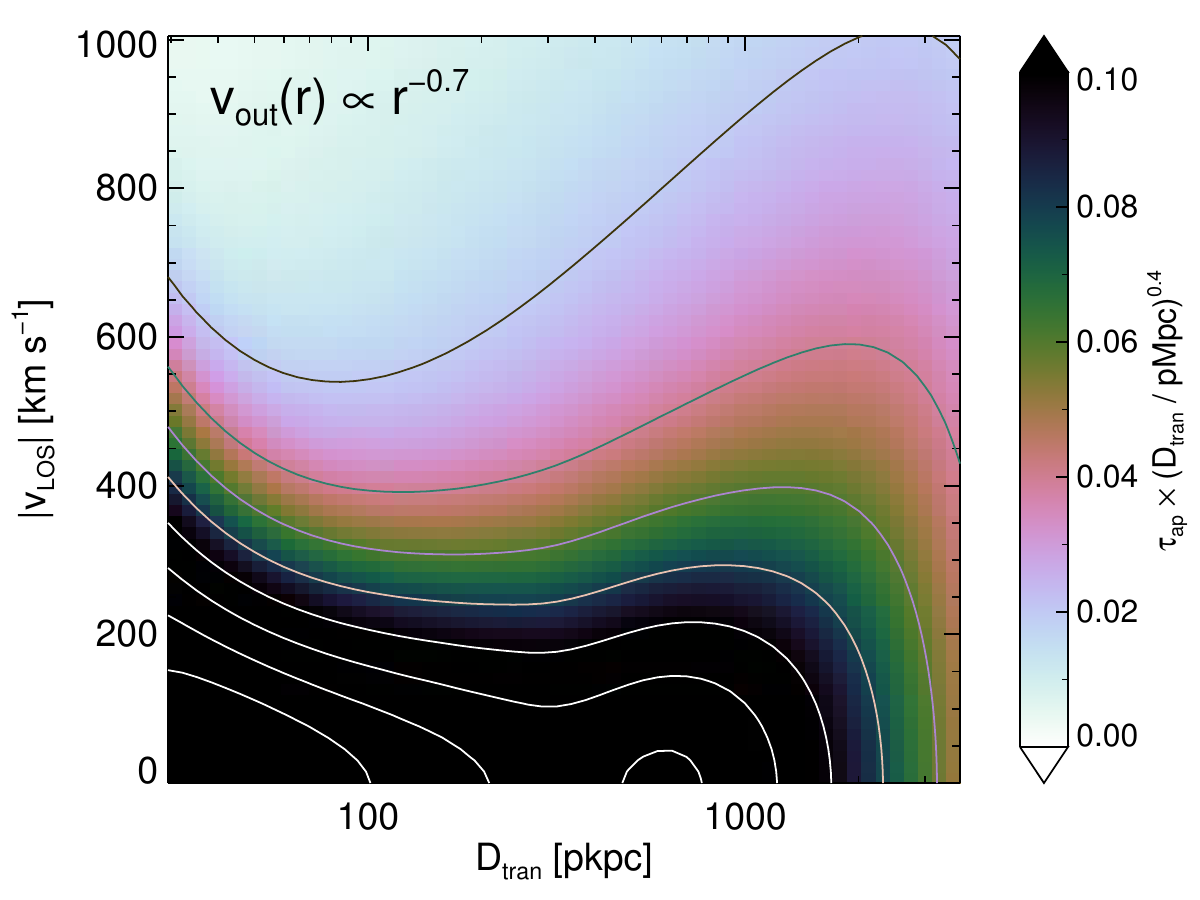}
\includegraphics[width=8.5cm]{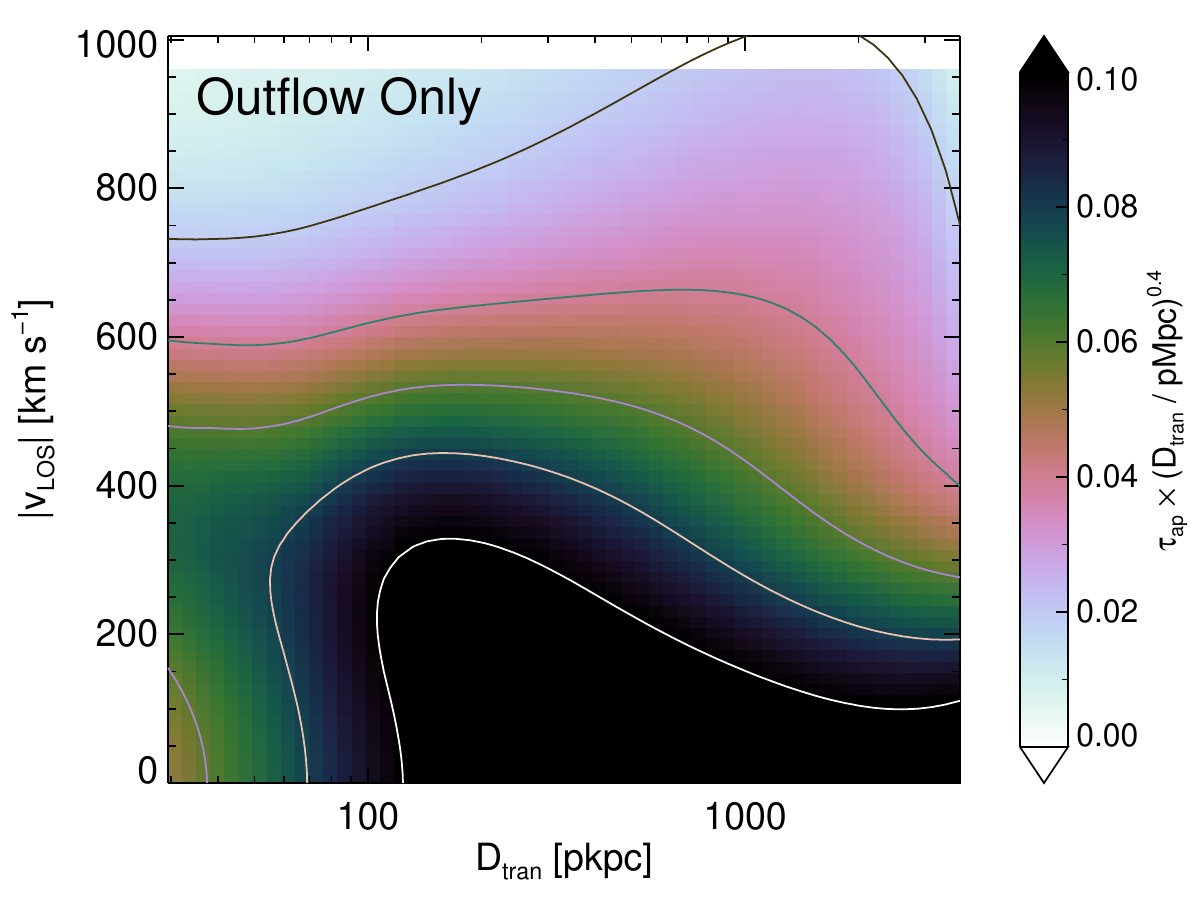}
\caption{The best fitting parametrisations of alternative models, to be compared with that of the fiducial model shown in Figure \ref{fig:model}. Both maps show the
quantity $\tau_\mathrm{ap} \times D_\mathrm{tarn}^{-0.4}$, as in the bottom panels of Figure \ref{fig:model}. ({\it Left:}) Same as fiducial, but with the outflow
velocity $v_\mathrm{out}(r)$ as power law. ({\it Right:}) A (ballistic) outflow-only model that eliminates the infall component completely. The power law model fits the
data similarly to the fiducial; the outflow-only model is a poor fit.}
\label{fig:altmodel}
\end{figure*}

As a test, we relaxed the ballistic assumption and considered a model in which the outflow velocity is simply a power-law
function of galactocentric radius, $v_{\rm out}(r) = v_1 \cdot r^{\beta}$. With this parametrisation, we found    
a best-fitting $\beta = -0.7_{-0.2}^{+0.5}$, while the $\alpha_\mathrm{out}(r)$ remains largely unchanged for the KGPS-Full sample (see Figure \ref{fig:altmodel}). 
The alternative model does result in a more rapid decline in the range of $v_{\rm LOS}$ at $D_{\rm tran} \sim 100$ pkpc
(and thus slightly closer to the observed map in the same range) but the overall fit has $\chi^2$ similar to that of our fiducial model. Moreover, the power
law model failed to converge in the case of the smaller KGPS-$z_{\rm zneb}$ sample.     

Another questionable assumption in the model is that all gas motion is radial with respect to the galaxy center of mass.
We acknowledge that with a more sophisticated parameterisation, one might be able to reproduce all of the observed features
with a combination of radial and non-radial motions, e.g. orbital or random motions. 
However, the observations measure line-of-sight components of velocity only, so the distinction between infall, random, and circular/orbital motion
based on an ensemble of sightlines would not be evident even with much higher quality data.  On the other hand,  
as long as the line of sight component of velocity is slowly varying with $D_{\rm tran}$ -- as expected for typical dark matter
halo mass distributions -- one can think of the ``inflow'' component as a generalized  
proxy for CGM gas whose motion is dictated by the galaxy potential only, whether that motion is radial, random, circular, or some combination.

To test the robustness of the model fits, we tried fitting the outflow component using constant weighting (rather
than inverse variance weighting) across
the map; the resulting best-fit model parameters 
parameters change by 
$\simlt 10$\% compared to the weighted fit. We also tried fitting the 2D maps of the ``red'' and ``blue'' halves
separately: the power of $\alpha_\mathrm{ap}(r)$ (i.e., $\gamma$) for inflow is $0.40_{-0.04}^{+0.05}$ on the blue side and $0.71_{-0.03}^{+0.03}$ on the red side, almost different by a factor of 2, while other parameters are consistent within 25\%. 

Finally, we attempted to fit the observed $\tau_{\rm ap}$ maps with single component gas distributions, i.e. with infall only or outflow only. 
The infall model fails to converge because a match to the observations requires an abrupt increase in velocity (by a factor of $\sim 3$) and in absorption coefficient (by a
factor of  $\sim 5$) at $r\sim 50 \mathrm{~pkpc}$; in any case, as discussed earlier, it is hard to account for $v_{\rm LOS} > 300$ \kms\ with
gravitationally-induced infall onto a halo of mass $10^{12}$ M$_{\odot}$. Moreover, we know unequivocally from the kinematics of strong interstellar absorption lines observed
in DTB 
spectra of the foreground galaxies that
outflows dominate the kinematics on scales of at least a few pkpc, and that they have maximum outflow velocities similar to the values of $v_1$ in our model outflows. 

Figure \ref{fig:altmodel} shows the best-fit outflow-only model to the KGPS-Full sample. The model includes the effects of Hubble expansion, $v_\textrm{out-only}(r) = v_\mathrm{out}(r) + H(z)r$, because the behaviour of the observed map would 
clearly require a substantial fraction of the outflowing gas 
to escape the halo, after which its kinematics would be dominated by Hubble expansion. 
For $D_\mathrm{tran} \lesssim 100 \mathrm{~pkpc}$, the model clearly fails to reproduce either the absorption strength or 
the kinematics of the central region in order to reproduce the behaviour at larger $D_{\rm tran}$. 

Despite the uncertainty in the parametrisation of the model, we argue that the observations require both outflowing and accreting components; this assertion is discussed
further in the next section.  

\subsubsection{Implications for Detailed CGM \ion{H}{I} Kinematics}
\label{sec:detailed}

Despite the ability of the model to reproduce the general features 
of the observed $\tau_\mathrm{ap}$ maps, there are some details present in the data that are not successfully captured. Most notable
is that the data exhibit contours of constant $\tau_{\rm ap}$ that change overall shape depending on the contour level, whereas
the model contours are relatively smooth and self-similar at different contour levels (Figure~\ref{fig:model})
-- i.e. the model lacks
kinematic structure within $D_\mathrm{tran} < 100 \mathrm{~pkpc}$ and $|v_\mathrm{LOS}|<600\mathrm{~km~s}^{-1}$, especially the observed
compression between 50 and 100 pkpc that characterizes the observed map.
Some of the discrepancy could be caused by the relatively small number of sightlines that comprise the measurements on small $D_{\rm tran}$ 
scales, with the correspondingly larger  
sample variance, and the lack of non-radial gas motion in the model. However, very similar structure at $D_{\rm tran} < 100$ pkpc is observed at 
both positive and negative $v_{\rm LOS}$ in the unfolded maps in Figure~\ref{fig:2dabs_unfold}, suggesting that the structure is likely to be real. 

The rightmost panels of Figure~\ref{fig:model} show the residuals between the observed KGPS-Full maps and the corresponding best-fit model. 
These illustrate that the model systematically under-predicts the apparent optical depth at small $D_{\rm tran}$, particularly for $v_{\rm LOS} \simgt 200$ \kms, 
suggesting that a single, smoothly varying outflowing component whose velocity depends only on $r$ is an oversimplification of the true situation. From
high-resolution QSO spectra, it is common to observe complexes of absorption systems at small galactocentric impact parameters, with a wide range
of \nhi\ and $v_{\rm LOS}$ (e.g., \citealt{rudie12a}), with many pixels reaching zero intensity. Disentangling these complexes often requires measurement of the
higher Lyman series lines even in spectra with resolution $\simlt 10$ \kms\ and very high S/N; it is therefore unsurprising that 
a simple model fails to capture the details. Nevertheless, the rapid fall-off in the maximum $v_{\rm LOS}$ at relatively high $\tau_{\rm ap}$ 
over the range $D_{\rm tran} \simeq 50-100$ pkpc, and the flattening of the profile out to $D_{\rm tran} \simeq 200-300$ pkpc, are not present
in our fiducial model or the alternatives discussed in the previous section.   

As discussed in \S\ref{sec:results}, a power-law parametrisation of the outflow component provides a fit to the $\tau_{\rm ap}$ map (at least,
for the KGPS-Full version) with similar $\chi^2$, and allows for a more rapid decrease in $v_{\rm out}(r)$ with radius compared to the ballistic model with no mass loading. However, it still  cannot
account for the changing contour shapes of $\tau_{\rm ap}$ present in the data.  Better matches might be
obtained by treating high-$\tau_{\rm ap}$ in the outflow component separately from the remainder, allowing it to experience more
rapid deceleration at $D_{\rm tran} \sim 50$ pkpc than more diffuse gas, or by a rapid transition from high to low-\nhi\ absorbers for fast-outflowing gas. 
Both scenarios are consistent with the results of \citet{rudie12a}, in which absorbers with $\nhi > 10^{14.5} \mathrm{~cm}^{-2}$ were found to occupy a smaller range in 
$v_{\rm LOS}$ than absorbers with $\nhi < 10^{14.5} \mathrm{~cm}^{-2}$ at $D_\mathrm{tran} \lesssim 2 \mathrm{~pMpc}$.  

At $50 \mathrm{~pkpc} \simlt D_\mathrm{tran} \simlt 100  \mathrm{~pkpc}$, where the $\tau_\mathrm{ap}$ profile is at its narrowest, 
the line profile (i.e., the distribution of $v_{\rm LOS}$ at a given $D_{\rm tran}$) is consistent with the 
effective velocity resolution of the map ($\sim 190 \mathrm{~km~s}^{-1}$), and changes 
very little out to $D_\mathrm{tran} \sim 100$--200 pkpc.  As best-seen in the bottom panels of Figure~\ref{fig:model}, beyond $D_{\rm tran} \sim 100$ pkpc, while the
optical depth continues to decrease roughly as $D_{\rm tran}^{-0.4}$, the range of $v_{\rm LOS}$ begins to increase again, by an amount that depends
on the contour level. The best fit radial and kinematic profiles shown in Figure~\ref{fig:rprofile} 
suggest that the minimum in the line-of-sight velocity field at $D_{\rm tran} \sim 100-200$ pkpc may mark a caustic where outflows and infall 
both reach minimum $v_\mathrm{LOS}$, perhaps with opposite sign. 
The location of this feature also corresponds to the clear change in slope of the relationship between $D_{\rm tran}$ and \wlya\ (Figure~\ref{fig:ew}),
and is just beyond the expected virial radius given a halo mass of $M_\mathrm{h}=10^{12} M_\odot$. Since the escape velocity at 100 pkpc for such a halo
is $v_{\rm esc} \simeq 440$ \kms, we expect that most of the neutral H at $r\simlt 100$ pkpc remains bound to the central galaxy. 

\subsection{Velocity Asymmetry and Emission Filling}
\label{sec:asymmetry}
One of the more puzzling features of the 2D maps of $\tau_{\rm ap}$ -- evident in Figure~\ref{fig:2dabs_unfold} -- 
is the apparent asymmetry of $\tau_\mathrm{ap}$ with respect to $v_\mathrm{LOS}$. 
In the KGPS-Full sample, for $D_{\rm tran}$ between $\sim 70$ and $150 \mathrm{~pkpc}$, the integrated $\tau_{\rm ap}$ for 
$v_{\rm LOS} > 0$ is $>1.5$ times that of the corresponding value for $v_{\rm LOS} < 0$, meaning that redshifted \lya\ absorption is
stronger than blueshifted \lya\ absorption.  The asymmetry can also be seen in 
Figure~\ref{fig:egstack}: for the composite \lya\ spectrum for the bin with $80 \le D_{\rm tran}/{\rm pkpc} \le 160$,  although the 
maximum depth in the absorption profile occurs at a velocity consistent with $v_{\rm LOS} = 0$, there is a red wing extending to 
$v_\mathrm{LOS} \sim +500 \mathrm{~km~s}^{-1}$, whereas the blue absorption wing reaches only $v_{\rm LOS} \sim -300$ \kms.  
Similar asymmetry is also present for the KGPS-$z_\mathrm{neb}$ sample, suggesting that residual systematic redshift errors are not likely to
be the principal cause. 

\begin{figure*}
\centering
\includegraphics[width=14cm]{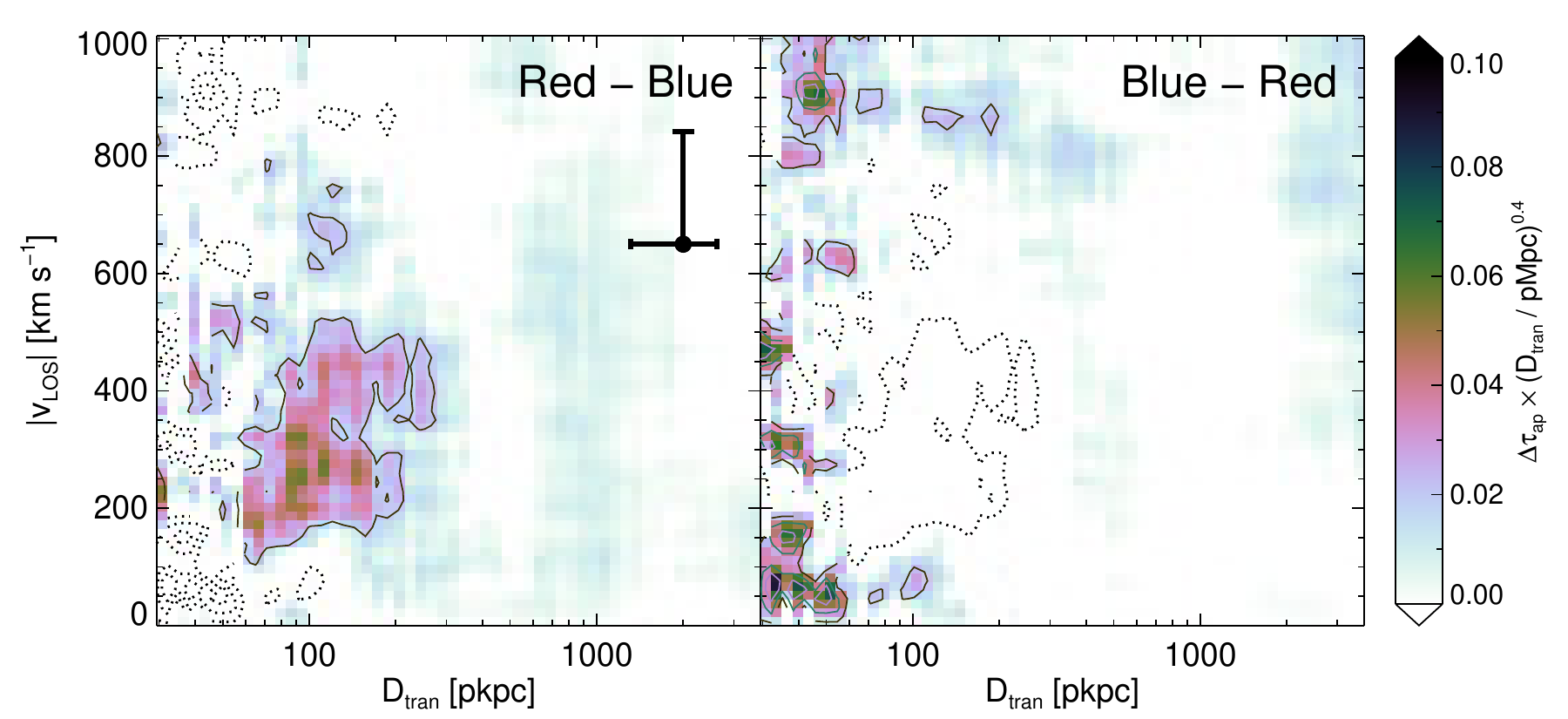}
\caption{ Same as the $\tau_\mathrm{ap} \times D_\mathrm{tran}^{0.4}$ map in Figure \ref{fig:2dabs}, except that instead of averaging the blue and red sides, the left (right) map is the \textit{subtraction} of the blue (red) side from the red (blue) side. Significant asymmetry can be seen at $D_\mathrm{tran}\simeq 100 \mathrm{~pkpc}$. Again, the solid (dotted) contours represent positive (negative) values. The excess blueshifted absorption at $D_\mathrm{tran} \lesssim 50 \mathrm{~pkpc}$ could be due to sample variation, while the excess redshifted absorption with $D_{\rm tran}= 50-200$ pkpc is likely real.  }
\label{fig:asymmetry}
\end{figure*}

Figure~\ref{fig:asymmetry} shows the result of subtracting the blue (red) component of the $\tau_\mathrm{ap} \times D_\mathrm{tran}^{0.4}$ map from the
red (blue) side for the KGPS-Full sample. 
The most prominent residual appears at $50 \lesssim D_\mathrm{tran}/{\rm pkpc} \lesssim 200$ and $200 \lesssim |v_\mathrm{LOS}|/\kms \lesssim 500$; 
residuals resulting from differencing of the KGPS-$z_\mathrm{neb}$ sample are similar. 
Although the residual has modest SNR ($\simeq 2$ per \AA), it extends over a contiguous region larger than the effective resolution of the map, and is therefore 
significant. There is a fractionally less-significant excess of blueshifted \lya\ absorption at $D_{\rm tran} \simlt 50$ pkpc (better shown in the righthand panel of
Figure~\ref{fig:asymmetry}), apparently extending over
the full range  $0 \le |v_{\rm LOS}|/\kms \le 1000$ that could be attributable to the relatively small number of galaxy pairs at small separations (i.e., to 
sample variance). It is harder to dismiss the excess redshifted absorption in the lefthand panel of Figure~\ref{fig:asymmetry}, which is based on a 10-times larger
sample ($\simeq 1000$) of galaxy pairs. On the other hand, there is no significant asymmetry in the \lya\ absorption profiles beyond $D_{\rm tran}\sim 200$ pkpc, 
where the S/N of the map is high.  

If the observed asymmetry is of astrophysical origin, it remains to be explained.  
We considered possible causes of the apparent excess redshifted \lya\ absorption, noting that it is confined to the foreground 
galaxy CGM within $D_{\rm tran} \simeq 200$ pkpc: 
possibilities include observational selection effects or biases, which
might have been introduced by particular properties favored among the foreground or background 
galaxies that were observed successfully (e.g., compactness, dust attenuation, orientation, etc.). 
We could not envision a plausible scenario that could explain the asymmetry in the absorption profile.

We also considered the real {\it temporal} difference between the time light from the background galaxy passes through the ``far'' and ``near'' sides of
the foreground galaxy CGM, but the relevant timescale would be very short, $\Delta t \simlt 400~{\rm pkpc}/c \sim 10^{5}$ yrs. 
As above, it is hard to explain why any time-dependent effect would systematically bias the kinematics of the absorbing gas.  
  
It is now well-known that \lya\ emission ``halos'' are a generic property of star-forming galaxies at high redshift (e.g., \citealt{steidel11,wisotzki16,erb18}) 
and that they extend to projected distances $\sim 5-10$ times larger than their UV continuum light.    
Of the possible explanations, we find the least implausible to be that the \lya\ absorption profile has been altered by \lya\ emission contamination
within the spectroscopic aperture used to record the background galaxy spectra.  
Most surveys of  
intervening absorption lines have used 
bright background point sources, such as QSOs, in which case this possible source of contamination can ordinarily be neglected. 
One usually thinks of \lya\ absorption systems in terms of equivalent width \wlya\ or column density \nhi, but one can also
think of a \lya\ absorption feature at $z=z_{\rm fg}$ as a record of the flux removed
from the beam of the background source at $z_{\rm bg}$, scattered out of the line of sight by \ion{H}{I} in the CGM of a foreground galaxy. 
As an example, 
for an observation of a $m=18$ QSO at $z_{\rm bg}$ whose sightline passes near to a typical $z_{\rm fg} = 2.2$ galaxy 
and produces an absorption feature with rest-frame \wlya$=1.0$ \AA\ at $z_{\rm fg}$,  
the flux removed from the QSO spectrum would be $\simeq 6\times10^{-16}$ ergs s$^{-1}$ cm$^{-2}$, which is large even in comparison
to the {\it total} \lya\ flux from the galaxy (\citealt{steidel11}), and hundreds of times larger than the \lya\ flux likely to
be collected by a slit located 5-10 arcsec away from the $z_{\rm fg}$ galaxy. 
For a typical $m=25$ background galaxy, on the other hand, the same absorption line equivalent width
would correspond to a flux smaller by a factor of $\sim 630$, or $\simeq 10^{-18}$ ergs s$^{-1}$ cm$^{-2}$, i.e., approaching
the flux expected from the low surface brightness \lya\ emission halo of the foreground galaxy. In the latter case, the presence
of the \lya\ emission could have a measurable effect on the apparent absorption strength if the velocity ranges overlap.  

To be more quantitative, 
we estimated the flux of \lya\ emission captured within the aperture used to obtain background galaxy spectra using the average \lya\ 
halo observed by \citet{steidel11} (S2011) for a sample of star forming galaxies with similar properties to those in the larger KBSS
sample, but with $\langle z \rangle = 2.65$. The mean \lya\ surface brightness profile 
was found to be reasonably well-described by an exponential, $S(b)=C_{\rm l}~{\rm exp}(-b/b_0)$,
where $b=D_{\rm tran}$, $C_{\rm l} = 2.4\times10^{-18}$ ergs s$^{-1}$ cm$^{-2}$ arcsec$^{-2}$, and $b_0 = 25$ pkpc.  
The mean halo is detected down to a surface brightness limit of $S \simeq 10^{-19}$ ergs s$^{-1}$ cm$^{-2}$ arcsec$^{-2}$ 
at $D_{\rm tran} \sim 80$ pkpc.  

By assuming that the intrinsic \lya\ luminosity and scale length of the mean \lya\ halo of the KGPS foreground 
galaxies are the same as those of the S2011 sample, we applied the redshift-dependent surface brightness correction to move from  
$\langle z \rangle=2.65$ to $\langle z \rangle = 2.25$, and  
extrapolated the \lya\ surface brightness profile to $D_{\rm tran} = 100$ pkpc. 
The predicted \lya\ surface brightness would be 
$\sim 7.2\times10^{-20}$ ergs s$^{-1}$ cm$^{-2}$ arcsec$^{2}$. Taking the typical extraction aperture for the LRIS spectra to be
of angular size 1.35 arcsec $\times$ 1.2 arcsec ($\sim 1.6$ arcsec$^{2}$), the integrated Ly$\alpha$ emission flux within the slit 
would be $F_{\mathrm{Ly}\alpha,\mathrm{em}} \sim 1.1\times 10^{-19} \mathrm{~erg~s}^{-1}\mathrm{~cm}^{-2}$. 

We estimated the mean background galaxy continuum flux density near 4000\AA\ (i.e., the wavelength of \lya\ at $z \sim 2.25$) by 
interpolating the flux density between 
the photometric $U_{\rm n}$ (3520/600) and $G$ (4730/1100) passbands, with the result $\langle m_{\rm AB}(4000\textrm{\AA})\rangle \simeq 25.3$, or
$\langle F_{\nu} \rangle  \simeq 0.26 \mu\mathrm{Jy}$ ($\langle F_\lambda \rangle \simeq 5 \times 10^{-19} \mathrm{~erg~s}^{-1}\mathrm{cm}^{-2}\textrm{\AA}^{-1}$.) 

\begin{figure}
\centering
\includegraphics[width=8.5cm]{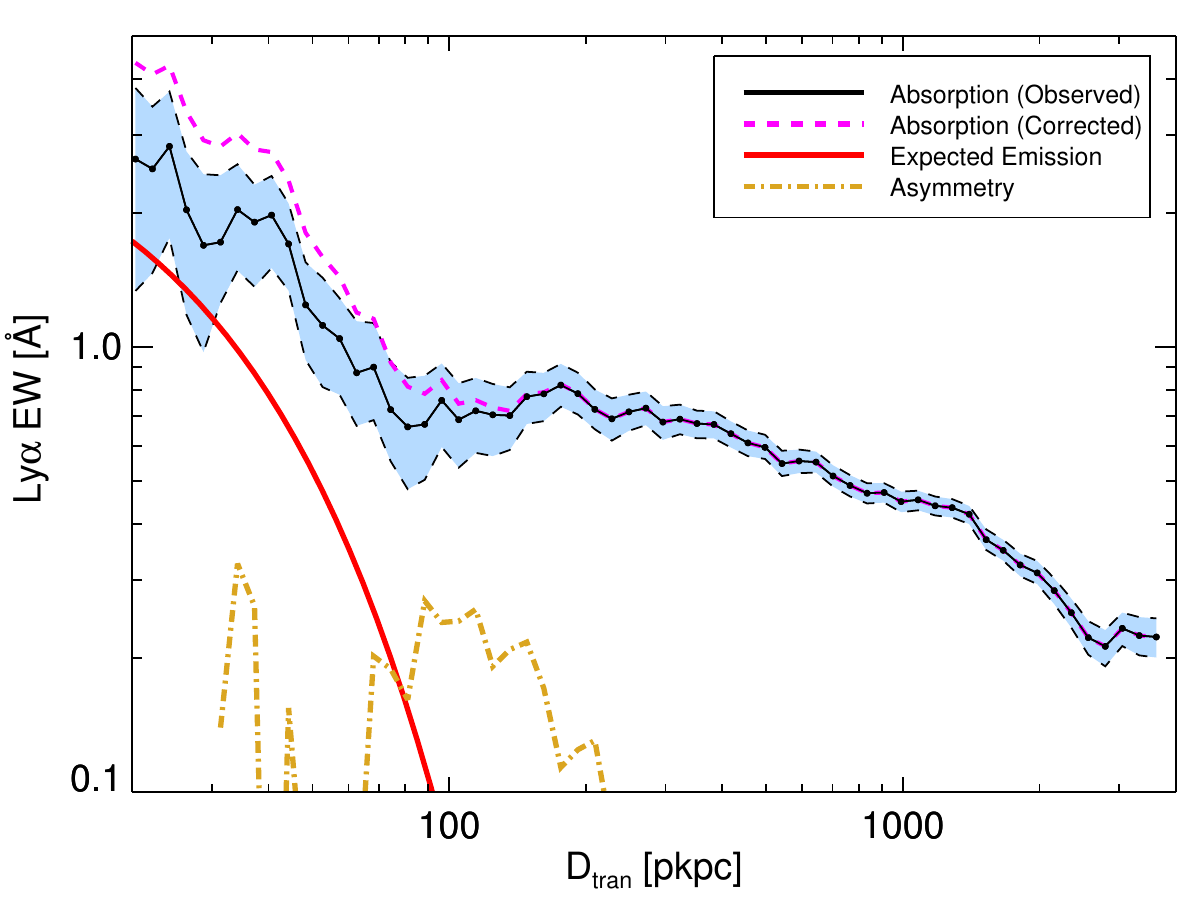}
\caption{ Expected contribution to \wlya\ by \lya\ {\it emission} surrounding foreground galaxies (red), 
compared to the observed \wlya\ in absorption (black, same as in Figure \ref{fig:ew}). 
The dashed magenta curve shows the absorption after correction for the estimated contribution from emission filling. 
The yellow dashed curve shows the fraction of \wlya\ contributed by asymmetry, calculated by integrating the blue/red halves of the absorption profiles
spectra within $|\Delta v_\mathrm{LOS}|<700 \mathrm{~km~s}^{-1}$ and subtracting one from the other. 
}
\label{fig:ew_emission}
\end{figure}

The observed \lya\ absorption equivalent width at $D_{\rm tran} \simeq 100$ pkpc is  $\langle \wlya_{\rm obs} \rangle \simeq 2.7$ \AA (assuming $\langle z \rangle = 2.25$), 
which removes an average flux
from the background galaxy spectrum of $2.7\times5\times10^{-19} \simeq 1.4\times10^{-18}$ ergs s$^{-1}$ cm$^{-2}$. The fractional perturbation of the total 
\lya\ absorption equivalent width by foreground galaxy \lya\ emission is then $1.1\times10^{-19}/1.4\times10^{-18} \simeq 0.08$ (8\%). 
Using the same arguments, emission filling
at $D_{\rm tran} \simeq 25$ pkpc -- where the mean \lya\ surface brightness is $\sim 20$ times higher (according to S2011) but the absorption rest equivalent width is
larger by a factor of only $\simeq 2.5$ (see Figure~\ref{fig:ew}) -- emission would be predicted to affect the observed \wlya\ at the $\simgt 50$\% level.   
Figure~\ref{fig:ew_emission} shows the predicted effect of diffuse emission on the observed \wlya\ absorption as a function of $D_{\rm tran}$.  
Also shown in Figure~\ref{fig:ew_emission} is the amplitude of the observed asymmetry converted to $\Delta\wlya$, the net equivalent width of the
\wlya\ residuals shown in Figure~\ref{fig:asymmetry}. 

The average equivalent width of the residual shown in the lefthand panel of Figure~\ref{fig:asymmetry} 
is $\Delta\wlya \simeq 0.2$ \AA\ for $70 \simlt D_{\rm tran}/{\rm pkpc} \simlt 150$, or $\simeq 25-30$\% of the total observed \wlya\ over the same
range of $D_{\rm tran}$ -- within a factor of $\sim 3$ of the estimated effect from emission filling, possibly consistent given the uncertainties.
However, if the emission were distributed symmetrically in velocity space with respect to the absorption, its effect would mostly likely have remained
unrecognized. 
The asymmetry in \lya\ absorption at $D_{\rm tran} \sim 100$ pkpc, if attributed to emission filling, would require that the 
bulk of \lya\ emission must be 
{\it blueshifted} by $\Delta v_{\rm LOS} \sim 200-400$ \kms\ with respect to the foreground galaxy systemic redshift. 
This is opposite to what is typically observed for \lya\ emission in DTB galaxy spectra, in which the dominant component of \lya\ emission is redshifted
by several hundred \kms\ for the galaxies in the KBSS sample (e.g., see the top and bottom panels of Figure~\ref{fig:zcal}.) 
In this context, it is possibly relevant that the excess emission (if that is indeed the cause) falls in the range of $D_{\rm tran}$ where
our kinematic modeling (\S\ref{sec:model}) suggested a transition between outflow-dominated and accretion-dominated flows.   
If the velocity field at $D_{\rm tran} \simeq 100$ pkpc were dominated by inflows, scattering of \lya\ photons in the observer's direction would tend to
be blueshifted (e.g., \citealt{faucher2010,dijkstra14}). 

Also qualitatively (but perhaps not quantitatively) consistent with this picture is that the asymmetry on smaller transverse scales 
($D_{\rm tran} \simlt 40$ pkpc) has the opposite sign, i.e., the {\it absorption} is stronger on the blueshifted side of $v_{\rm sys}$, which 
might be attributable to excess redshifted emission, as for typical DTB spectra. However, as shown in Figure~\ref{fig:ew_emission}, in order
to be consistent with the expected effect of emission filling at small $D_{\rm tran}$, the kinematic asymmetry of the \lya\ emission would need
to comprise only a fraction of the total emission, since the observed net effect on \wlya\ from the asymmetry is only $\sim 10-15$\%.    
On the other hand, recent observations using the Keck Cosmic Web Imager (KCWI; \citealt{morrissey18}) for a subset of the KBSS galaxy sample 
suggests that the ratio between the blueshifted and redshifted components of \lya\ emission from the CGM increases with projected galactocentric distance 
(\citealt{erb18}; Chen et al, in prep.), and approaches 1:1 by $r\sim 50$ pkpc; unfortunately, the observations of individual galaxies with KCWI (or MUSE) 
are not yet sensitive enough to evaluate at $D_{\rm tran} > 50$ pkpc\footnote{Over the range of $D_{\rm tran}$ in common, the mean \lya\ halo in the KCWI
data is consistent in both shape and intensity with the mean \lya\ halo presented by S2011.}. 

An obvious observational test of the hypothesis that emission filling significantly alters the strength of \wlya\ measured along background 
galaxy sightlines 
could be made using
QSO sightlines, since QSOs are at least $\sim 6$ magnitudes brighter than the typical $z\sim 2.5$ background galaxy, so that
foreground galaxy \lya\ emission is expected to be negligible compared to the absorbed flux from the background source. 
Unfortunately, the sample size of QSO-galaxy pairs for $z_{\rm fg} > 2$ is very small in comparison. 
We examined the velocity distribution of \ion{H}{I} for QSO sightlines within $D_{\rm tran} \simeq 200$ pkpc of KBSS galaxies in
\citet{rudie12a, rudie2019}. In the range of $50 < D_\mathrm{tran}/\mathrm{pkpc} < 200$, there are slightly more redshifted absorption components 
by number, but the excess is not statistically significant. When the high resolution QSO spectra are analyzed in the same way as the 
galaxy spectra by smoothing to reduce the spectral resolution and averaging the spectral regions near \lya\ at the redshift of foreground
galaxies  -- as done by \citet{turner14} (see Figure~\ref{fig:ew}) -- \wlya\ is larger by $\simeq 50$\% for the QSO sightlines for the
bin at $D_{\rm tran} \simeq 90$ pkpc
compared to the observed  \wlya\ for the KGPS sightlines. Though the two measurements are still consistent with one another
at the $\sim 2\sigma$ level, the difference is in the direction expected if the KGPS sample has been affected by \lya\ emission. At smaller
$D_{\rm tran}$, where the expected effect of emission filling in the galaxy sightlines is larger, the QSO sightline samples are too small to provide a meaningful comparison
(e.g., there are no QSO-galaxy pairs with $D_{\rm tran} < 50$ pkpc in the KBSS sample used by \citealt{turner14}.)   

In any case -- whether or not it is responsible for the asymmetry in the current data --  it is probably important to account for the effects of emission filling on \wlya\ in absorption, as a function of
$D_{\rm tran}$, which will depend on the relative brightness of the foreground and background sources. When we apply our estimate from above, 
the corrected \wlya\ vs. $D_{\rm tran}$ (Figure~\ref{fig:ew_emission}) accentuates the existence of two distinct ``zones'' in the behavior of 
\lya\ absorption around KBSS galaxies: the inner zone, at $D_{\rm tran} \simlt 100$ pkpc, where $\wlya\propto D_{\rm tran}^{-1.1}$, and an outer
zone with $100 \simlt D_{\rm tran}/{\rm pkpc} \simlt 300$ kpc, over which \wlya\ remains remarkably flat.  

\subsection{Baryon Escape}

A significant fraction of gas associated with galaxy-scale outflows is expected to be retained 
by relatively massive galaxies. Even if some gas does manage to escape to beyond the halo virial radius, 
theoretical expectations are that, for galaxies with halo masses $\simeq 10^{12}$ M$_{\odot}$ at $z \simgt 2$, most
will eventually be re-accreted by the galaxy through a process known as ``recycling'' (e.g., \citealt{oppenheimer10,muratov15}). 
It is of interest to ask whether the observed ensemble 
kinematics of circumgalactic \ion{H}{I} we have presented in previous sections suggests the presence of neutral gas
capable of escaping the potential well of the central galaxy. The most straightforward indicator would be significant
absorption components
of the CGM with $v_{\rm LOS}(D_{\rm tran})  > v_{\rm esc}(r)$ where both are measured relative to the central galaxy systemic redshift
(e.g., \citealt{adelberger05,rudie2019}.)  

\begin{figure}
\centering
\includegraphics[width=8.5cm]{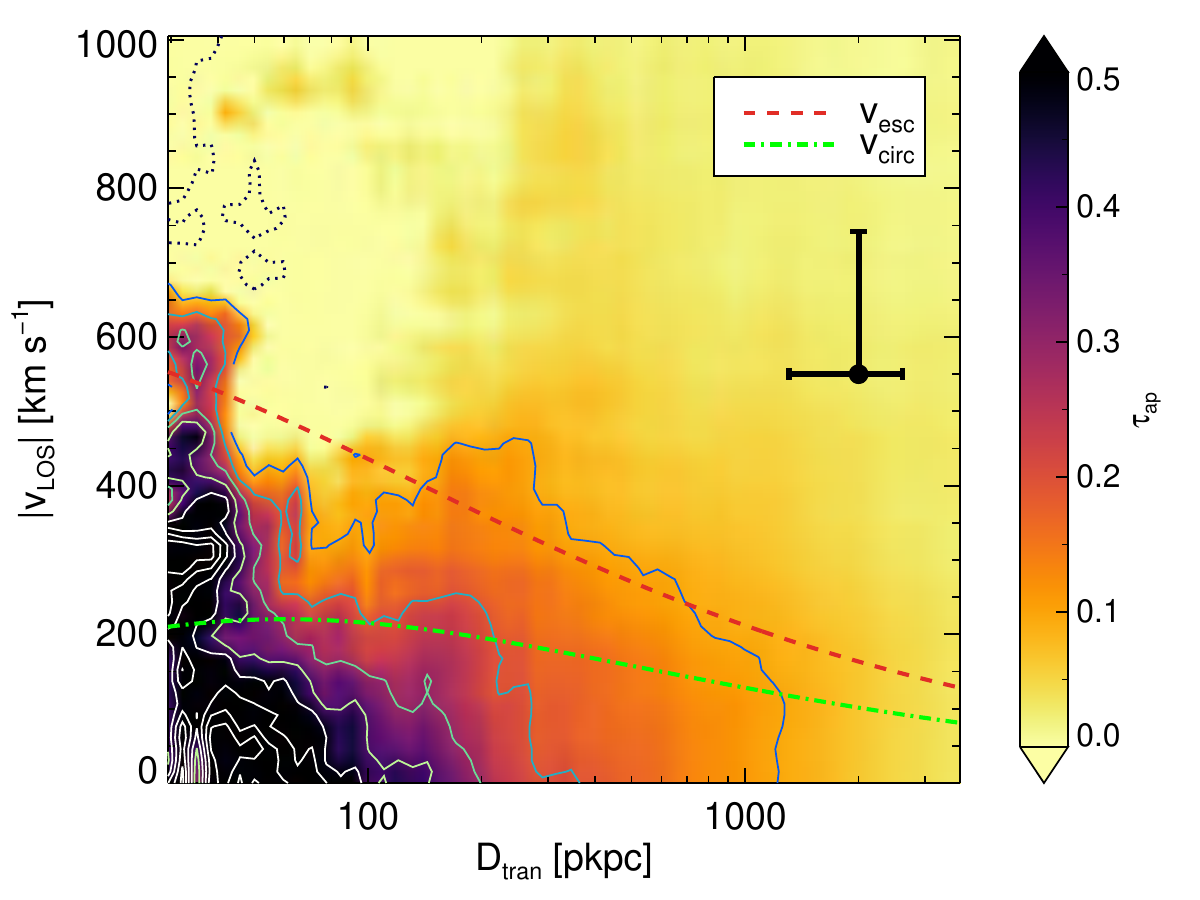}
\caption{Same as the $\tau_\mathrm{ap}$ map in Figure \ref{fig:2dabs}, with the radial dependence 
of the escape velocity ($v_\mathrm{esc}$, red dashed) and circular velocity ($v_\mathrm{circ}$, green dash-dotted) 
for an NFW halo with $M_\mathrm{h}=10^{12} M_\odot$ superposed (assuming that $D_\mathrm{tran}$ is equivalent to the galactocentric distance $r$.)}
\label{fig:vesc}
\end{figure}

Figure \ref{fig:vesc} reproduces the folded map of $\tau_{\rm ap}$, with curves denoting the 3-D escape velocity ($v_\mathrm{esc}$) 
for a $M_\mathrm{h} = 10^{12} M_\odot$ NFW halo: since the map shows only projected distances and line-of-sight velocities, 
the actual galactocentric radius and 3-D space velocities of gas may be greater than $D_\mathrm{tran}$ and $v_{\rm LOS}$, respectively. Considering
only gravity as in \S\ref{sec:model}, any gas with $v_{\rm LOS} > v_\mathrm{esc}$ would be capable of escaping the halo.  
Conversely, gas with $v_{\rm LOS} < v_\mathrm{esc}$ 
is not guaranteed to be bound to the halo, but the chance of escape rapidly decreases as $|v_\mathrm{LOS}| \rightarrow 0$. 
Clearly, judged on this basis, most of the relatively-high-$\tau_{\rm ap}$ \ion{H}{I} within $D_{\rm tran} \simlt 100$ pkpc 
is unlikely to escape the galaxy potential. 

Nevertheless, it remains likely that some gas {\it does} escape, given the presence of significant $\tau_{\rm ap}$ with $v_{\rm LOS}$ close to or
exceeding $v_{\rm esc}$ 
It is also likely that \ion{H}{I} is not the best tracer of the fastest-moving gas, particularly at large galactocentric
distances, 
based on both observations of high-ionization
metals in nearby and high-redshift starburst galaxies (e.g., \citealt{shc+04,strick09,turner15,rudie2019}) and on simulations such as those presented
in \S\ref{sec:discussion_zoomin}.

\subsection{Redshift-Space Distortions}
\label{sec:discussion_zdistortion}

\begin{figure}
\centering
\includegraphics[width=8.5cm]{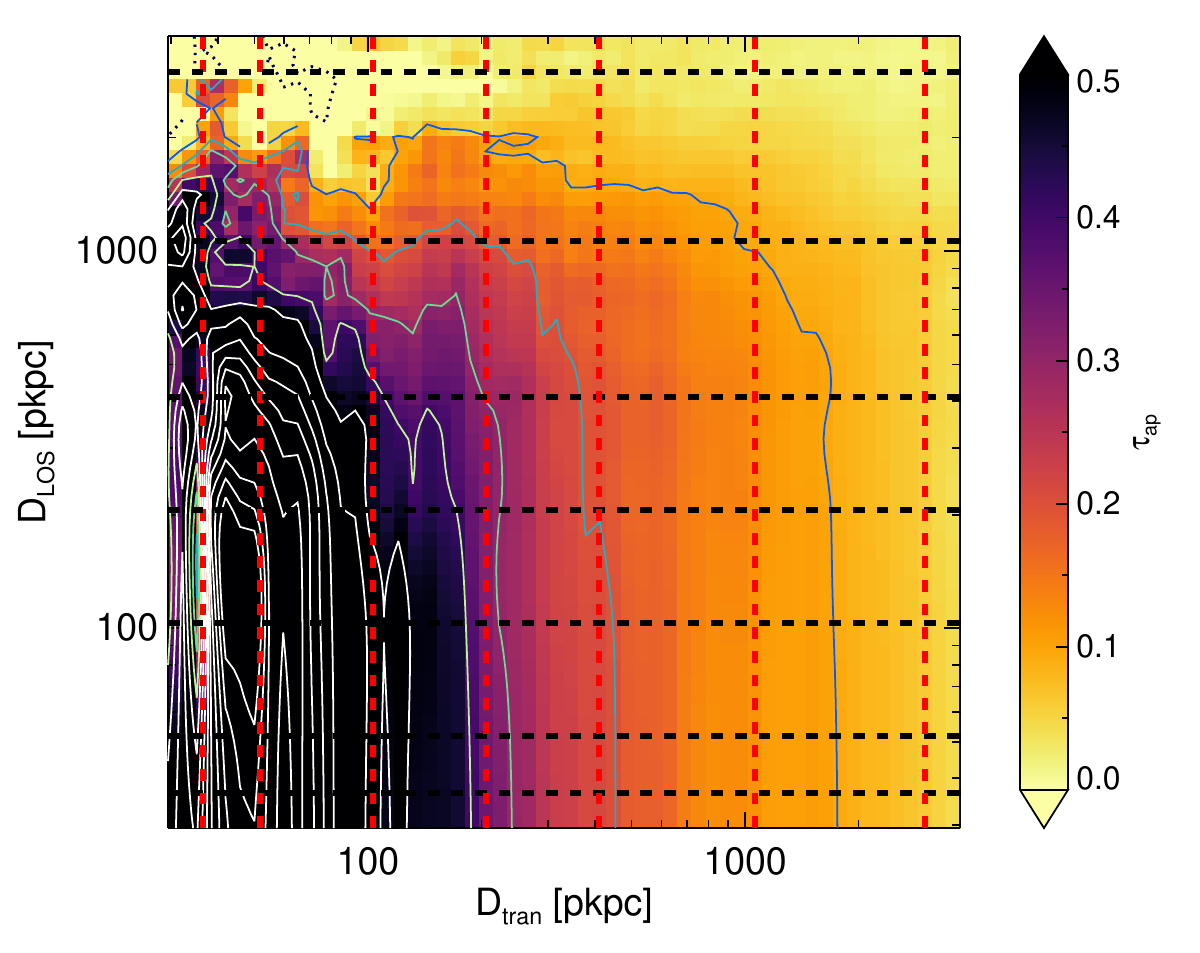}
\caption{Same as the top-left plot of Figure \ref{fig:2dabs}, except here the y-axis has been converted to line-of-sight distance assuming that $v_\mathrm{LOS}$ 
is entirely due to Hubble expansion. 
The two axes have been adjusted so that any departures from symmetry indicate the presence of peculiar motions of gas with respect to the Hubble flow. 
The red vertical dashed lines and black horizontal dashed lines correspond to the locations of extracted profiles in Figure \ref{fig:dtrandlos}.}
\label{fig:2dabs_log}
\end{figure}
\begin{figure}
\centering
\includegraphics[width=8.5cm]{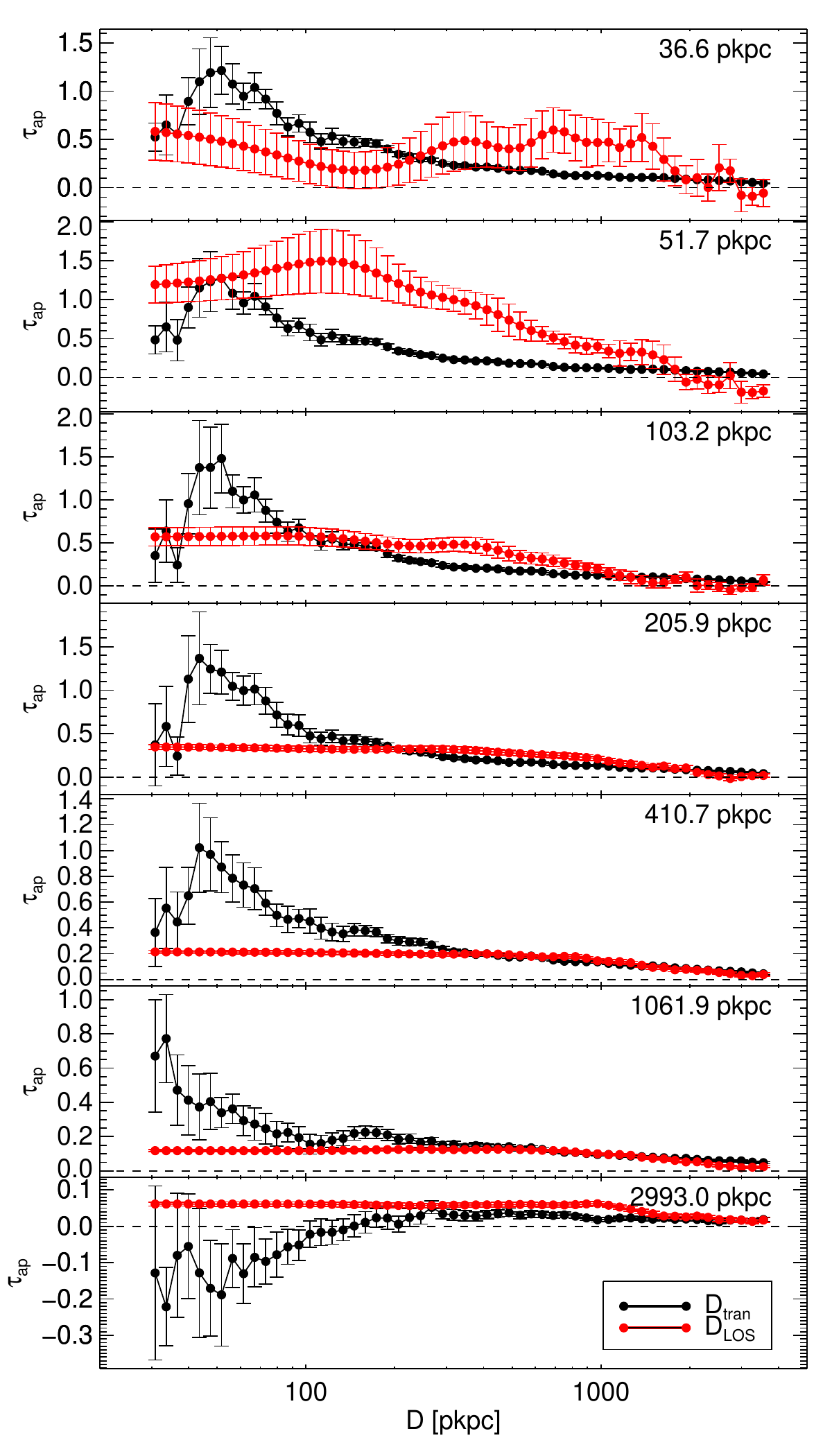}
\caption{Comparison of the extracted $\tau_\mathrm{ap}$ profiles (see Figure~\ref{fig:2dabs_log}) in the $D_\mathrm{tran}$ (black) and $D_\mathrm{LOS}$ (red) directions. Numbers on the top-right corner of the plots are the distances in pkpc to the center of the 
galaxy in the direction orthogonal to that over which the profile is extracted.}
\label{fig:dtrandlos}
\end{figure}

In order to highlight the effects of peculiar velocities on the observations, we resampled the $\tau_\mathrm{ap}$ map in the line-of-sight direction, 
assuming that $v_\mathrm{LOS}$ is due entirely to Hubble expansion, i.e., that $D_\mathrm{LOS} = v_\mathrm{LOS}/H(z)$. Figure \ref{fig:2dabs_log} shows 
the re-sampled $\tau_\mathrm{ap}$ map of the KGPS-Full sample with matching $D_\mathrm{LOS}$ and $D_\mathrm{tran}$ axes. 
Regions with $D>100 \mathrm{~pkpc}$ closely resemble similar results compiled using QSO sightlines from KBSS \citep{rudie12a, rakic12, turner14}.

Figure \ref{fig:dtrandlos} shows direct comparisons of $\tau_\mathrm{ap}$ profiles extracted along the two axes (line of sight and transverse to the line of sight) for seven representative distances, randomly chosen from the figure. 
In the absence of peculiar velocities and redshift errors,  
the $\tau_{\rm ap}$ profiles at fixed $D_\mathrm{LOS}$ and $D_\mathrm{tran}$ would be identical.   
In reality, the $D_{\rm LOS}$ and $D_{\rm tran}$ profiles differ significantly: e.g., 
at $D\lesssim 500 \mathrm{~pkpc}$, the $D_\mathrm{tran}$ cuts have similar shapes, while the corresponding $D_\mathrm{LOS}$ cuts on the same scales
vary considerably. 
Most obvious is the ``finger of God'' elongation, known as the ``Kaiser effect'' \citep{kaiser87}, along the line of sight due to peculiar velocities of gas with $D_{\rm tran} \simlt 100$ pkpc. However,
there is also a more subtle signature of infall in Figure \ref{fig:2dabs_log} that is evident only at the lowest two contours in $\tau_{\rm ap}$, 
manifesting most clearly as a compression in
the $D_{\rm LOS}$ direction between
$D_{\rm tran} \simeq 100$ and $D_{\rm tran} \simeq 50$ pkpc  
as also noted previously. 

At $D \gtrsim 500 \mathrm{~pkpc}$ (upper-right corner of 
Figure \ref{fig:2dabs_log}), the $D_\mathrm{tran}$ and $D_\mathrm{LOS}$ cuts begin to match, 
indicating diminishing redshift-space distortion. 

In addition to the KBSS-based studies mentioned above, there have been a number of other studies, focused
on establishing the cross-correlation between ``absorbers'' along QSO sightlines and galaxies in the surrounding volumes at $z < 1$ (e.g., 
\citealt{chen05,ryanweber06,chen09,tejos14}), $0.7 \simlt z \simlt 1.5$ (\citealt{shone10}), and $z \simeq 2-3$ (\citealt{adelberger03,adelberger05,tummuangpak14,bielby17}).  Care must be exercised in comparing the results
of these studies with those presented in this work, for two reasons: first, the aforementioned studies have cast the results
in terms of auto- and cross-correlation functions, in co-moving coordinates, so that all length scales in the present work
must be suitably adjusted\footnote{In comoving
coordinates, the transverse scale for the 2D map in (e.g.) Figure~\ref{fig:2dabs_unfold} would need to be multiplied
by a factor of $\simeq 3.2$ for direct comparison, i.e., our measurements extend to transverse a transverse scale of $D_{\rm tran, com} \simeq 12.8 h_{\rm 70}^{-1}$ cMpc.}; second, most of these studies were
not well-suited to measuring galaxy-gas correlations 
on scales smaller than $\sim 1$ cMpc ($\sim 300$ pkpc at $\langle z \rangle = 2.2$) either due to redshift errors ($z > 2$ studies), paucity of absorber-galaxy pairs, or both. We note that the redshift space distortion in the KGPS map 
 -- both the ``finger of God'' at $D_{\rm tran} \simlt 50$ pkpc and the ``compression'' on scales of
$50 \simlt D_{\rm tran} \simlt 200$ pkpc, would probably have gone unrecognized. Otherwise, on scales larger than $\sim 1$ cMpc, there
is reasonable agreement: for example, one of the conclusions of \cite{tejos14} is that there is little evidence for either infall or
outflows of gas with peculiar velocities larger than $\sim 120$ \kms; our results are in agreement - recall that we found that infall
velocities of $\sim 100$ \kms\ gave the best fits for our simple model. However, there are much larger peculiar velocities acting on
scales smaller than $\simeq 150$ pkpc (0.5 cMpc), within the CGM.

Finally, we note that the values of $\tau_\mathrm{ap}$ for the $D_{\rm tran}$ profile evaluated    
at $D_\mathrm{LOS} = 2993 \mathrm{~pkpc}$ (black points and curve in the bottom panel of Figure~\ref{fig:dtrandlos}) are 
consistently negative for $D_\mathrm{tran} \lesssim 100 \mathrm{~pkpc}$. 
Because most of the points are correlated and each is individually consistent with $\tau_{\rm ap} = 0$, we believe the most likely culprits are 
sample variance and continuum uncertainties exacerbated by the relatively small sample size. Meanwhile, as discussed in \S \ref{sec:asymmetry}, it is possible that \lya\ emission from the CGM of the foreground galaxy has a significant effect on the measured strength of \lya\ absorption in the
spectrum of faint background continuum sources, and that the magnitude of the effect would be largest at small $D_{\rm tran}$. 
Scattered \lya\ emission in down-the-barrel spectra of the galaxies tends to be dominated by a redshifted component, 
and it is not unusual for the red wing of \lya\ to extend well beyond $v_{\rm LOS} \sim 600$ \kms; if the \lya\ emission strength exceeds
the flux removed by absorption against the continuum of the background source, it is possible in principle to have ``negative'' net \lya\ absorption. 
  
\section{Summary} 
\label{sec:summary}

In this paper, we assembled 2862 spectroscopically-identified galaxies from KBSS ($1.9 \simlt z \simlt 3.3$; $\langle z \rangle = 2.51$) 
into $\sim 200,000$ unique angular pairs of physically-unrelated galaxies;  
we then used the spectra of the background galaxies to probe the \ion{H}{I} content of the CGM/IGM of the foreground galaxies  
as a function of projected physical distance over the range of $30 \simlt D_{\rm tran}/{\rm pkpc} \simlt 4000$. 

To maximize the utility of composite spectra for mapping the strength and kinematics of  \lya\ absorption surrounding galaxies, 
we used the $\simeq 45$\% of galaxies with precise and accurate nebular emission line measurements to re-calibrate 
the relationship 
between the galaxy systemic redshift from $z_{\rm neb}$ and redshifts measured using spectral 
features in the rest-frame far-UV, $z_{\mathrm{Ly}\alpha}$ and/or 
$z_\mathrm{IS}$, which are biased by the effects of outflowing gas. 
We created composite spectra, stacked in bins of $D_{\rm tran}$, of background galaxy spectra shifted to the rest frame of the corresponding foreground galaxy in each pair. The very large number of distinct galaxy-galaxy pairs allowed us to construct a well-sampled ensemble map of neutral H surrounding the average
foreground galaxy in the sample. In particular, the improved sampling within $D_\mathrm{tran} \simlt 100 \mathrm{~pkpc}$ is crucial in probing the effects of galaxy-scale outflows on the \ion{H}{I} kinematics. 

We compared the observed Ly$\alpha$ map with cosmological zoom-in simulations and with a simple analytic model of outflows and infall surrounding
a galaxy hosted by a dark matter halo of mass $M_{\rm h} \simeq 10^{12}$ M$_{\odot}$. 
The principal results are summarized below:  

\begin{enumerate}

\item The Ly$\alpha$ equivalent width as a function of impact parameter $D_{\rm tran}$ can be approximated as  
a power law,  $\mathrm{\wlya\ } \propto D_\mathrm{tran}^{-0.4}$ over the full range observed, but there are at least three
distinct impact parameter zones for the run of \wlya\ vs. $D_{\rm tran}$: $D_{\rm tran} < 100$ pkpc (slope $\simeq -1.0$), $100 < D_{\rm tran}/{\rm pkpc}  \simlt 300$ (slope $\simeq 0$), and 
$300 \simlt D_{\rm tran}/{\rm pkpc}  \simlt 2000$ (slope $\simeq -0.5$) (\S\ref{sec:ew}).

\item The 2-D map of apparent Ly$\alpha$ optical depth $\tau_{\rm ap}$ (Figures~\ref{fig:2dabs_unfold}, \ref{fig:2dabs})
in $v_\mathrm{LOS}$-$D_\mathrm{tran}$ space exhibits a dense ``core'' at $|v_\mathrm{LOS}|< 500 \mathrm{~km~s}^{-1}$ and $D_\mathrm{tran} < 100 \mathrm{~pkpc}$,
that transitions to a diffuse component that becomes broader with increasing $D_{\rm tran}$.  
The maps using the full KGPS sample and the sub-sample for which nebular redshifts are available for the foreground galaxy show consistent features. (\S\ref{sec:kinematics})

\item Comparison of the $\tau_\mathrm{ap}$ map with the projected $N_\mathrm{HI}$ map of a simulation with similar halo mass to the observed sample from the FIRE project shows that the dense ``core'', and the outer ``envelope''  match remarkably well in both $v_\mathrm{LOS}$ and $D_\mathrm{tran}$. 
(\S\ref{sec:discussion_zoomin})

\item A simple, two-component analytic model with radial inflow and outflow 
can reproduce the general features of the observed 2-D $\tau_\mathrm{ap}$ map; however, the model fails to fit 
abrupt features in the \lya\ absorption kinematics at particular values of 
$D_{\rm tran}$, which clearly indicate a level of complexity 
that is not captured by the adopted model parametrisation. 
(\S\ref{sec:model})

\item The $\tau_\mathrm{ap}$ map exhibits significant asymmetry in velocity relative to the galaxy systemic redshifts, the strongest of which is at 
projected distances $50\lesssim D_\mathrm{tran}/\mathrm{pkpc}\lesssim 200$ and $200 \kms \lesssim |v_\mathrm{LOS}| \lesssim 500 \kms$. 
The asymmetry is significant, and is unlikely to be explained by unaccounted-for
systematic errors in galaxy redshifts. We suggest that the most plausible explanation is contamination of the \lya\ absorption signal by diffuse \lya\ {\it emission} associated with the extended \lya\ halo
of the foreground galaxy scattering into the slit apertures used to measure the spectra of the background galaxies. Estimates of the expected effect of \lya\ emission
contamination on measurements of \wlya\ suggest that it should be non-negligible for any sample that uses background sources that are comparably bright to the foreground galaxies being probed. (\S\ref{sec:asymmetry})
 
\item Matching the $\tau_\mathrm{ap}$-$D$ profile in $D_\mathrm{LOS}$ and $D_\mathrm{tran}$ axes shows strong redshift-space distortion at small $D$, and similar $\tau_\mathrm{ap}$ profiles in the two directions at large $D$, 
suggesting that the redshift-space distortion becomes less prominent as $D\gtrsim 500 \mathrm{~pkpc}$. (\S\ref{sec:discussion_zdistortion})

\item The range of projected distance $50 \lesssim D_\mathrm{tran}/\mathrm{pkpc} \lesssim 150$ marks a transition in both the \wlya-$D_\mathrm{tran}$ 
relation, and the 2-D $\tau_\mathrm{ap}$ map, suggesting that outflows gradually cede to infall as the dominant source of absorbing gas within that range.  
This inference is also supported by the narrow velocity profile of absorption in the same range of $D_{\rm tran}$, consistent with 
the effective resolution of the observed map, indicating a local minimum dispersion in $v_{\rm LOS}$ where the confluence of infall, outflow, and Hubble expansion create a
caustic-like feature in $v_\mathrm{LOS}$ space. (\S\ref{sec:model})
\end{enumerate}

Our results for the spatial distribution and kinematics of \ion{H}{I} could be compared with cosmological zoom-in simulations 
to test additional physical effects or feedback prescriptions \citep[e.g., ][]{hummels2013}, and it is certainly possible to devise more realistic semi-analytic CGM models. Meanwhile, the clear distinction between outflow and inflow in the $v_\mathrm{LOS}$-$D_\mathrm{tran}$ space and the transition $D_\mathrm{tran}$ between the two provides vital information on the interaction between galaxies and their surroundings during the periods of rapid galaxy growth. 

Similar observations using metal-line absorption from CGM gas can provide information on gas with a wider range of physical conditions, 
further constraining the distribution, kinematics, and physical conditions of baryons around galaxies. 
It will also be intriguing to compare the \ion{H}{I} kinematics in absorption and emission {\it for the same galaxies} with aid of deep IFU spectroscopy,
which will provide more nuanced view of the structure and kinematics of the CGM and how it affects the radiative transfer of \lya. We
are pursuing both of these approaches in forthcoming work. 

\section*{Acknowledgments}
Based on data obtained at the W. M. Keck Observatory, which is operated as a scientific partnership among the California Institute of Technology, the University of California, and the National Aeronautics and Space Administration. The Observatory was made possible by the generous financial support of the W. M. Keck Foundation. 

This paper has included data obtained using Keck/LRIS \citep{oke95,steidel04,rockosi10} and Keck/MOSFIRE \citep{mclean10,mclean12}. We thank the W.M. Keck Observatory staff for their assistance with the observations over two decades.

The following software packages have been crucial to preparing for this paper: the IDL Astronomy User's Library\footnote{\href{https://idlastro.gsfc.nasa.gov/}{https://idlastro.gsfc.nasa.gov/}}, the Coyote IDL library\footnote{\href{http://www.idlcoyote.com/}{http://www.idlcoyote.com/}}, Astropy \citep{price18}, Emcee \citep{foreman13}, the yt project \citep{turk11}, and Trident \citep{hummels2017}.  

This work was supported in part by grant AST-1313472 from the U.S. NSF, and by a grant from the Caltech/JPL President's and Director's Program. CAFG was supported by NSF through grants AST-1517491, AST-1715216, and CAREER award AST-1652522, by NASA through grant 17-ATP17-0067, by STScI through grants HST-GO-14681.011, HST-GO-14268.022-A, and HST-AR-14293.001-A, and by a Cottrell Scholar Award from the Research Corporation for Science Advancement.

We would like to thank the anonymous referee for providing valuable feedback. YC would like to thank Hongjie Zhu, for her continuous encouragements when preparing for this paper. We would like to acknowledge Yiqiu Ma, E. Sterl Phinney, and Mateusz Matuszewski, for their constructive discussions. Finally, we thank collaborators Kurt L. Adelberger, Matthew P. Hunt, David R. Law, Olivera Rakic, and Monica L. Turner for their contributions to the KBSS survey over the course of nearly two decades. 

\section*{Data Availability}
The processed data underlying this article are available on the KBSS website (\href{http://ramekin.caltech.edu/KBSS}{http://ramekin.caltech.edu/KBSS}) with DOI: \href{http://dx.doi.org/10.22002/D1.1458}{10.22002/D1.1458}.

\bibliographystyle{mnras}
\bibliography{main.bib} %

\appendix

\section{Stacking Method}
\label{app:stack}

\begin{figure*}%
\centering
\includegraphics[width=0.3\textwidth]{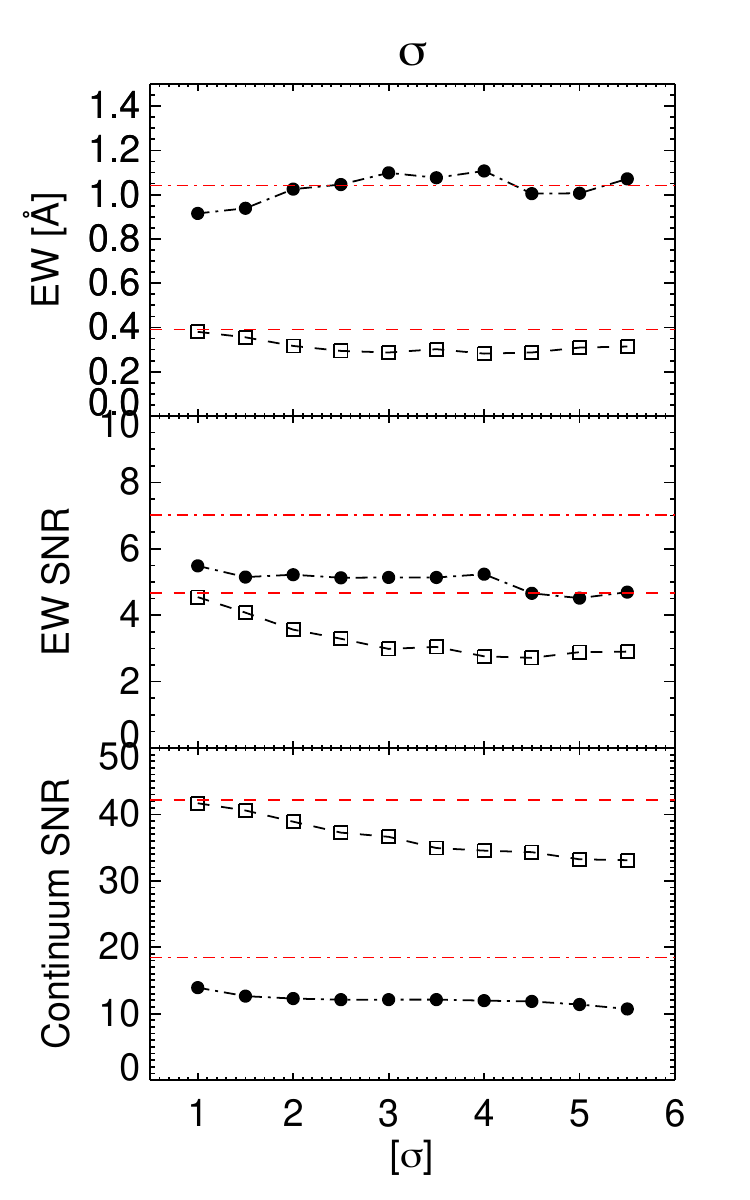}
\includegraphics[width=0.3\textwidth]{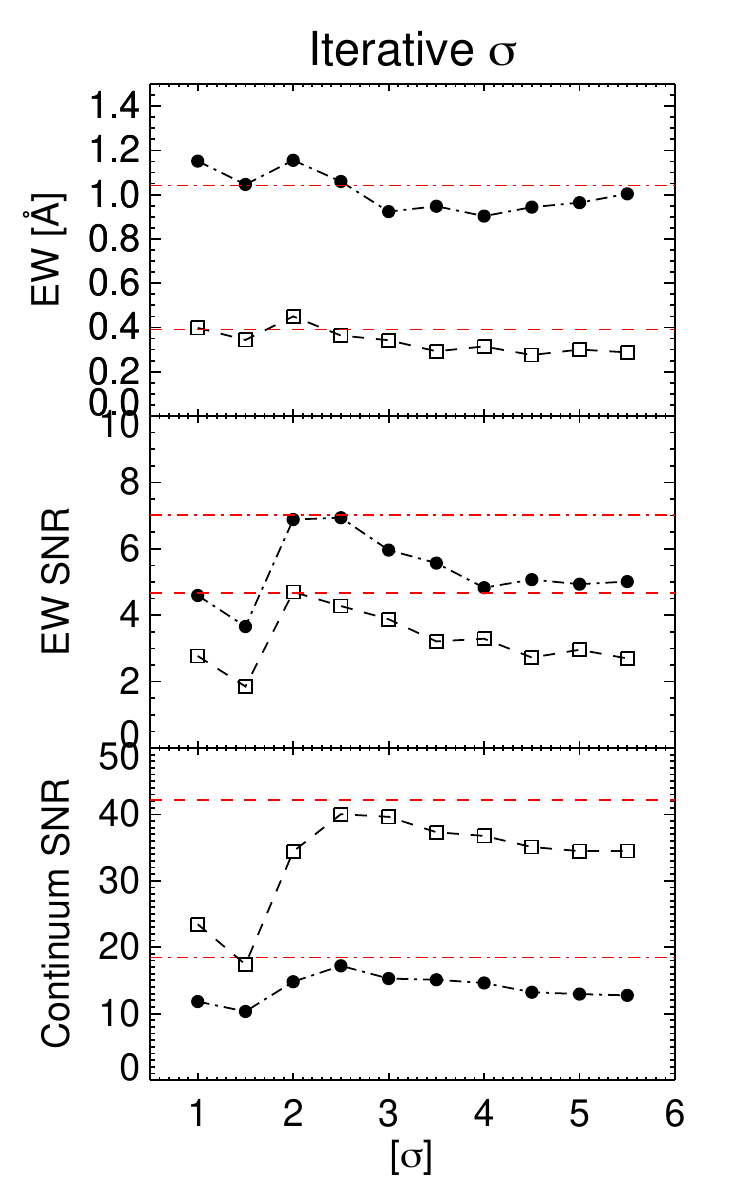}
\includegraphics[width=0.3\textwidth]{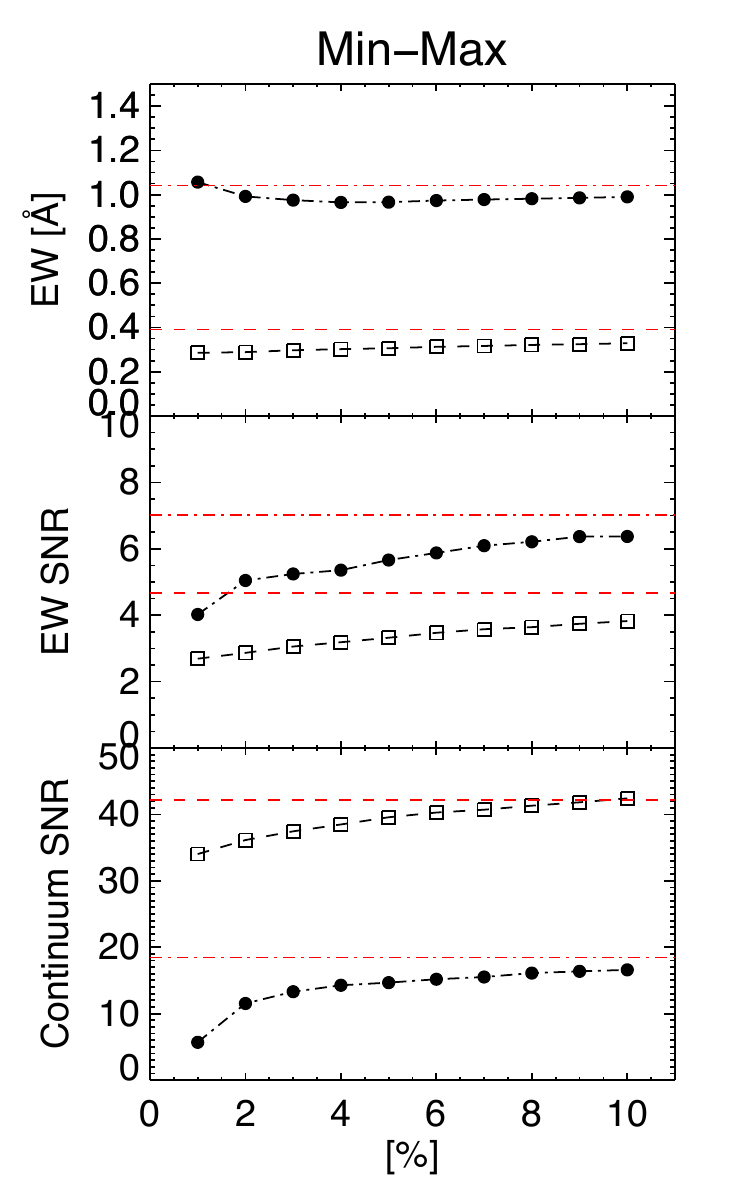}
\caption{ The impact of different stacking methods on \wlya\ measurements: each panel shows \wlya ({\it top});  the SNR of \wlya ({\it middle});  
the SNR  of the continuum near \lya ({\it bottom}). The 3 panels (left to right) show the results for sigma clipping, \textit{iterative} sigma clipping, 
and min-max rejection. For min-max rejection, the x-axis shows the fraction of data points rejected from each side of the sample distribution. 
Within each panel, the black filled points correspond to a sample with $(D_{\rm tran}/{\rm pkpc}) \le  100$ and the skeletal boxes to a sample with 
$500 \le (D_{\rm tran}/{\rm pkpc}) \le 550$; the horizontal red lines show the corresponding values for a median stack with no other rejection algorithm applied.}
\label{fig:snr}
\end{figure*}

To optimise the SNR of stacked spectra, we tested several different stacking methods: 1) sigma-clipped mean, 2) \textit{iterative} sigma-clipped mean, 3) min-max clipped mean, and 4) median. For all cases, the error was estimated from bootstrap resampling of the sightlines within each bin using 2000 realisations. 
We select two $D_\mathrm{tran}$ bins -- one with $(D_{\rm tran}/{\rm pkpc}) \le  100$ and the other with $500 \le (D_{\rm tran}/{\rm pkpc}) \le 550$ 
from the KGPS-Full sample to demonstrate the effect of clipping on strong and weak \lya\ absorption in Figure~\ref{fig:snr}. \wlya\ was measured in the same 
way as in \S\ref{sec:ew}. The SNR of the continua is defined as the median flux density divided by the median error within two windows of rest-wavelength, 
[1207,1211] \AA ~and [1220,1224] \AA. 

While keeping most of the data points in the mean stacks does not alter the value of \wlya\ for either strong or weak lines, it significantly affects the SNR. 
On the other hand, achieving SNR similar to that obtained for median stacks requires more aggressive rejection ($\sim 2 \sigma$ or $\sim 8\%$) for the 
sigma clipping and min-max rejection, respectively. 
For iterative sigma clipping, we find that the optimal SNR is achieved at $\sim 2.5 \sigma$, where the SNR of \wlya\ for weak 
lines and that of the continuum near strong lines are close to those of the median stack. 
Otherwise, the SNR remains $\sim 10\%$ smaller than for spectra combined using a median stack; this suggests that the data values at each wavelength pixel 
are not normally distributed, and therefore outliers are not easily removed by sigma clipping. Given that the spectral continuum of the pre-stacked spectra are not normalised, part of this is contributed by the variations of the stellar continuum in the spectra of the background galaxies. 
In addition, we find that in some regimes, the systematic effects of clipping methods have the opposite signs for strong and weak \wlya\ (top panel of Figure~\ref{fig:snr}). 
Since the median stack consistently returns values close to the optimal clipped mean, both in terms of measured \wlya\ and SNR(\wlya), 
we subsequently adopted the median stack for producing all composite spectra used in this paper.

\section{Effective Spectral Resolution in Composite Spectra} 
\label{app:resolution}

Having a reliable measurement of the effective spectral resolution of the stacked spectra is crucial because some of the features in the 2D map are only marginally resolved. We stack spectra from the KGPS sample in the rest frame of $z_\mathrm{bg}$ and use the strong down-the-barrel UV features to estimate the effective resolution. We used the \ion{C}{ii} $\lambda$1334 line since it does not suffer from significant contamination from other lines, and has the most consistent $W_\lambda$ (RMS $\lesssim 5\%$) in the stacks of different subsamples.  

Assuming that the observed line width can be expressed as,
\begin{eqnarray}
\label{eq:resolution}
\sigma_\mathrm{obs} & = & \sqrt{\sigma_0^2+\sigma_\mathrm{eff}^2} \\
 & = & \sqrt{\sigma_0^2 + \sigma_\mathrm{inst}^2 + f_\mathrm{zuv}\sigma_\mathrm{zuv}^2},
\end{eqnarray}
where $\sigma_0$ is the intrinsic width, $\sigma_\mathrm{eff}$ is the effective velocity resolution, $\sigma_\mathrm{inst}$ is the  resolution of the spectrograph, and $f_\mathrm{zuv}$ is the fraction of spectra in stacks that use calibration from $z_{\mathrm{Ly}\alpha}$ and $z_\mathrm{abs}$ as $z_\mathrm{sys}$, and $\sigma_\mathrm{zuv}$ is the uncertainty of the calibration in velocity space. In this case, $\sigma_\mathrm{inst}$ and $\sigma_\mathrm{zuv}$ can be estimated separately. We assume that the redshift uncertainty is negligible when using $z_\mathrm{neb}$ as $z_\mathrm{sys}$.

\begin{figure}
    \centering
    \includegraphics[width=\columnwidth]{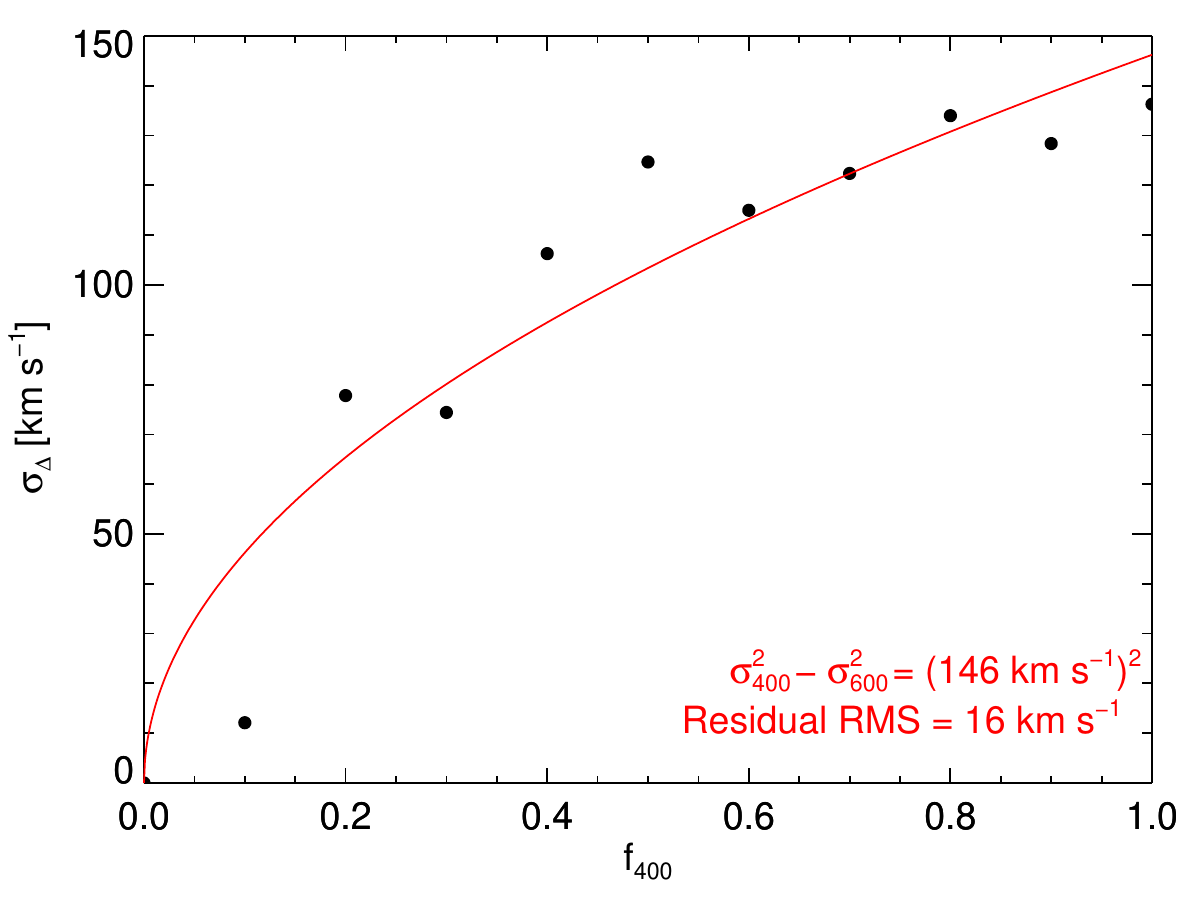}
    \caption{Width of the Gaussian kernel ($\sigma_\Delta$) used in convolution to match the line profile of \ion{C}{ii} $\lambda$1334, between stacks with pure 600/4000 spectra and ones with a fraction of the 400/3400 grism spectra ($f_{400}$). The red line is the best-fit model using Equation \ref{eq:sigdelta1}.}
    \label{fig:f400}
\end{figure}

Since essentially all our LRIS observations used either the 400/3400 or 600/4000 grisms, $\sigma_\mathrm{inst}$ can be further divided into, 
\begin{eqnarray}
\label{eq:sigmainst}
    \sigma_\mathrm{inst} = \sqrt{f_{400}\sigma_{400}^2+f_{600}\sigma_{600}^2}, 
\end{eqnarray}
where $\sigma_{400}$ and $\sigma_{600}$ are the instrument resolution for the 400/3400 and 600/4000 grisms, and $f_{400}$ and $f_{600}$ are the fractions of spectra observed with the two grisms in the stack. To measure the absolute values of $\sigma_{400}$ and $\sigma_{600}$, we constructed 11 samples. Each has different ratio of $f_{400}/f_{600}$, ranging from 100\% of the spectra observed by the 400/3400 grism to 100\% observed by the 600/4000 grism. All are based on exactly the same 95 objects, which have been observed with both 400/3400 and 600/4000 grisms, and have their systemic redshift measured from nebular lines. We fit the 10 stacks with non-zero contribution of 600/4000 spectra by convolving the 100\% 600/4000 stack with a gaussian kernel, whose standard deviation ($\sigma_\Delta$) is the only free parameter. The best-$\chi^2$ fits of $\sigma_\Delta$ are summarized in Figure \ref{fig:f400}. Based on Equation \ref{eq:sigmainst}, 
\begin{eqnarray}
\label{eq:sigdelta1}
    \sigma_\Delta = \sqrt{f_{400}(\sigma_{400}^2-\sigma_{600}^2)}.
\end{eqnarray} 
Therefore, $\sqrt{\sigma_{400}^2-\sigma_{600}^2} = 146 \mathrm{~km~s}^{-1}$, obtained by fitting the equation above to the measured points in Figure \ref{fig:f400}. 

\begin{figure}
    \centering
    \includegraphics[width=\columnwidth]{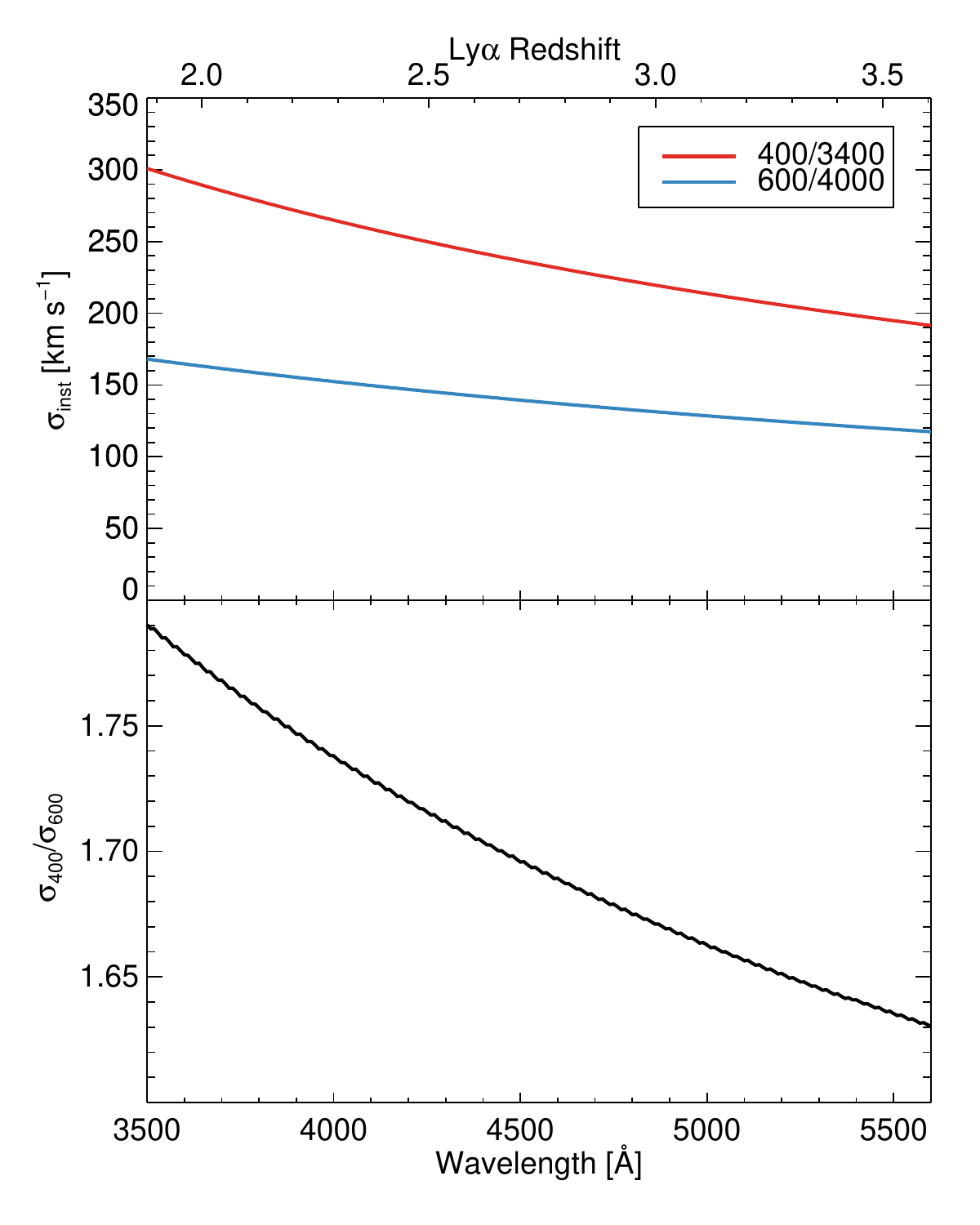}
    \caption{Spectroscopic resolution for 400/3400 and 600/4000 grisms plotted as 1-$\sigma$ error in velocity space. Top: Absolute value estimated from arc spectra with 1.2-arcsec slit width. Bottom: Ratio between $\sigma_{400}$ and $\sigma_{600}$, which remains unchanged with varying object size in slits.}
    \label{fig:resolution}
\end{figure}

Another ingredient for estimating the absolute value of $\sigma_{400}$ and $\sigma_{600}$ is the ratio of $\sigma_{400}/\sigma_{600}$. Figure \ref{fig:resolution} shows the instrument resolution of the two grisms for an object with uniform illumination of a 1.2 arcsec slit, estimated from the arc spectra taken during afternoon calibrations. For a median redshift of 2.4 in our DTB stacks, the observed wavelength of \ion{C}{II} $\lambda$1334 is 4559 \AA, which gives $\sigma_{400}/\sigma_{600} \sim 1.69$. Therefore, $\sigma_{400}=181 \mathrm{~km~s}^{-1}$, $\sigma_{600}=107 \mathrm{~km~s}^{-1}$. This suggests a typical galaxy size of FWHM $=0.9$ arcsec. We then convert this measurement to the resolution that would be obtained for Ly$\alpha$ at $z_\mathrm{med} = 2.2$. Assuming the $\lambda$-dependence of $\sigma$ as in Figure \ref{fig:resolution}, the velocity resolutions are $\sigma_{400}(\mathrm{Ly}\alpha) = 211 \mathrm{~km~s}^{-1}$ ($R=604$), and $\sigma_{600}(\mathrm{Ly}\alpha) = 121 \mathrm{~km~s}^{-1}$ ($R=1055$). 

To test the reliability, we also made a stack with the same objects and combined all 400/3400 and 600/4000 spectra together, weighted by the number of objects instead of the number of spectra to reproduce the scenario in \S \ref{sec:stacking}. With 45\% of the weight given to 400/3400 spectra, this yields a $\sigma_\Delta$ of 98 $\mathrm{~km~s}^{-1}$, consistent with Equation \ref{eq:sigdelta1} within 10\%. For our typical foreground stacks in \S \ref{sec:stacking}, $\sim 70\%$ of weight is contributed by 400/3400 spectra, and 600/4000 spectra make up the remaining 30\%. Therefore, the effective resolution, $\sigma_\mathrm{inst}(\mathrm{Ly}\alpha) = 189 \mathrm{~km~s}^{-1}$.

\begin{figure}
    \centering
    \includegraphics[width=\columnwidth]{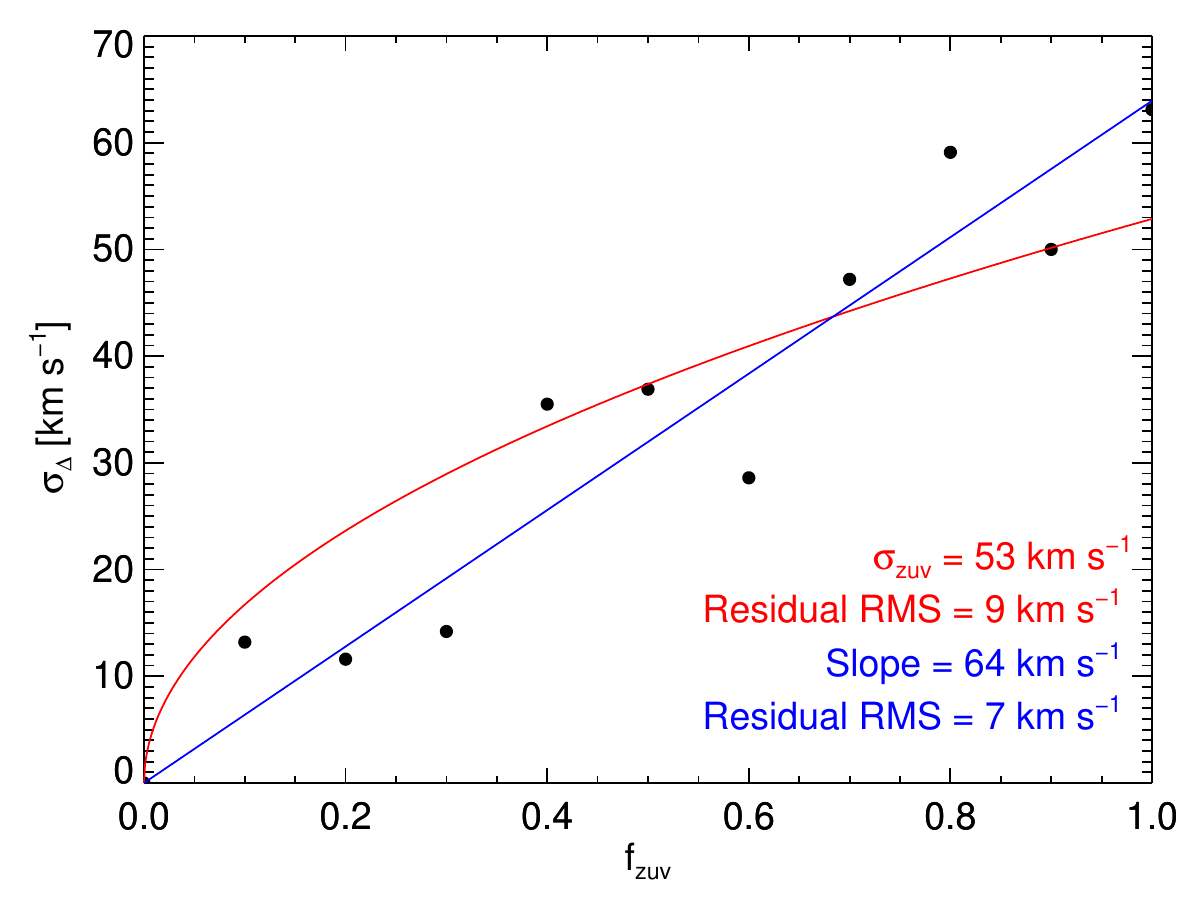}
    \caption{Similar to Figure \ref{fig:f400}, but the x-axis is the fraction of objects in the stacks whose $z_\mathrm{sys}$ is determined using rest-UV spectral features ($f_\mathrm{zuv}$). The red curve is the best-fit of Equation \ref{fig:resolution}, and the blue line is a simple linear fit. }
    \label{fig:fzuv}
\end{figure}

The absolute value of $\sigma_\mathrm{zuv}$ in Equation \ref{eq:resolution} can be obtained using a similar method. We constructed 11 stacks with 382 objects that have redshift measurements with both nebular and rest-UV features. In these stacks, we randomly select $f_\mathrm{zuv}$ fraction of objects and use the calibrated UV redshift ($z_\mathrm{UV}$) instead of the precise $z_\mathrm{neb}$. We use a similar method to fit the 10 stacks with non-zero $f_\mathrm{zuv}$ by convolving a Gaussian kernel with the 100\% $z_\mathrm{neb}$ stack. The resulting kernel width as a function of $f_\mathrm{zuv}$ is shown in Figure \ref{fig:fzuv}. As can be seen in the figure, the relationship between $\sigma_\Delta$ and $f_\mathrm{zuv}$ can be better represented by a linear function rather than a square-root function, and the redshift errors associated with the use of UV features to estimate $z_\mathrm{sys}$ are smaller than derived in \S \ref{sec:redshift}. 
We suspect that the reason for the apparent discrepancy is that a large fraction of the derived $\sigma_{\rm z}$ in \S\ref{sec:redshift} 
is caused by noise in the measurement of individual UV features in the spectra of individual galaxies, whereas the calibration between
\zneb\ and $z_{\rm UV}$ effectively averages out such random noise. Nevertheless, we provide results for the two separate fits and use both in our final estimation. For a typical $f_\mathrm{zuv} = 50\%$ in our foreground stacks for the KGPS-Full sample, $\sigma_\Delta = 37$ (square-root) or 32 (linear) $\mathrm{km~s}^{-1}$. Putting all components together in Equation \ref{eq:resolution}, the effective resolution for the KGPS-Full sample is $\sigma_\mathrm{eff} (\mathrm{Ly}\alpha) = 192 \mathrm{~km~s}^{-1} $ for both the square-root and the linear fit. 

In summary, the effective spectral resolution for the KGPS-$z_\mathrm{neb}$ and KGPS-Full sample is nearly identical: 
$\sigma_\mathrm{zneb} = 189 \mathrm{~km~s}^{-1}$ ($R=676$) and $\sigma_\mathrm{full} = 192 \mathrm{~km~s}^{-1}$ ($R=665$), respectively. 
The uncertainty associated with the estimates is $\sim 15\mathrm{~km~s}^{-1}$, based on the rms residuals of the fits for $\sigma_\Delta$-$f_{400}$ and $\sigma_\Delta$-$f_\mathrm{zuv}$ relations. 


\bsp	%
\label{lastpage}

\end{CJK*}
\end{document}